# Stoichiometric cluster learning for few-shot property prediction of multi-ionic integrated energetic materials

Ming-Yu Guo, Wei-Jia Zou, Yu Shang*, Wei-Xiong Zhang*

*MOE Key Laboratory of Bioinorganic and Synthetic Chemistry, School of Chemistry, IGCME, Sun Yat-sen University, Guangzhou, 510275, China*

*Corresponding authors are shangy8@mail.sysu.edu.cn and zhangwx6@mail.sysu.edu.cn*

## Abstract

Multi-ionic materials pose a distinct representational challenge in machine learning-driven materials design. Different from single-molecule or composition-based materials, their properties arise from how charged building blocks aggregate into specific assemblies. Here, we show how pretrained machine-learned interatomic potentials (MLIPs) can bypass full crystal-structure prediction and support pre-synthesis screening from stoichiometric ionic clusters using multi-ionic integrated explosives (MIXs) as a synthesis-facing example. This strategy combines a stoichiometric ionic-cluster representation, which represents each candidate material by a non-periodic, stoichiometry-preserved formula-unit cluster, with multi-task fine-tuning (MT-FT), which adapts a pretrained atomistic backbone while retaining the energy–force objective as physical regularization for the sparse detonation-velocity labels. With the pretrained backbone regularized by MT-FT, this surrogate provides a cross-validated screen across only 25 structurally curated perovskite-type energetic materials (PEMs) with experimentally derived Kamlet–Jacobs (K–J) detonation velocities. Representation probes show that the learned descriptors implicitly retain site-aware ionic organization, density information, and coarse packing compatibility, implying why non-periodic clusters can remain predictive before full crystal structures are known. The surrogate extends known PEMs chemistry to three newly synthesized $ABX_4$ materials with both unseen $ABX_4$ stoichiometry and an unseen ethylenediammonium B-site cation, yielding three-point concordance with K–J reference velocities and a mean absolute error (MAE) of 92 $m \cdot s^{-1}$ without retraining. Together, these results establish stoichiometry-preserved cluster learning as a synthesis-facing screening strategy for data-scarce multi-ionic materials.

## Introduction

Machine-learning representations for materials usually encode the level at which a candidate is specified. Typical input levels include elemental compositions for alloy-like or structure-agnostic screening, isolated molecular graphs for molecular discovery, and periodic graphs for crystalline solids whose structures are already known or proposed.[1–10] The multi-ionic materials considered here are crystalline solids after growth, but their chemically meaningful identity is not captured by any single ionic component. It depends on which charged building blocks are paired, how they are assigned to site roles, and how they co-assemble locally. Composition-level inputs preserve the overall formula but collapse molecular-ion identity and site assignment. Isolated molecular graphs preserve each covalent component but remove the inter-ionic relations that distinguish one material from another. Periodic graphs preserve these relations only after a crystal structure has been resolved or proposed. Generating such a periodic candidate remains a nontrivial per-composition bottleneck for molecular ionic crystals.[11–14] Given that ionic components, stoichiometric clusters, and periodic crystals are successive aggregation states of the same material, the key question is whether the stoichiometric-cluster level can retain enough relational information for prediction before the final periodic structure is known.

This representation problem is especially consequential for energetic materials (EMs), where useful candidates must balance detonation performance, thermal stability, and sensitivities to impact and friction. Recent property-conditioned generative models for single-component energetic molecules show that molecular-level property targeting can steer enthalpy-of-formation distributions, but these workflows do not address the assembly-level representation problem posed by multi-ionic materials.[15] Multi-ionic integrated explosives (MIXs) place this multi-property design problem in a modular ionic setting, because each material is assembled from fuel-rich cations, oxidizing anions, and smaller cationic coformers rather than from one covalent energetic molecule.[16–25] Their design logic therefore shifts from optimizing one molecular graph to choosing charged building blocks and arranging them into stoichiometry- and geometry-constrained assemblies.[26] Reported members such as DAP-4 reach detonation performance comparable to conventional in-service EMs such as hexogen, also known as research department explosive (RDX), placing MIXs in a chemically relevant regime for synthesis-facing screening.[16,19,25,27] MIXs therefore provide a synthesis-facing test case in which candidates are proposed from ionic components before growth, whereas detonation performance depends on co-assembly in the eventual crystal.

Label scarcity makes this representation choice more consequential, because a small target set gives the model little opportunity to compensate for inputs that discard site identity or co-assembly information. Since the first perovskite-type MIX family was reported in 2018,[16] fewer than 40 related materials have been described, and only 25 in our curated set possess reliable crystal structures paired with consistent Kamlet–Jacobs (K–J) detonation velocities. We use K–J detonation velocity as the standardized endpoint because it remains closely tied to experimentally refined crystal data, while providing a cleaner cross-literature benchmark than properties such as friction sensitivity, decomposition temperature, or reported heats of explosion, which often vary strongly with morphology, handling history, and measurement protocol. Nor can this scarcity be removed simply by importing larger existing energetic-crystal datasets, which mainly cover organic molecular explosives rather than multi-ionic integrated energetic materials. In practice, models trained on the public Davis *et al.* 2024 detonation-property energetic-crystal release processed here (denoted as Davis2024) systematically underestimate detonation velocity on these multi-ionic integrated energetic materials by more than 1 $km \cdot s^{-1}$ on average (Extended Data Fig. 1 and Supplementary Table S1), because their building-block chemistry lies outside the training distribution.[28] This perovskite-type MIX regime therefore provides a focused test case in which both the input representation and the label scale are constrained.

These challenges motivate a screening strategy that can operate before crystal growth while transferring from larger, self-consistent atomistic datasets to sparse detonation labels. Pretrained machine-learned interatomic-potential (MLIP) backbones, both universal,[29–31] and domain-specific,[26,32] encode local chemical environments from broad energy–force supervision, allowing the same machinery to process non-periodic stoichiometric clusters as atomistic assemblies. We developed DeepEMs-LAM as a domain-specific MLIP with DPA3 backbone pretrained on the DeepEMs-25 density functional theory (DFT)-labeled high-temperature and high-pressure decomposition-trajectory dataset covering 20 energetic materials, giving an energetic-material prior over local chemical environments.[26,33] The multi-task fine-tuning (MT-FT) strategy then adapts this backbone to sparse detonation-velocity labels while retaining the auxiliary energy–force objective as physical regularization, a strategy supported by cross-property and multi-task studies showing that inductive transfer can remain useful in small-data or ultra-low-data regimes.[34–36] Whether the intermediate cluster level retains enough co-assembly signal for useful prediction without a resolved crystal remains open.

Here we propose a two-stage workflow to test this representation-sufficiency question (Fig. 1). First, we introduce a stoichiometric ionic-cluster representation that represents each candidate material as a non-periodic, charge-balanced formula-unit cluster, preserving the chemical identity of each A, B, and X building block together with their local coordination and mutual orientation, while reducing reliance on per-candidate full crystal-structure prediction. Second, we apply the established MT-FT protocol,[36] adapting DeepEMs-LAM[26] to detonation-velocity prediction in the data-scarce PEM regime. We evaluate the resulting surrogate through a ladder of transfer tests, first on literature OOD holdouts spanning $ABX_3$, $A_2BX_5$, and known $ABX_4$ MIXs, and then through wet-laboratory synthesis and single-crystal X-ray characterization of three new $H_2en^{2+}$-based $ABX_4$ materials, PEP, PEP-M, and PEP-H. Because the same modular material can be described at progressively higher structural resolution, from ionic components through stoichiometric clusters to periodic crystals, this workflow provides a bounded route for prioritizing related A–B–X combinations, stoichiometric branches, and coformer-ratio variants before crystal growth, provided that candidate materials preserve recognizable charged building blocks and local packing compatibility.

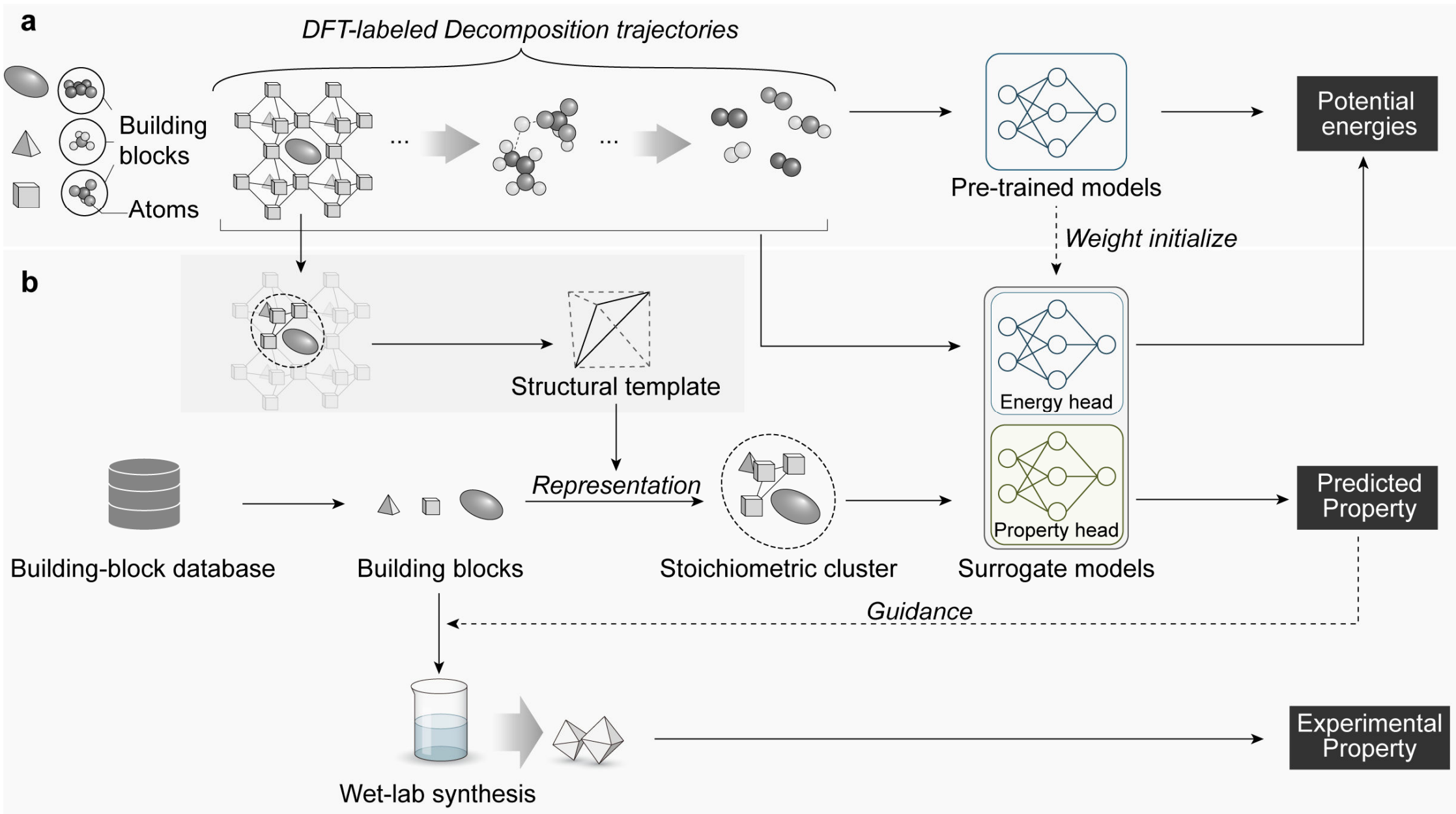

**Figure 1.** Overview of the study workflow. (a) Domain-specific pretraining. A DPA3 graph neural network,[33] is trained on the DeepEMs dataset,[26] i.e., DFT-labelled decomposition trajectories of energetic crystals to yield the DeepEMs-LAM backbone, which predicts potential energies and forces on energetic-material potential-energy surfaces. (b) Property prediction in the data-scarce PEM/MIX regime, in which each PEM crystal is reduced to a stoichiometric ionic cluster from a building-block database, and the cluster is fed into the established MT-FT surrogate that shares the pretrained backbone between an energy head (regularization) and a property head (calculated detonation velocity). Predicted properties guide wet-lab synthesis (dashed arrow) of new $ABX_4$ materials, whose experimentally-derived K–J reference velocities close the validation loop (see Methods, "Wet-lab experiments for synthesis and characterization of $ABX_4$ compounds").

# Results

Throughout the Results, $V_{det}$ denotes detonation velocity, the propagation speed of the detonation front. The K–J $V_{det}$ values used here are empirical endpoints computed from experimentally refined crystal densities and thermochemical inputs. In the K–J framework, detonation velocity is estimated by[37]

$$D = 1.01\ \Phi^{1/2}(1 + 1.30\rho)$$
$$\Phi = 31.68\ N(MQ)^{1/2} \qquad (1)$$

where $D$ represents detonation velocity (km·s$^{-1}$), $\rho$ is the density of the explosive (g·cm$^{-3}$), $N$ is the molar amount of gaseous detonation products per gram of explosive (mol·g$^{-1}$), $M$ is their average molecular weight (g·mol$^{-1}$), and $Q$ is the heat of detonation (kcal·g$^{-1}$).

## Chemical space of MIXs

We first delineate the chemical space of MIXs. Within MIXs, perovskite-type energetic materials (PEMs) are a site-resolved subfamily in which perovskite-style A, B, and X sites assign distinct chemical roles to the ionic building blocks. In this notation, A-site organic cations primarily act as fuel components, B-site cations modulate structure and packing, and X-site anions supply oxidizing groups, defining an ABX chemical space across the reported PEM families.[16–25] This site-resolved modularity makes PEMs a natural starting point for defining the MIX chemical space on which the cluster surrogate is trained and tested. The current PEM literature spans about 40 reported compounds.[25] PEM compound names follow a uniform three-letter prefix plus numerical suffix convention,[25] in which the first letter labels the A-site organic cation family (D for $H_2dabco^{2+}$, P for $H_2piperazine^{2+}$), the third letter labels the X-site oxidizer family (P for $ClO_4^-$, N for $NO_3^-$, I for $IO_4^-/H_4IO_6^-$), and the numerical suffix labels the B-site cation (1 for $Na^+$, 2 for $K^+$, 3 for $Rb^+$, 4 for $NH_4^+$). For example, DAP-4 denotes $(H_2dabco)(NH_4)(ClO_4)_3$ and DAN-2 the $K^+/NO_3^-$ analogue sharing the same A-site. An additional letter before the numerical suffix flags a sub-variant within a family (M for methyl substitution on the A-site as in PAN-M2, H for homologue ring expansion as in PAN-H2, X for an alternative X-site oxidizer state as in DAI-X1), with a complete glossary provided in Supplementary Table S3. Among the reported PEMs, 25 currently have both reliable single-crystal structures and $V_{\mathrm{det}}$ values suitable for benchmarking (Supplementary Table S2). This dataset therefore samples only a small portion of the accessible combinatorial space. Within this chemically diverse but sparsely populated $ABX_3$ perovskite space, a descriptive one-way analysis of variance (ANOVA) of $V_{\mathrm{det}}$ by A-, B-, and X-site labels identifies X-site identity as the dominant compositional factor, consistent with the previously proposed design hierarchy,[25] with quantitative effect-size estimates and rare-site robustness deferred to the representation analysis below (Fig. 2a–c). We refer to this curated 25-material in-distribution set as **IND**, with each material scored only by the held-out fold during cross-validation.

Beyond the $ABX_3$ core dataset, non-PEM MIXs such as $ABX_4$ and $A_2BX_5$ materials still follow the PEM-inspired A/B/X convention, i.e., A denotes the larger organic cation, B a smaller cationic coformer, and X an oxidizing anion, with both A and B able to contribute fuel content and structural modulation. We therefore include non-PEM MIXs as natural out-of-distribution (OOD) tests because their stoichiometries shift from the $ABX_3$ PEM core to $A_2BX_5$ (EAP-4)[27] or $ABX_4$ (DEP, previously reported as "SY"),[38,39] and also include single-site $ABX_3$ holdouts (DAC-4 with rare $ClO_3^-$ X-site and TAP-2 with held-out A-site). We also tested DAI-$1_{0.5}4_{0.5}$ (previously reported as "DPPE-1"), a double-perovskite failure case with ordered $Na^+/NH_4^+$ B sites, and report it separately to mark the boundary of the current surrogate (Supplementary Note S5.2). This sparse $ABX_3$ set defines the IND benchmark, while the related $ABX_4$ and $A_2BX_5$ branches define the transfer tests used later.

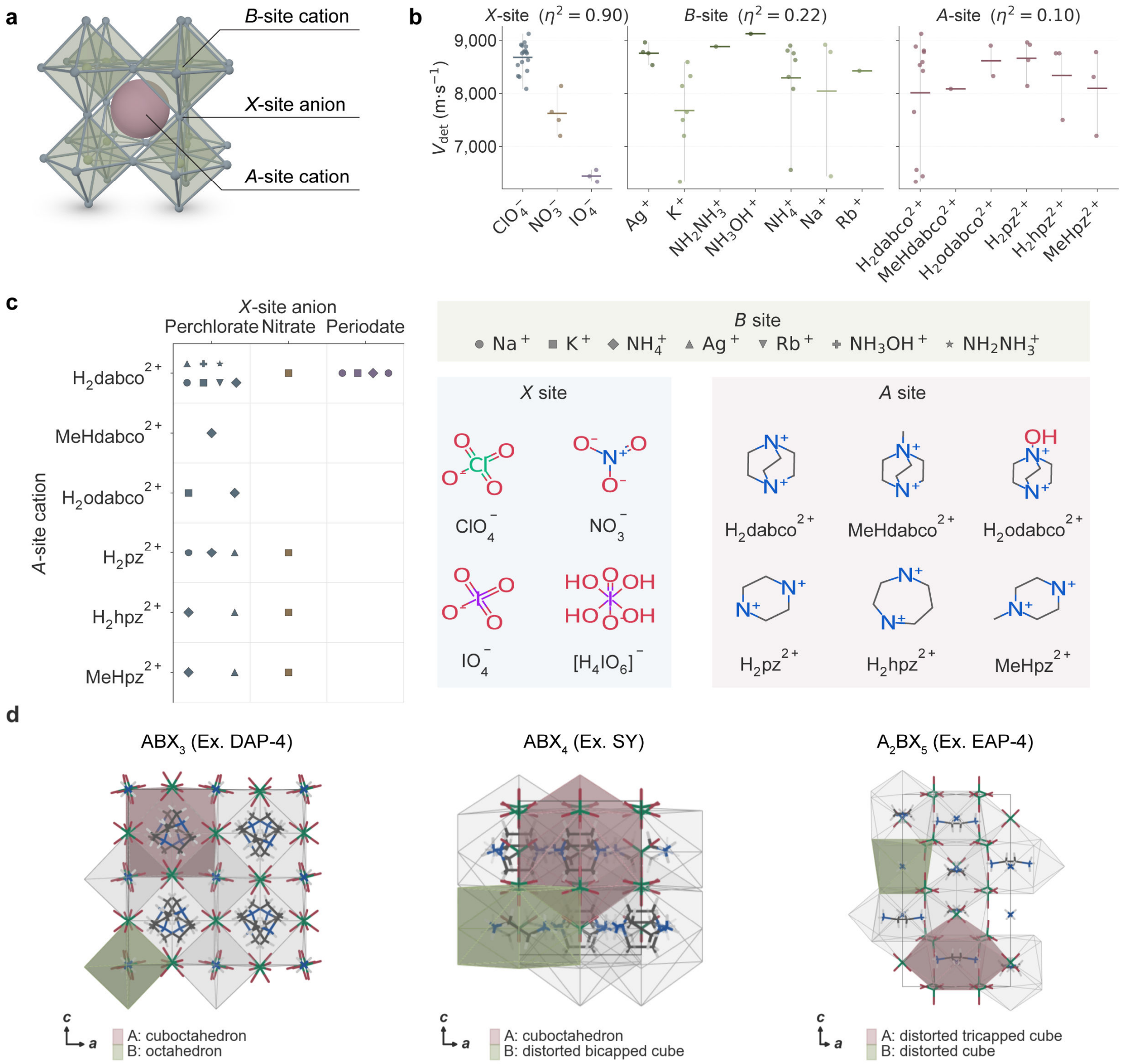

**Figure 2.** Chemical space of MIXs. (a) $ABX_3$ perovskite architecture. (b) One-way ANOVA decomposition of $V_{det}$ across the three ionic sites, with X-site identity dominant and smaller contributions from A and B sites. (c) Compositional matrix of the 25 known materials, illustrating sparse but chemically diverse combinatorial coverage. (d) Schematic organization of the currently observed MIX stoichiometric branches into a small set of coarse polygonal packing motifs, highlighting $ABX_3$, $ABX_4$, and $A_2BX_5$ as related A/B/X assembly patterns.

## IND accuracy of stoichiometric clusters

We next evaluated whether stoichiometric clusters provide a useful IND surrogate before testing transfer beyond the curated 25-material set. The five folds were mixed at the coarse X-site level instead of being concentrated in a single oxidizer family, with every validation fold containing at least two X-site families where possible and every training split retaining perchlorate, nitrate, and periodate/orthoperiodate examples (Supplementary Table S4). Under the stoichiometric-cluster representation, the established MT-FT baseline (defined with the other model variants in Extended Data Table 1) reaches a five-fold **IND** MAE of 273 $m \cdot s^{-1}$, below the 690 $m \cdot s^{-1}$ global-mean predictor (Fig. 3a,b and Extended Data Table 2). This corresponds to an approximately 60% reduction in material-level error relative to the composition-blind global-mean predictor, with per-fold MAEs spanning 201–333 $m \cdot s^{-1}$ and a bootstrap 95% confidence interval of 201–350 $m \cdot s^{-1}$. Per-fold MAEs are computed after averaging the three cluster realizations of every material into a single prediction (see Methods, "Cluster representation"), and the mean within-material cluster-variant standard deviation remains below 17 m·s$^{-1}$ in every fold. The IND result therefore establishes minimum predictive viability of the cluster representation relative to a composition-blind baseline. Single-task fine-tuning (ST-FT) and single-task training from scratch (ST-TFS) are used as adaptation controls. It shows that the ST-FT and ST-TFS controls remain in the same broad few-hundred m·s$^{-1}$ IND error regime, so the stronger distinction among adaptation strategies comes from the OOD and representation analyses below rather than from IND MAE alone.

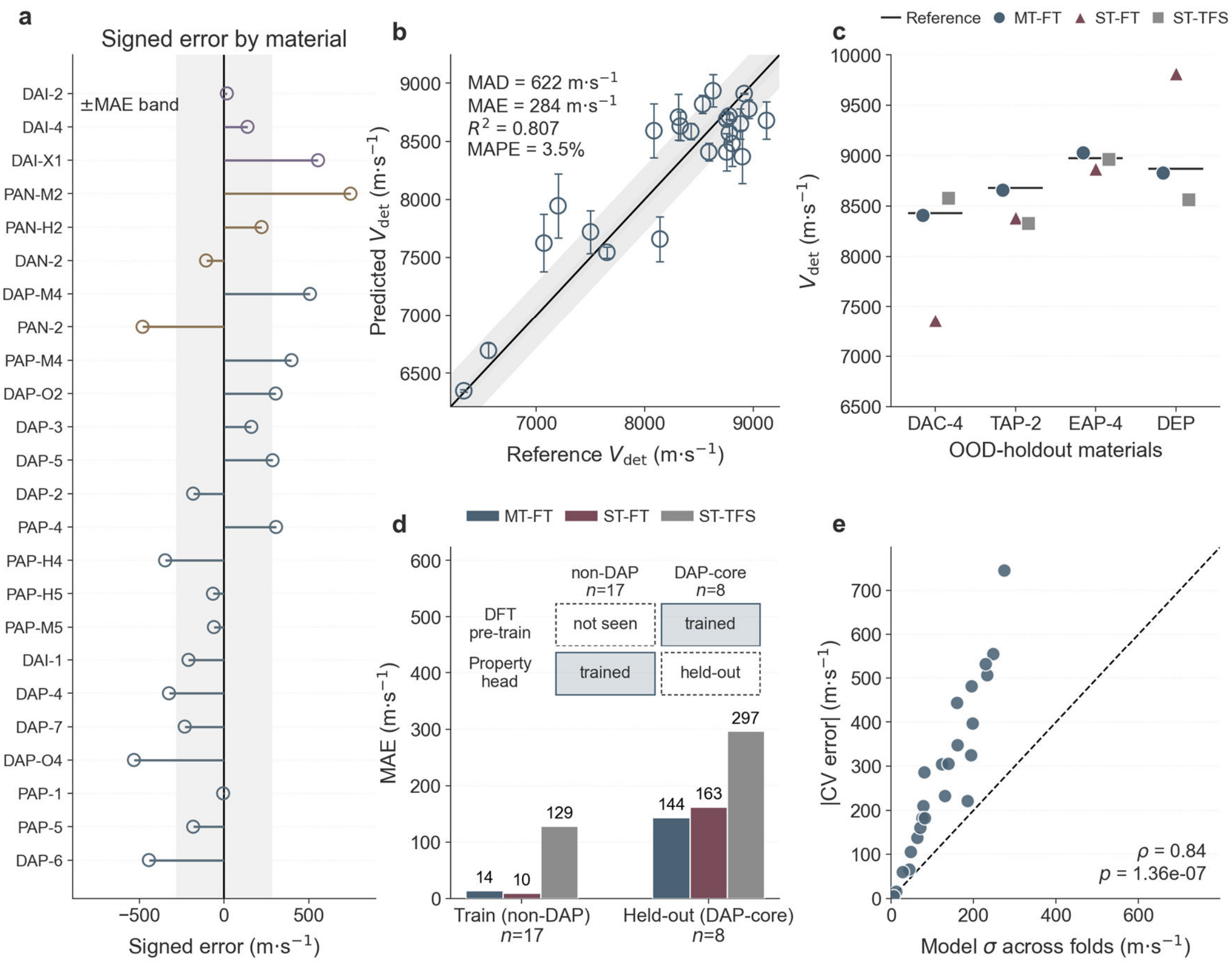

**Figure 3.** Accuracy and transfer behavior of the cluster-based MT-FT surrogate. (a) Material-level signed held-out errors for the 25 PEMs. (b) Held-out parity against K–J reference velocities. (c) Literature OOD holdout comparison across MT-FT, single-task fine-tuning (ST-FT), and single-task training from scratch (ST-TFS) variants. (d) Pretrained-domain DAP-core holdout test. (e) Per-material absolute error versus inter-fold standard deviation.

The MT-FT auxiliary-head variant (MT-FT-aux, Extended Data Table 1) tests whether a larger organic energetic-crystal K–J dataset can further regularize the sparse PEM property head by adding a third K–J velocity head trained on 12,498 Davis2024 systems. This variant slightly lowers IND MAE from 273 to 265 m·s$^{-1}$, but the paired improvement is not statistically resolved at this sample size (Wilcoxon $p$=0.13; Extended Data Table 2). It also increases the literature OOD-holdout MAE from 35 to 67 m·s$^{-1}$, suggesting that the organic CHNO auxiliary crystals may introduce a domain mismatch for ionic MIX transfer targets. We therefore retain the two-head MT-FT surrogate as the production model and treat MT-FT-aux as a domain-mismatch ablation. Within IND, material-level absolute error correlates with across-fold ensemble spread (Spearman $\rho$=0.84, Fig. 3e). We therefore use inter-fold variability as a candidate-prioritization cue, while noting below that it cannot detect biases shared across all folds.

## Representation analysis of stoichiometric clusters

Having established a usable IND surrogate, we next ask what information the cluster representation encodes and which adaptation choices support it at $n$=25. Because the intended use case is pre-synthesis screening, the key representation question is not whether clusters outperform resolved crystals but whether cluster-trained descriptors remain usable when periodic structures are unavailable. To benchmark the cluster surrogate against a higher-information periodic-crystal baseline, the same training setup was also tested as an MT-FT periodic-crystal variant (MT-FT-crystal, Extended Data Table 1) using resolved PEM crystal inputs. As expected, this periodic-crystal baseline gives the lower matched-input **IND** error, 201 m·s$^{-1}$ versus 273 m·s$^{-1}$ for cluster-to-cluster validation. We then swapped the evaluation inputs between the two representation-trained models. We refer to these swaps as **cross-representation** tests. The cluster-trained model degraded from 273 to 405 m·s$^{-1}$ when evaluated on periodic crystals, whereas the crystal-trained model degraded from 201 to 1314 m·s$^{-1}$ on fold-matched clusters and to 1624 m·s$^{-1}$ on the three new $ABX_4$ cluster inputs discussed below (Extended Data Fig. 2 and Supplementary Fig. S14). This asymmetric test shows that the cluster input is not a higher-fidelity substitute for a resolved crystal. It is a lower-information representation that retains enough local structural signal for pre-synthesis screening.

The DeepEMs-LAM backbone was pretrained on DFT trajectories that overlap with **IND** only in the DAP-core perovskite family (8 of 25 materials). We next use ablations to separate which parts of the established adaptation protocol support the cluster representation. The DAP-core holdout test (Fig. 3d) isolates the contribution of the pretrained DFT prior, with the 8 DAP-core materials removed from the property head while their DFT trajectories remain in the energy head. On this split MT-FT and ST-FT remain close (144 and 163 m·s$^{-1}$), whereas ST-TFS degrades to 297 m·s$^{-1}$. The DAP-core holdout indicates that the pretrained energetic-material prior supplies most of the gain over training from scratch, while the energy head acts as a stabilizing constraint in this $n$=25 setting.

Accuracy alone cannot distinguish a surrogate that has learned chemically meaningful features from one that exploits fold-specific shortcuts. We therefore probe the learned representation from three complementary viewpoints. These are frozen-descriptor ridge probes that ask which physical quantities are linearly readable without retraining the backbone, uniform manifold approximation and projection (UMAP) and silhouette analyses that test whether atom embeddings organize by ionic role and element, and a perturbation ladder that moves from shell-preserving changes to fragment swaps and geometry-destroying atom-level scrambles.

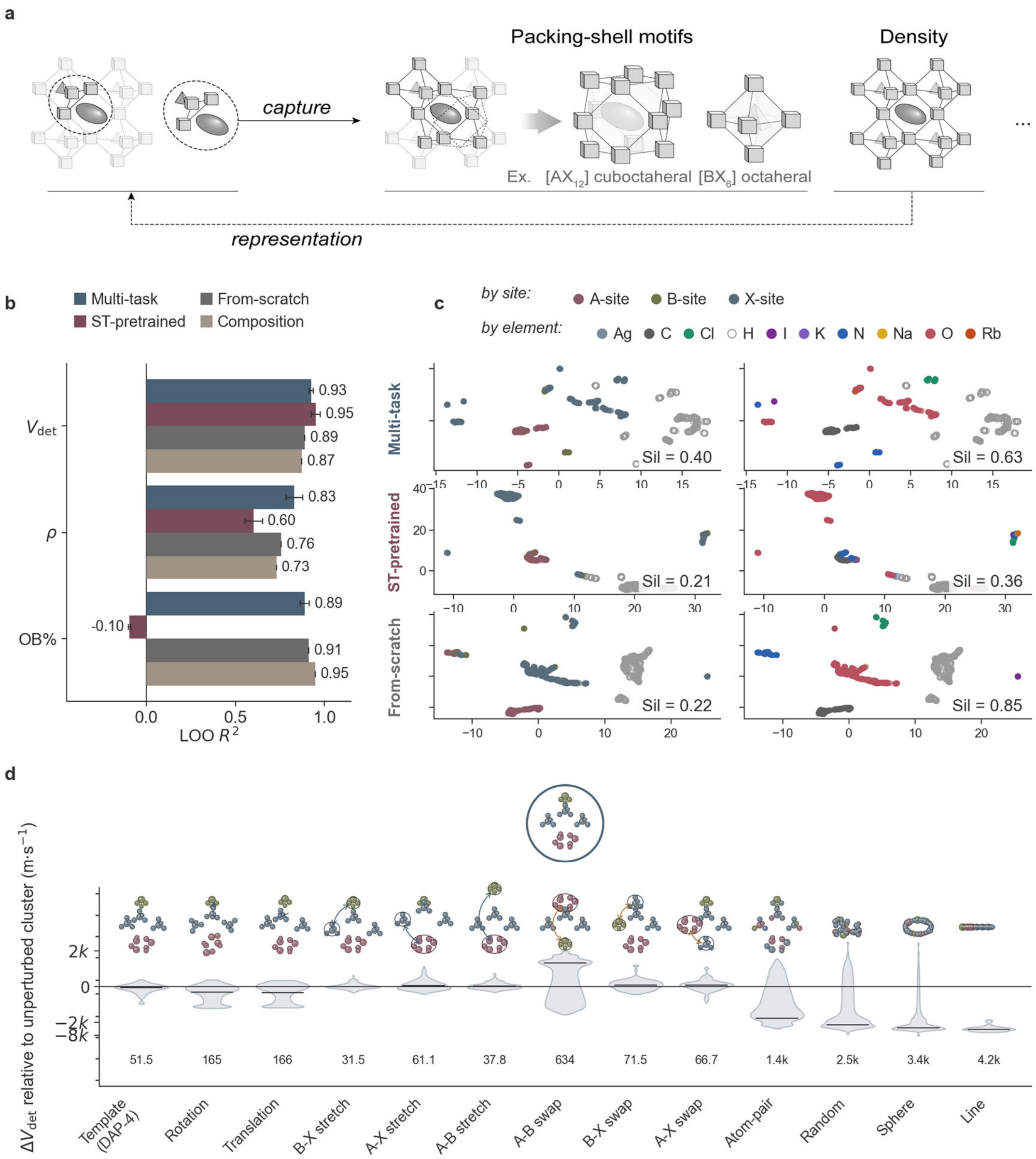

**Figure 4.** Frozen-embedding analysis of the cluster-based surrogate. (a) Schematic cluster density representation of stoichiometric ionic clusters. (b) Leave-one-out $R^2$ for ridge probes of $V_{\mathrm{det}}$, density, and oxygen balance across MT-FT, ST-FT, ST-TFS, and composition-only representations. (c) Atomic-level UMAP of 256-dimensional descriptors coloured by ionic site and element. (d) A single $ABX_3$ reference cluster perturbed by 13 mechanisms spanning template substitution, rigid-body transformations, A–X/B–X/A–B distance stretches, fragment-level swaps, and atom-level scrambles, ordered by increasing mean absolute shift. Vertical kernel density estimate (KDE) densities show the resulting $\Delta V_{\mathrm{det}}$ distributions, with overlaid numbers denoting mean absolute shifts in m·s$^{-1}$.

In the frozen-probe protocol (Fig. 4b), we fit leave-one-out ridge regressions from each variant's locked per-cluster embedding onto $V_{\mathrm{det}}$, crystal density, and oxygen balance (see Methods, “Site-resolved compositional analysis”). Site-resolved one-way ANOVA on $V_{\mathrm{det}}$ and a one-hot composition baseline benchmark how much signal is carried by site identity alone. ANOVA assigns the dominant compositional effect to X-site identity ($\eta^2$=0.90), with much smaller A- and B-site contributions (Fig. 2b). Site-resolved ridge probes confirm that the X-site-pooled descriptor captures most of this signal ($R^2$=0.85), only slightly above the composition-only baseline (Supplementary

Fig. S13). Fold-to-fold variability at $n$=25 is large, so the hierarchy is best read qualitatively. To test the robustness of the ANOVA hierarchy reported in Fig. 2b, we introduce a leave-one-site-out (LOSO) diagnostic that withholds all materials sharing a given A-, B-, or X-site identity (Extended Data Fig. 6 and Methods, "Site-resolved compositional analysis"). LOSO and Type II partial $\eta^2$ both confirm that only X-site dominance is robust, while the apparent A- vs B-site ordering of one-way $\eta^2$ is not, because A–X and B–X co-occurrences in the 25-material design are highly non-factorial. Density and oxygen balance are physically meaningful axes for assessing candidate $ABX_4$ materials because the K–J endpoint depends explicitly on crystal density and composition-derived thermochemical terms. A representation encoding these axes remains useful when only the $V_{det}$ head is updated downstream.

At the atomic level (Fig. 4c), the multi-task baseline organizes its element embeddings into more sharply site-resolved subgroups than the single-task pretrained variant, consistent with higher site-grouped silhouette scores (0.40 vs 0.21) and with multi-task supervision retaining site-aware atomic environments that narrow under single-task fine-tuning at $n$=25. The material-level descriptor encodes crystal density without direct supervision ($R^2$=0.83), supporting the interpretation that the surrogate has internalized packing constraints rather than memorized geometric shortcuts. This site-resolved separation provides a plausible representational basis for transferring within the related $H_2en^{2+}$-based $ABX_4$ branch below.

Controlled perturbations (Fig. 4d) organize the response into a ladder of increasing chemical disruption. At the low-disruption end, rigid motions, template substitution, and A–X/B–X/A–B distance stretches mostly preserve fragment identity and local shell topology. Fragment swaps exchange the centre-of-mass positions of A-, B-, or X-site fragments while keeping each fragment internally rigid, and atom-scrambling operations destroy the original fragment organization altogether. Together with the density probe, this ladder supports a coarse packing-compatibility interpretation. The representation preserves information about chemically meaningful local organization, while responding most strongly when that organization is grossly violated. This perturbation analysis is defined on a single reference cluster and should not be interpreted as a direct ranking of dataset-level site importance, which is addressed separately by ANOVA across materials.

Taken together, the three probes answer the central representation question posed above. The frozen descriptors retain physically meaningful axes relevant to detonation performance, the atom-level embeddings preserve site-aware ionic organization under multi-task fine-tuning, and the perturbation ladder shows stability to shell-preserving transformations but strong response when fragment identity or local organization is destroyed. These diagnostics support a restricted transfer hypothesis for MIX branches that preserve recognizable A, B, and X building blocks and local packing compatibility.

## OOD transfer across held-out MIX branches

These representation diagnostics motivate OOD tests on held-out materials that alter one site, the stoichiometry, or both while remaining within the MIX design logic. We refer to this literature-based set as **OOD-holdout** (Fig. 3c). The four-material aggregate comprises literature-reported, externally synthesized energetic materials outside **IND** with experimentally-derived K–J detonation velocities. It includes two $ABX_3$ PEMs, DAC-4 with rare $ClO_3^-$ X-site and TAP-2 with held-out A-site, and two non-$ABX_3$ MIX representatives, EAP-4 of $A_2BX_5$ stoichiometry[27] and DEP of $ABX_4$ stoichiometry.[38,39] For DEP, the reference used here is the K–J value recalculated from the remeasured and refined 298(2) K single-crystal structure, whose refined density is 1.869 g·cm$^{-3}$, giving $V_{det}$=8.867 km·s$^{-1}$. On this four-material aggregate MT-FT reaches an MAE of 35 m·s$^{-1}$, compared with 67 m·s$^{-1}$ for MT-FT-aux, while both single-task variants degrade substantially (Fig. 3c and Extended Data Fig. 3). ST-FT errors concentrate on DAC-4 and DEP, a pattern that matches representational narrowing under the $n$=25 constraint (Supplementary Note S3.3). We retain MT-FT as an enabling adaptation recipe because it provides stable transfer in this low-data setting, not because the current 25-material benchmark establishes statistical superiority over all single-task alternatives.

We also tested DAI-$1_{0.5}4_{0.5}$, a $Na^+$/$NH_4^+$ ordered double perovskite with $IO_4^-$ X-site, on which all four variants overestimate the experimental reference by approximately 1300 m·s$^{-1}$ (Extended Data Fig. 3). This failure marks a boundary of the present representation and illustrates why inter-fold variability cannot detect systematic bias shared across all training folds.

## Synthesis validation in the $H_2en^{2+}$-based $ABX_4$ branch

As the final test, we applied the cluster surrogate to the $H_2en^{2+}$-based $ABX_4$ branch, which preserves known PEM building blocks but introduces an unseen B-site cation and $ABX_4$ stoichiometry. The target branch is motivated by the site-resolved design hierarchy and by the known DEP $ABX_4$ member.[38,39] We define **OOD-new** as three newly synthesized $ABX_4$ materials of formula $(A)(H_2en)(ClO_4)_4$, namely PEP (A = $H_2pz^{2+}$, inherited from PAP-4), PEP-M (A = $MeHpz^{2+}$, from PAP-M4), and PEP-H (A = $H_2hpz^{2+}$, from PAP-H4), with full names listed in Supplementary Table S3. This design keeps the A-site cations related to known PEM cousins and the X-site perchlorate chemistry fixed, while introducing an unseen $H_2en^{2+}$ B-site and one extra X-site $ClO_4^-$ relative to the $ABX_3$ PEM stoichiometry. Synthesis, single-crystal characterization, and K–J reference velocities were carried out independently of the surrogate, providing a stringent OOD test that probes both unseen B-site chemistry and stoichiometry. Figure 5a,b first summarize the ionic topologies and refined single-crystal structures of the three OOD-new $ABX_4$ materials, before panels c–f compare their diffraction signatures, energetic performance, predictive accuracy, and embedding-space placement. The three new materials were confirmed as phase-pure crystalline solids by powder X-ray diffraction (Fig. 5c, with the remeasured DEP reference overlay reproduced in Supplementary Fig. S19), thermally stable up to their measured onset decomposition temperatures (Supplementary Fig. S20), with impact and friction sensitivities reported in Supplementary Table S20. Single-crystal refinement confirms dense ionic solids suitable for K–J-based validation (Extended Data Table 3).

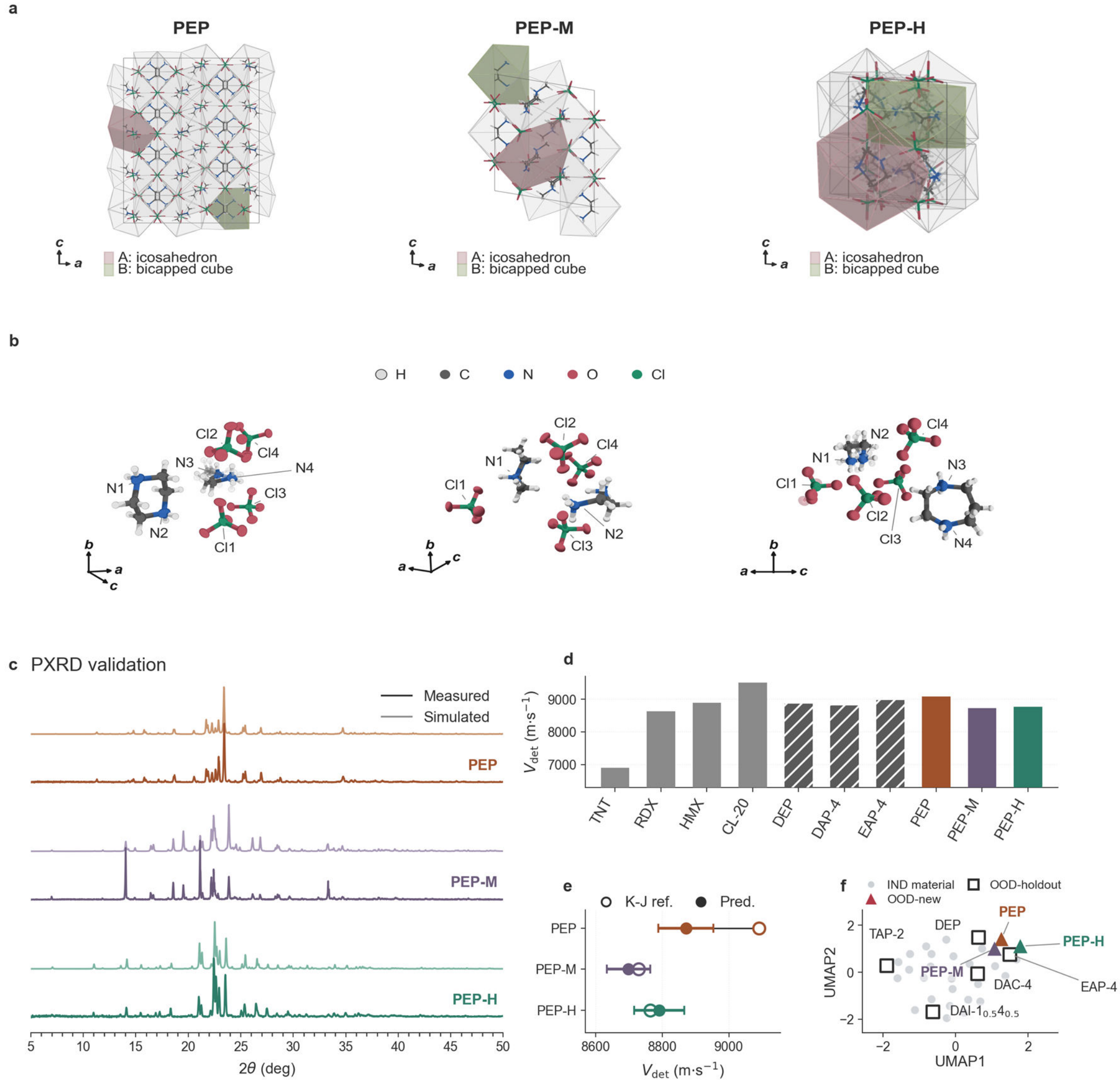

**Figure 5.** Synthesis, characterization, and predicted detonation performance of the OOD-new $H_2en^{2+}$-based $ABX_4$ branch. (a) Ionic-topology projections of PEP, PEP-M, and PEP-H. (b) Single-crystal X-ray structures. (c) Measured and simulated powder X-ray diffraction patterns. (d) Detonation-velocity comparison against reference explosives. (e) Predicted versus K–J reference detonation velocities. (f) Atomistic-embedding UMAP centroids of the OOD-new $ABX_4$ materials against the 25 IND materials.

Compared with established energetic compounds, the three new $ABX_4$ velocities lie above RDX and octogen (also known as high melting explosive, HMX) and approach hexanitrohexaazaisowurtzitane (also known as China Lake compound 20, CL-20), placing them at the upper end of representative MIX members such as DEP, DAP-4, and EAP-4[27] (Fig. 5d). Using these experimentally refined crystallographic information files (CIFs) as direct inputs, MT-FT predicts all three materials in the high-performance regime and remains tightly clustered across folds (Table 1, Fig. 5e). Against K–J reference velocities computed from the same refined densities, the three-material MAE is 92 $m \cdot s^{-1}$, with individual absolute errors of 219, 31, and 27 $m \cdot s^{-1}$ for PEP, PEP-M, and PEP-H (Table 1). The same ranking places PEP at the strongest calculated performance in this branch (Extended Data Table 3). Their cluster embeddings project to a region of the MT-FT atomistic UMAP that is not densely populated by **IND** materials (Fig. 5f, Supplementary Fig. S15).

**Table 1.** Predicted versus K–J reference detonation velocities for the three OOD-new $H_2en^{2+}$-based $ABX_4$ materials. Predictions are five-fold ensemble means (three cluster realizations pooled within each fold, with the reported inter-fold standard deviation taken across $n$=5 folds). Reference values are K–J detonation velocities computed from the same refined crystal densities.

| Material | Predicted $V_{det}$ (m·s$^{-1}$) | Inter-fold std (m·s$^{-1}$) | K–J reference $V_{det}$ (m·s$^{-1}$) |
|---|---|---|---|
| PEP | 8870.8 | 82.3 | 9090 |
| PEP-M | 8698.5 | 65.0 | 8729 |
| PEP-H | 8790.9 | 75.3 | 8764 |

Together, the OOD-new three-point comparison establishes concordance with K–J reference velocities. This comparison does not constitute direct detonation-test validation, which is deferred to future work. We further test a template-based pre-synthesis variant of the same prediction, in which each OOD-new target is built on a DAP-4 $ABX_3$ scaffold by substituting in its A-site molecule and the $H_2en^{2+}$ B-site, without using the experimental $ABX_4$ crystal. The three-point mean absolute deviation against the K–J reference remains at approximately 100 m·s$^{-1}$ (Extended Data Fig. 4), supporting template-based screening prior to crystal growth. This $ABX_4$ extrapolation should therefore be read not as universal zero-shot transfer, but as family-level transfer enabled by a representation that remains stable to coarse shell-preserving perturbations while retaining physically meaningful density and composition axes. The $H_2en^{2+}$-based $ABX_4$ branch is thereby expanded from one to four structurally characterized members. The result supports family-level transfer within related MIX branches.

# Discussion

**Contribution.** This work contributes a representation strategy for few-shot modular ionic assemblies. By collapsing each crystal to a stoichiometry-preserved cluster and fine-tuning a pretrained atomistic backbone under multi-task regularization, we bridge the gap between periodic-crystal atomistic models and data-scarce multi-ionic materials whose properties emerge from fragment co-assembly. Alongside this methodological contribution, we synthesized and single-crystal characterized three new $H_2en^{2+}$-based $ABX_4$ MIXs (PEP, PEP-M, and PEP-H), thereby expanding this branch of MIXs.[38,39]

**Limitations.** With only 25 in-domain materials spanning non-factorial site coverage, cross-validation uncertainty is large and extrapolation across a genuinely new X-site oxidizer family remains untested. The cross-fold ensemble UQ detects per-material variance but is blind to systematic bias shared across all folds, as the DAI-$1_{0.5}4_{0.5}$ failure case demonstrates. The $ABX_4$ validation is anchored on K–J reference velocities rather than directly measured detonation velocities, placing it at the empirical-equation endpoint level. The cluster representation deliberately removes long-range periodic electrostatics, so its reliability may decrease for materials whose target property is dominated by lattice-scale fields rather than local stoichiometric coordination.

**Outlook.** Despite the small 25-material target set and the K–J-anchored validation endpoint, the $H_2en^{2+}$-based $ABX_4$ result suggests that the cluster-based MT-FT pipeline could help prioritize underexplored stoichiometric branches within the broader MIX class. With appropriate caution, the same workflow could also be extended to new stoichiometries within fixed component sets and related energetic-material assemblies in which the coformer identity or ratio is varied (e.g., in energetic cocrystals). Within the MIX space, broader labels across X-site oxidizer families would help constrain systematic transfer bias and support extension to other A–B–X combinations. For the energetic-materials community, part of the next bottleneck may shift from screening workflows to clean, comparable target labels collected under aligned experimental protocols, for example friction-sensitivity measurements made with the same testing strategy. A larger multi-ionic-material dataset would also enable controlled downsampling studies that compare composition-level, molecular-ion, stoichiometric-cluster, and periodic-crystal representations under matched label budgets. Such benchmarks would clarify when the stoichiometric ionic-cluster representation is sufficient, when resolved periodic structures are needed, and how far this workflow can be treated as a reference strategy for data-scarce, structurally modular energetic and multi-ionic property-prediction tasks.

# Methods

## Data curation

### Crystal structures and target properties of PEMs

The target-property dataset was curated from published PEM crystallographic reports and contains 25 materials with reported K–J-derived detonation velocities in $m \cdot s^{-1}$, listed in Supplementary Table S2. CIFs with partial occupancy or disorder were converted to ordered structures before use. Ordered crystal structures were retained for periodic-crystal controls, and the main workflow used the stoichiometric cluster representation described below.

The 25-material target set was partitioned into five folds for cross-validation. Each fold holds out five materials as the validation set and trains on the remaining twenty. Fold assignments were made by random sampling with seed 42, with full fold membership given in Supplementary Table S4.

### Davis2024 auxiliary dataset

DPA-based atomistic models require explicit three-dimensional inputs, so each Davis2024 record was paired with a crystal representation before training or with a molecule representation for zero-shot controls. The public Davis *et al.* dataset reports each material's Cambridge Structural Database (CSD) refcode, chiral SMILES and calculated properties, including K–J detonation velocity.[28] We reused the published detonation-velocity labels directly, retrieved crystal structures from the CSD using those refcodes, and generated molecule-input controls from the reported chiral SMILES with RDKit. After structure and label filtering, 12,498 auxiliary systems remained and were split 85/15 by random assignment (seed 42), yielding 10,623 training and 1,875 validation systems.

## Cluster representation

For the main representation, each ordered PEM crystal was converted to a non-periodic stoichiometric vacancy cluster, and the model input was the resulting atom coordinates and species labels. Each structure was parsed as a molecular crystal using element-specific bond-distance thresholds, expanded using the smallest supercell in the schedule (2×2×2), (3×3×3), (2×3×3), (3×3×4), (4×4×4) that yielded at least three seed candidates, and reduced to the simplest stoichiometric unit. Vacancy construction centred on a maximally spread seed molecule returned the removed formula-unit cluster, which was minimum-image wrapped, centred in a 100 Å cubic box, and evaluated without periodic boundary conditions. Stoichiometric construction enforces overall charge neutrality, and no explicit Ewald summation or polarization correction is applied.

Up to three cluster variants were generated per material by selecting maximally spread seed positions (random seeds 101, 202, and 303 with offsets 0, 1, and 2, respectively). High-symmetry crystals with fewer than three structurally distinct seed molecules can produce duplicate variants. These variants were used both to average predictions and to assess sensitivity to the choice of cluster realization. Construction parameters, pseudocode, variant counts and cluster galleries are provided in Supplementary Algorithm S1, Supplementary Tables S5–S7, and Supplementary Figs. S1–S5.

## Model training

The backbone was DeepEMs-LAM,[26] a DPA3 atomistic model pretrained on energetic- and ionic-material DFT trajectories.[33] Finetuning was performed with DeePMD-kit[40] following the published MT-FT recipe.[36] MT-FT jointly optimized a K–J detonation-velocity head and the inherited energy–force head on the shared DPA3 descriptor. The same hyperparameters were used for the five cross-validation folds and the full-data model trained on all 25 materials. $ABX_4$ predictions used the corresponding trained checkpoints without additional retraining. Adaptation variants and learning-rate schedules are summarized in Extended Data Table 1, with additional details in Supplementary Note S3.1.

## Site-resolved compositional analysis

For each encoded cluster, per-atom 256-dimensional descriptors were assigned to A, B or X sites using the molecular-crystal decomposition and mean-pooled to obtain site-level vectors. For a cluster with $N$ atoms, the backbone outputs one descriptor $h_i \in \mathbb{R}^{256}$ for each atom $i$.

$$z_A = \frac{1}{|A|} \sum_{i \in A} h_i,\ z_B = \frac{1}{|B|} \sum_{i \in B} h_i,\ z_X = \frac{1}{|X|} \sum_{i \in X} h_i \quad (2)$$

with $z_A, z_B, z_X \in \mathbb{R}^{256}$. The corresponding all-atom summary is

$$z_{all} = \frac{1}{N} \sum_{i=1}^{N} h_i. \quad (3)$$

Site-pooled descriptors were computed after full-cluster encoding, and isolated-site structures were not evaluated. Atom-to-site assignments were inherited from the topology-aware MolCrysKit molecular-crystal decomposition, which identifies molecular building blocks and maps them onto the A-, B-, and X-site roles before descriptor pooling.[41] One-way ANOVA on material-level $V_{det}$ values grouped by site identity yields the descriptive effect-size estimates ($\eta^2$) cited in the main text. Because the 25-material design is sparse and non-factorial, Type II partial $\eta^2$ and leave-one-site-out splits were used to check site co-occurrence confounding.

Ridge regression probes were fitted from each site-specific embedding pool to $V_{det}$ by leave-one-out cross-validation. The per-site leave-one-out $R^2$ values are aggregated across the five fold models (mean and across-fold sample standard deviation). Leave-one-site-out split definitions, errors, and confound checks are summarized in Supplementary Tables S15–S17. Atomic-level UMAP projections in Fig. 4c were computed on per-atom descriptor vectors extracted from MT-FT, ST-FT, and ST-TFS, with identical UMAP hyperparameters and a fixed random seed across models. Silhouette scores are computed in the original 256-dimensional descriptor space as the primary representation metric, with the corresponding 2D UMAP-space values reported parenthetically as visual checks. Both calculations use heavy atoms only. Additional perturbation, probe and descriptor-stability diagnostics are summarized in Extended Data Fig. 5, described in Supplementary Notes S4 and S4.4, and tabulated in Supplementary Tables S9–S13. The corresponding perturbation-grid and probe-parity figures are shown in Supplementary Figs. S6–S13.

## Synthesis and characterization of $ABX_4$ compounds

SAFETY NOTE: Energetic perchlorate salts were handled on small scale under institutional energetic-materials safety procedures. PEP, PEP-M and PEP-H were prepared by protonating the corresponding A-site diamine in aqueous perchloric acid followed by ethylenediamine addition. Precipitated products were isolated by filtration and vacuum drying, with yields of approximately 80%. Colorless block crystals suitable for single-crystal X-ray diffraction grew from the mother liquor over several days at room temperature.

Single-crystal X-ray diffraction data were collected on a Rigaku XtaLAB P300DS diffractometer with Cu K$\alpha$ radiation ($\lambda$=1.54184 Å) at 298(2) K for PEP, PEP-M, and DEP (for re-measurement) and at 293(2) K for PEP-H.

Structure-solution and refinement details are reported in Supplementary Table S19. Phase purity of the bulk products was confirmed by comparison of measured powder X-ray diffraction patterns with patterns simulated from the single-crystal models.

K–J detonation parameters were calculated from DFT-computed heats of formation and refined crystal densities. Crystallographic details, thermal stability, mechanical sensitivity, K–J inputs, and validation descriptors are reported in Extended Data Table 3, Supplementary Figs. S19–S21, and Supplementary Tables S19–S22.

# Data availability

The curated target-property table will be made available upon publication. The associated dataset is available at https://doi.org/10.6084/m9.figshare.32967614.

## Code availability

The main preparation, training, inference, and analysis scripts are contained in the project repository. The code is available at https://github.com/SchrodingersCattt/cluster-representation-research.

## AI usage disclosure

Generative AI tools were used to assist in language polishing and grammar checking of the manuscript, as well as algorithm implementation and code debugging. The human authors have reviewed and edited the content as needed and take full responsibility for the contents.

## Author contributions statement

M.-Y.G. contributed conceptualization, methodology, software, computational investigation, data curation, formal analysis, visualization, and writing the original draft. W.-J.Z. contributed experimental investigation, validation, and crystallography and characterization data curation. Y.S. contributed supervision, experimental investigation, review, and editing. W.-X.Z. contributed supervision, conceptualization, funding acquisition, review, and editing.

## Additional information

**ORCID.**

Ming-Yu Guo: 0009-0008-3744-1543

Wei-Xiong Zhang: 0000-0003-0797-3465.

**Competing interests.**

The authors declare no competing interests.

# Extended Data

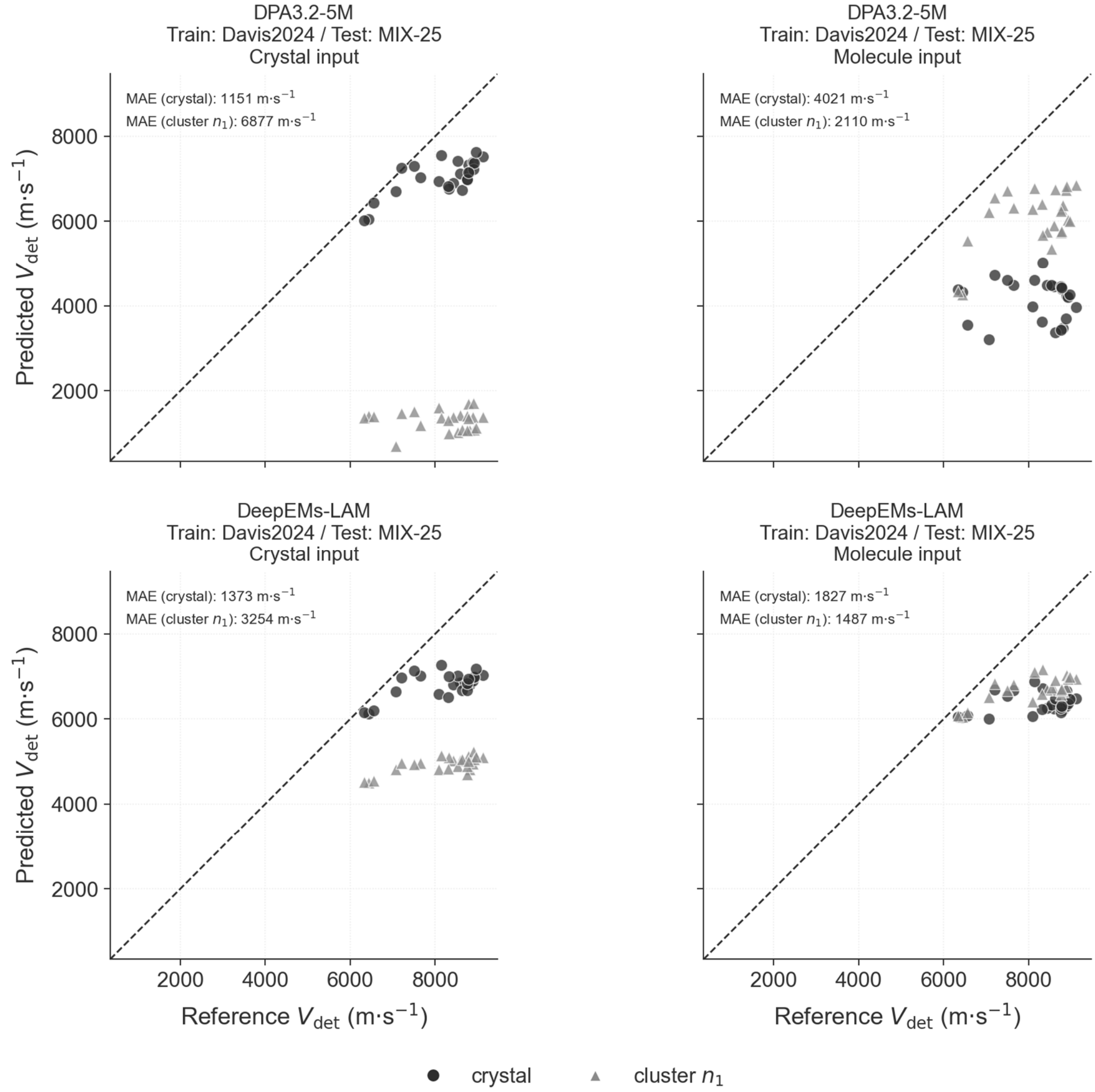

**Extended Data Fig. 1.** Davis2024-trained zero-shot transfer to the 25 PEM targets. Each panel compares periodic-crystal and cluster predictions against K–J reference detonation velocities for a CHNO-trained baseline evaluated without PEM labels, illustrating the systematic domain gap between organic energetic crystals and ionic PEMs.

**Extended Data Table 1.** Adaptation strategies and representation variants compared. "Heads" counts the simultaneously trained objectives; "Input" indicates whether the model uses the non-periodic stoichiometric vacancy cluster from the main text or a periodic crystal cell. MT-FT-highLR and MT-FT-full share identical hyperparameters but differ in training split: the former uses five-fold cross-validation for ablation evaluation, whereas the latter trains on all 25 PEMs without held-out data and serves as the production model for combinatorial screening.

| Variant | Heads | Pretrained | Input | Training split | Start LR | Stop LR | Decay step |
|---|---|---|---|---|---|---|---|
| MT-FT | 2 | yes | cluster | 5-fold CV | $2\times10^{-5}$ | $10^{-8}$ | 5 000 |
| MT-FT-highLR | 2 | yes | cluster | 5-fold CV | $1\times10^{-4}$ | $10^{-8}$ | 1 000 |
| MT-FT-aux | 3 | yes | cluster | 5-fold CV | $2\times10^{-5}$ | $10^{-8}$ | 7 500 |
| ST-FT | 1 | yes | cluster | 5-fold CV | $2\times10^{-5}$ | $10^{-8}$ | 5 000 |
| ST-TFS | 1 | no | cluster | 5-fold CV | $2\times10^{-5}$ | $10^{-8}$ | 5 000 |
| MT-FT-crystal | 2 | yes | crystal (PBC) | 5-fold CV | $2\times10^{-5}$ | $10^{-8}$ | 5 000 |
| MT-FT-full | 2 | yes | cluster | all 25 PEMs | $1\times10^{-4}$ | $10^{-8}$ | 1 000 |

**Extended Data Table 2.** Five-fold cross-validation summary for the adaptation strategies. MAEs are reported in $m\cdot s^{-1}$; confidence intervals are bootstrap 95% intervals from per-material absolute errors.

| Variant | F0 | F1 | F2 | F3 | F4 | Mean (95% confidence interval, CI) | Wilcoxon *p* |
|---|---|---|---|---|---|---|---|
| MT-FT | 264.9 | 333.1 | 318.3 | 201.1 | 248.6 | 273.2 [201, 350] | reference |
| MT-FT-aux | 254.7 | 315.7 | 323.3 | 192.4 | 237.7 | 264.8 [197, 337] | 0.13 |
| ST-FT | 172.4 | 276.6 | 233.0 | 161.5 | 773.3 | 323.4 [198, 506] | 0.83 |
| ST-TFS | 298.6 | 413.1 | 324.6 | 215.3 | 234.5 | 297.2 [225, 376] | 0.56 |

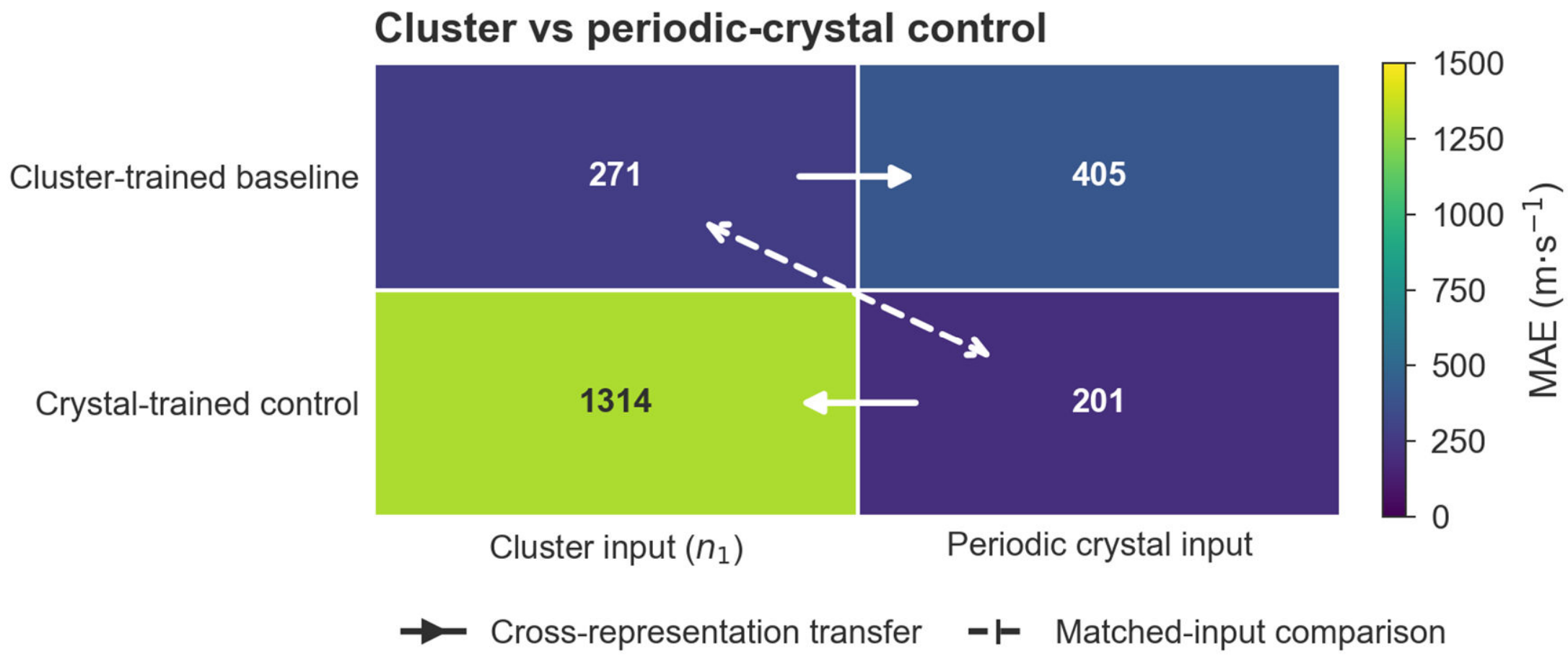

**Extended Data Fig. 2.** Periodic-crystal control and cross-representation specificity. Rows compare the same two-head MT-FT recipe trained either on stoichiometric clusters or on resolved periodic crystals; columns denote the evaluation input used in a paired cluster $n_1$ versus periodic-crystal diagnostic. The dashed arrow compares matched-input validation, where periodic-crystal validation is more accurate than cluster validation; solid arrows compare cross-representation transfer, where the crystal-trained model fails more severely on cluster inputs than the cluster-trained model does on crystal inputs.

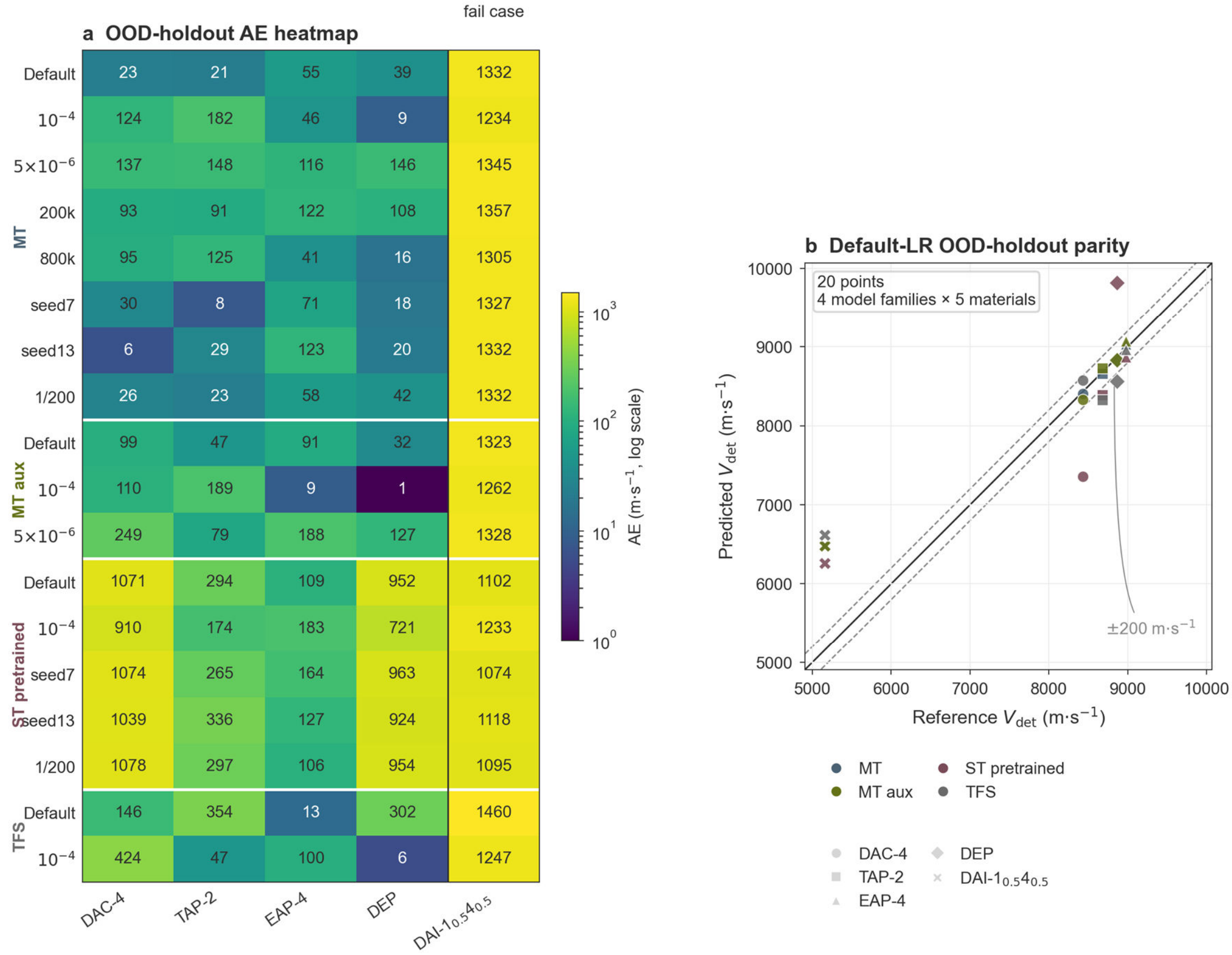

**Extended Data Fig. 3.** OOD-holdout evaluation and the DAI-1$_{0.5}$4$_{0.5}$ failure case (previously reported as "DPPE-1"). The heatmap shows per-material absolute errors across ablation families, with DAI-1$_{0.5}$4$_{0.5}$ separated because all variants overestimate this low-velocity outlier; the parity panel compares the default model families on the same held-out materials, including DEP, previously reported as "SY" by Ma, Zhu, and co-workers. The DEP reference is the K–J value recalculated from the remeasured and refined structure ($D_c$=1.869 g cm$^{-3}$ and $V_{det}$=8867 m s$^{-1}$). Grey dashed lines in panel b mark ±200 m s$^{-1}$ tolerance around the y=x reference.

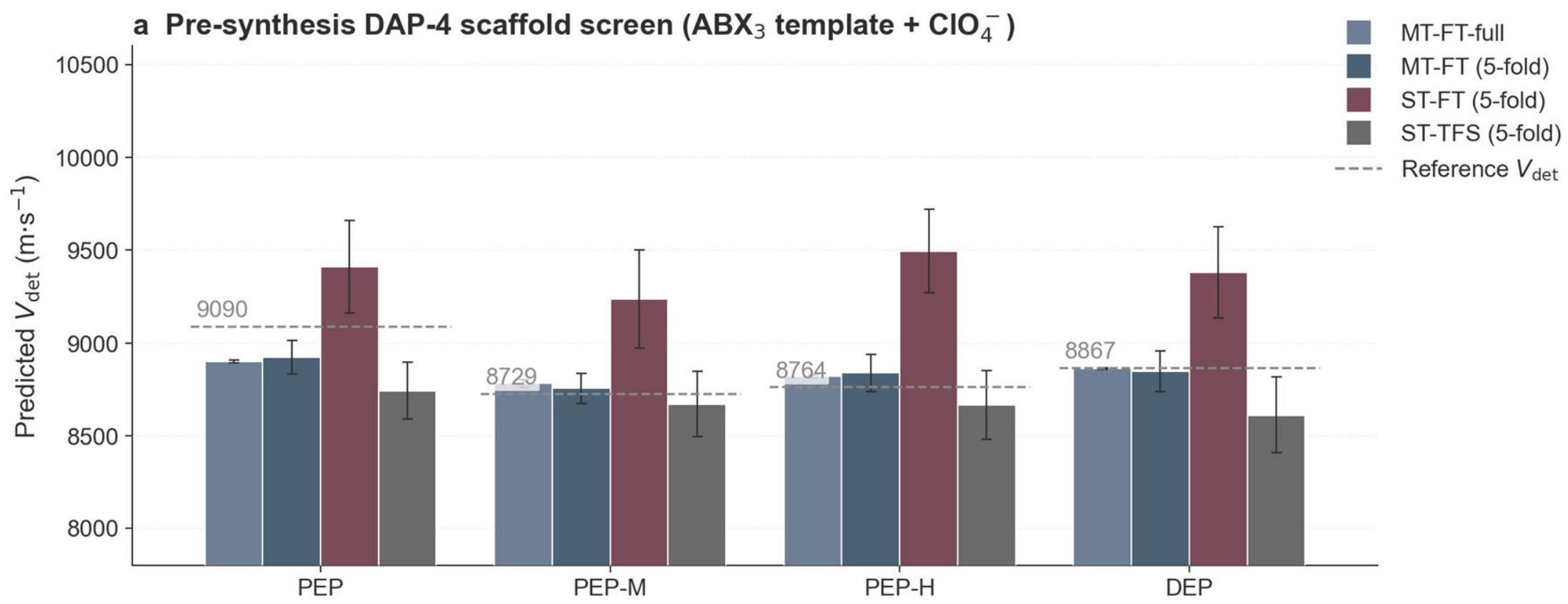

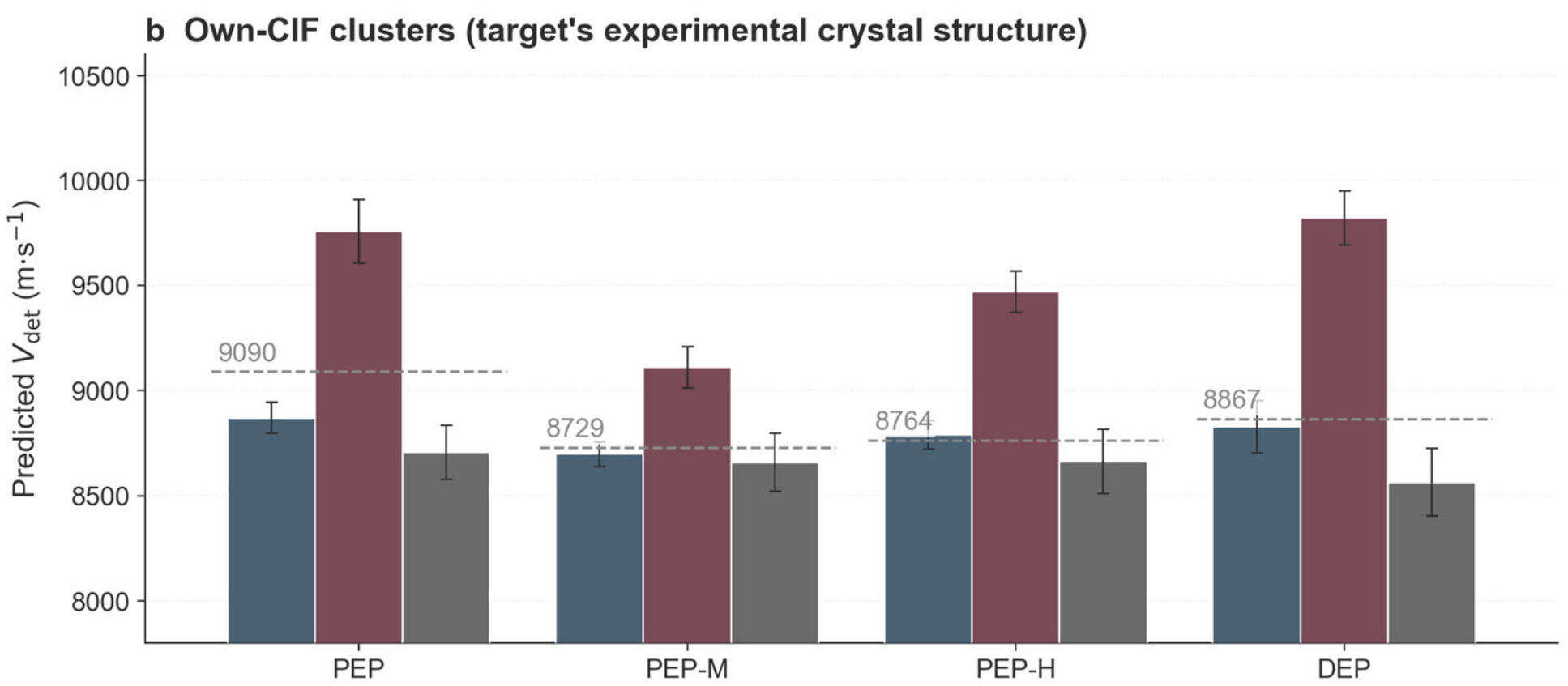

**Extended Data Fig. 4.** Template-based and own-CIF OOD-new evaluation for the $H_2en^{2+}$-based $ABX_4$ branch. Template-derived clusters built from a DAP-4 scaffold preserve useful ranking and accuracy for PEP, PEP-M, and PEP-H, while own-CIF clusters compare the three five-fold model families against the same K–J references.

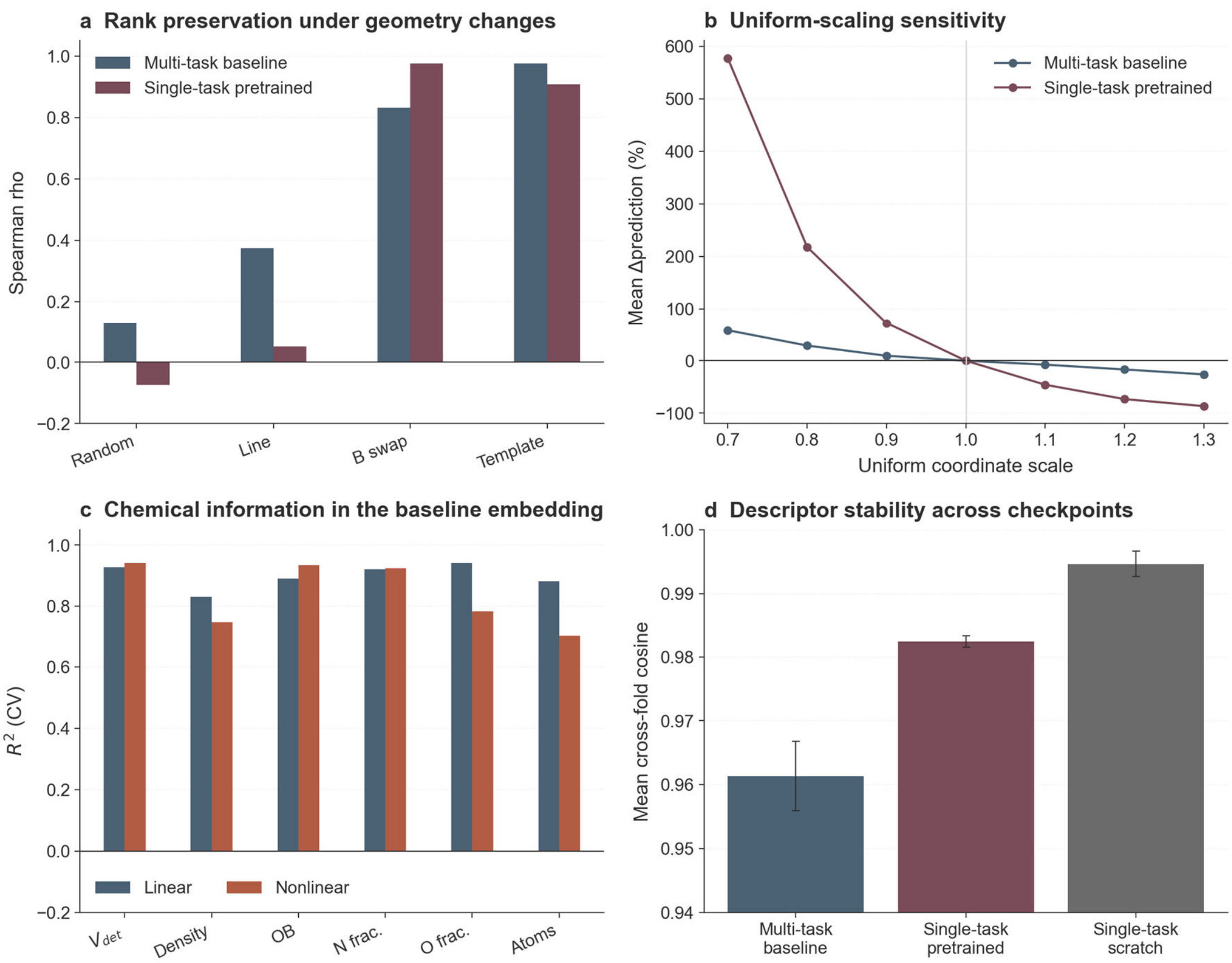

**Extended Data Fig. 5.** Representation diagnostics supporting the cluster surrogate. Panel a shows rank preservation under geometry changes, panel b shows uniform-scaling sensitivity, panel c compares linear and nonlinear probe recovery of chemical information in the baseline embedding, and panel d reports descriptor stability across fold models. Together these diagnostics test whether the learned descriptor retains chemically relevant signals rather than relying on incidental coordinate shortcuts.

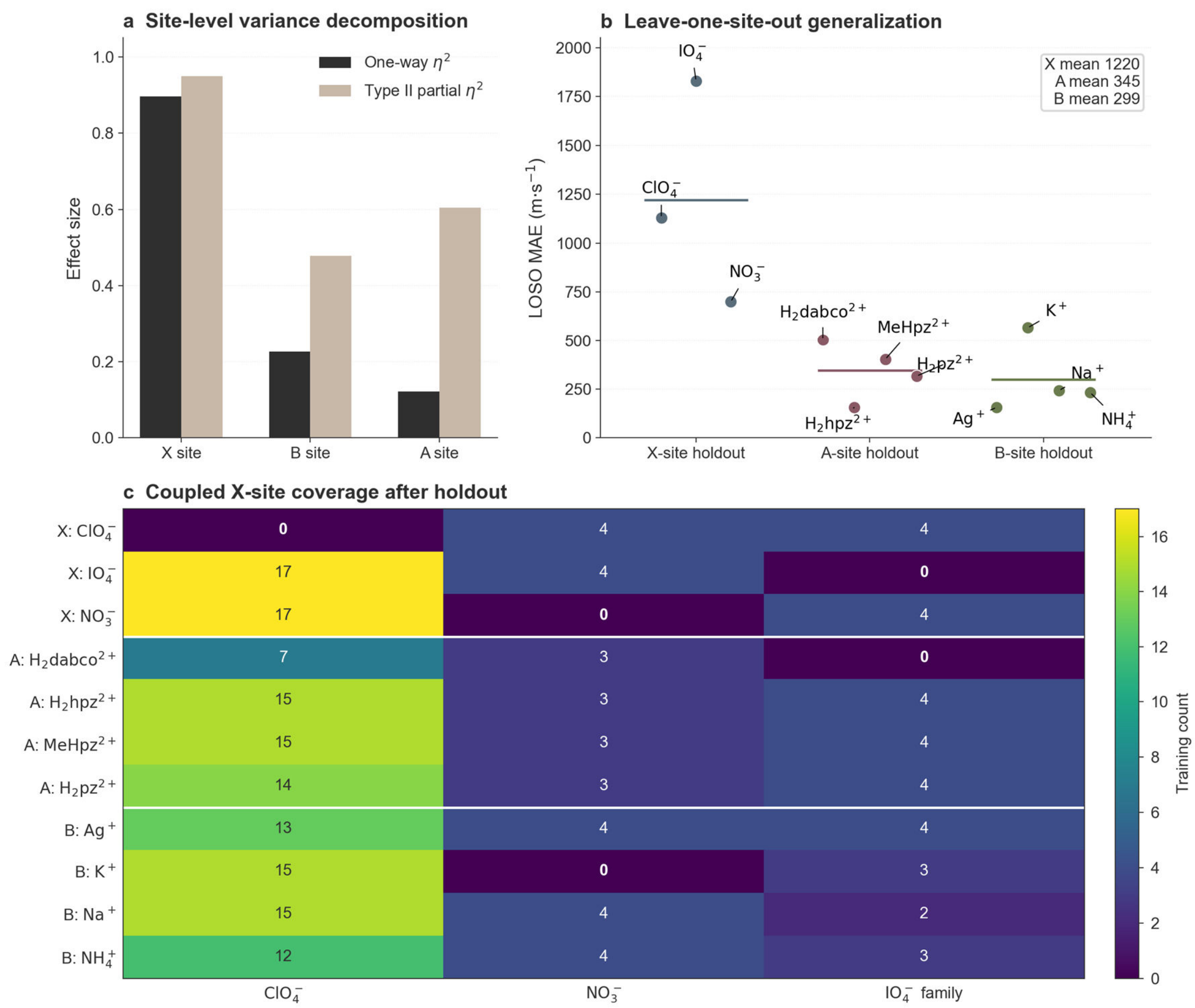

**Extended Data Fig. 6.** Site-hierarchy robustness under sparse leave-one-site-out splits. The X-site-dominant ANOVA signal is compared with leave-one-site-out MAEs and with the coupled X-site coverage that remains after each holdout, motivating a qualitative rather than fully factorial interpretation of the site hierarchy.

**Extended Data Table 3.** Minimal $ABX_4$ descriptors used for the OOD-new K–J validation. Densities are single-crystal-refined values; predicted velocities are five-fold MT-FT ensemble means.

| Material | Space group | Density (g cm⁻³) | Predicted $V_{det}$ (m·s⁻¹) | K–J $V_{det}$ (m·s⁻¹) | $\|\Delta V_{det}\|$ (m·s⁻¹) | K–J $P$ (GPa) |
|---|---|---|---|---|---|---|
| PEP | *Pbca* | 1.874 | 8870.8 | 9090 | 219.2 | 37.6 |
| PEP-M | $P2_1/c$ | 1.806 | 8698.5 | 8729 | 30.5 | 34.0 |
| PEP-H | $P2_1/c$ | 1.824 | 8790.9 | 8764 | 26.9 | 34.4 |

*Supplementary Information*

# Stoichiometric cluster learning for few-shot property prediction of multi-ionic integrated energetic materials

Ming-Yu Guo, Wei-Jia Zou, Yu Shang*, Wei-Xiong Zhang*

*MOE Key Laboratory of Bioinorganic and Synthetic Chemistry, School of Chemistry, IGCME, Sun Yat-sen University, Guangzhou, 510275, China*

*Corresponding authors are shangy8@mail.sysu.edu.cn and zhangwx6@mail.sysu.edu.cn*

## Dataset

### CHNO zero-shot baseline

To test how well existing energetic-material representations transfer to the PEM target domain, we evaluated Davis2024-trained zero-shot models on the 25 PEMs without using any PEM labels. The controls focus on two available sources of transfer signal: the Davis2024 CHNO detonation-property dataset and the public DPA-3.2-5M atomistic foundation model.

The Davis2024 dataset contains 12,498 CHNO energetic crystals, but it does not cover the ionic building-block chemistries that dominate the PEM target domain.[28] Models using Davis2024-based detonation-property supervision therefore give km $s^{-1}$-scale errors on PEMs (Table S1; Extended Data Fig. 1), showing that CHNO labels alone do not provide a PEM-ready detonation-velocity prior.

DPA-3.2-5M provides the public foundation-model control. It is a publicly released DPA3 large atomistic model based on the graph-neural-network architecture[33] and trained with a multi-task strategy on the OpenLAM datasets V2. (Details: https://www.aissquare.com/models/detail?pageType=models&name=DPA-3.2-5M&id=392, accessed on 2026-07-05.) Its zero-shot crystal MAE on PEMs remains 1151.3 m $s^{-1}$, indicating that a public general-purpose atomistic model plus CHNO detonation labels is not sufficient for reliable PEM transfer. Analogous CHNO-only training with the energetic-material DeepEMs-LAM backbone also remains above 1 km $s^{-1}$ error on periodic PEM crystals (1372.6 m $s^{-1}$). These results show that chemically distinct energetic-material subdomains still require dedicated fine-tuning for transfer, here PEM-labelled cluster MT-FT rather than CHNO-only transfer. As the in-domain reference, adding the 25 PEM target labels through the cluster-based multi-task fine-tuning (MT-FT) pipeline (Methods) reduces the MAE to 273.2 m $s^{-1}$.

The zero-shot settings are:

- **CHNO-only DPA crystal / molecule**: DPA-3.2-5M backbone trained only on Davis2024 CHNO labels and evaluated on either periodic PEM crystals or non-periodic molecular inputs, with no PEM labels.
- **CHNO-only DeepEMs crystal / molecule**: DeepEMs-LAM backbone under the same CHNO-only single-task training/evaluation protocol.
- **CHNO+DFT+experiment transfer**: DeepEMs-LAM multi-head transfer model trained with Davis2024 labels, DFT energy–force supervision, and the experimental-property head used for the Davis2024 task, still without PEM labels.
- **CHNO+DFT transfer**: DeepEMs-LAM two-head transfer model trained with Davis2024 labels and DFT energy–force supervision, again evaluated zero-shot on PEMs.

**Table S1.** Davis2024-based zero-shot baselines on the 25-material PEM set. Crystal MAE evaluates each model on PEM crystal CIFs; cluster $n_1$ MAE evaluates on the stoichiometric vacancy-cluster representation used by the main-text MT-FT pipeline. All entries use the saved model at the specified training step of the corresponding run (200k steps for single-task baselines, 400k steps for multi-head transfer baselines) and are taken without retraining on PEM labels.

| Baseline | Configuration | Crystal MAE (m $s^{-1}$) | Cluster_n1 MAE (m $s^{-1}$) |
| --- | --- | --- | --- |
| CHNO-only DPA crystal | DPA-3.2-5M, CHNO labels, crystal training input | 1151.3 | 6876.6 |
| CHNO-only DPA molecule | DPA-3.2-5M, CHNO labels, molecular training input | 4021.1 | 2110.0 |
| CHNO-only DeepEMs crystal | DeepEMs-LAM, CHNO labels, crystal training input | 1372.6 | 3253.9 |
| CHNO-only DeepEMs molecule | DeepEMs-LAM, CHNO labels, molecular training input | 1827.4 | 1486.7 |
| CHNO+DFT+experiment transfer | DeepEMs-LAM, CHNO labels + DFT + experimental-property head | 2543.5 | 1706.7 |
| CHNO+DFT transfer | DeepEMs-LAM, CHNO labels + DFT | 2364.8 | 1417.8 |

## Training materials: nomenclatures and glossaries

Table S2 lists the 25 materials used for model training and cross-validation.

**Table S2.** Ionic-site assignments and experimentally-derived detonation velocities for the 25 training materials. $V_{det}$: detonation velocity (m s$^{-1}$), converted from km s$^{-1}$ reported in the cited literature.

| Material | *A*-site | *B*-site | *X*-site | $V_{det}$ (m s$^{-1}$) | Reference |
|---|---|---|---|---|---|
| DAI-1 | $H_2dabco^{2+}$ | $Na^+$ | $IO_4^-$ | 6438 | S1 |
| DAI-2 | $H_2dabco^{2+}$ | $K^+$ | $IO_4^-$ | 6331 | S1 |
| DAI-4 | $H_2dabco^{2+}$ | $NH_4^+$ | $IO_4^-$ | 6558 | S1 |
| DAI-X1 | $H_2dabco^{2+}$ | $Na^+$ | $[H_4IO_6]^-$ | 7070 | S2 |
| DAN-2 | $H_2dabco^{2+}$ | $K^+$ | $NO_3^-$ | 7651 | S3 |
| DAP-1 | $H_2dabco^{2+}$ | $Na^+$ | $ClO_4^-$ | 8781 | 16 |
| DAP-2 | $H_2dabco^{2+}$ | $K^+$ | $ClO_4^-$ | 8591 | 16 |
| DAP-3 | $H_2dabco^{2+}$ | $Rb^+$ | $ClO_4^-$ | 8425 | 16 |
| DAP-4 | $H_2dabco^{2+}$ | $NH_4^+$ | $ClO_4^-$ | 8806 | 16 |
| DAP-5 | $H_2dabco^{2+}$ | $Ag^+$ | $ClO_4^-$ | 8534 | S4 |
| DAP-6 | $H_2dabco^{2+}$ | $NH_3OH^+$ | $ClO_4^-$ | 9123 | S5 |
| DAP-7 | $H_2dabco^{2+}$ | $NH_2NH_3^+$ | $ClO_4^-$ | 8883 | S5 |
| DAP-M4 | $MeHdabco^{2+}$ | $NH_4^+$ | $ClO_4^-$ | 8085 | 18 |
| DAP-O2 | $H_2odabco^{2+}$ | $K^+$ | $ClO_4^-$ | 8327 | S6 |
| DAP-O4 | $H_2odabco^{2+}$ | $NH_4^+$ | $ClO_4^-$ | 8900 | 18 |
| PAN-2 | $H_2pz^{2+}$ | $K^+$ | $NO_3^-$ | 8140 | S7 |
| PAP-1 | $H_2pz^{2+}$ | $Na^+$ | $ClO_4^-$ | 8917 | S6 |
| PAP-4 | $H_2pz^{2+}$ | $NH_4^+$ | $ClO_4^-$ | 8629 | 18 |
| PAP-5 | $H_2pz^{2+}$ | $Ag^+$ | $ClO_4^-$ | 8961 | S4 |
| PAN-H2 | $H_2hpz^{2+}$ | $K^+$ | $NO_3^-$ | 7500 | S7 |
| PAP-H4 | $H_2hpz^{2+}$ | $NH_4^+$ | $ClO_4^-$ | 8756 | 18 |
| PAP-H5 | $H_2hpz^{2+}$ | $Ag^+$ | $ClO_4^-$ | 8756 | S4 |
| PAN-M2 | $MeHpz^{2+}$ | $K^+$ | $NO_3^-$ | 7200 | S7 |
| PAP-M4 | $MeHpz^{2+}$ | $NH_4^+$ | $ClO_4^-$ | 8311 | 18 |
| PAP-M5 | $MeHpz^{2+}$ | $Ag^+$ | $ClO_4^-$ | 8778 | S4 |

**Table S3.** Glossary of A-, B-, and X-site ions used throughout the main text and Supplementary Information.

| Site | Abbreviation | IUPAC name | Formula |
|---|---|---|---|
| A | $H_2dabco^{2+}$ | 1,4-diazoniabicyclo[2.2.2]octane | $C_6H_{14}N_2^{2+}$ |
| A | $MeHdabco^{2+}$ | 1-methyl-1,4-diazoniabicyclo[2.2.2]octane | $C_7H_{16}N_2^{2+}$ |
| A | $H_2odabco^{2+}$ | 1-hydroxy-1,4-diazabicyclo[2.2.2]octane-1,4-diium | $C_6H_{14}ON_2^{2+}$ |
| A | $H_2pz^{2+}$ | piperazine-1,4-diium | $C_4H_{12}N_2^{2+}$ |
| A | $MeHpz^{2+}$ | 1-methylpiperazine-1,4-diium | $C_5H_{14}N_2^{2+}$ |
| A | $H_2hpz^{2+}$ | 1,4-diazepane-1,4-diium | $C_5H_{14}N_2^{2+}$ |
| A | $Huru^{2+}$ | 1,3,5,7-tetraazatricyclo[3.3.1.1$^{3,7}$]decane-1,3-diium | $C_6H_{14}N_4^{2+}$ |
| B | $Na^+$ | sodium | $Na^+$ |
| B | $K^+$ | potassium | $K^+$ |
| B | $Rb^+$ | rubidium | $Rb^+$ |
| B | $Ag^+$ | silver(I) | $Ag^+$ |
| B | $NH_4^+$ | ammonium | $NH_4^+$ |
| B | $NH_3OH^+$ | hydroxylammonium | $NH_3OH^+$ |
| B | $NH_2NH_3^+$ | hydrazinium | $N_2H_5^+$ |
| B | $H_2en^{2+}$ | ethane-1,2-diaminium | $C_2H_{10}N_2^{2+}$ |
| X | $ClO_4^-$ | perchlorate | $ClO_4^-$ |
| X | $ClO_3^-$ | chlorate | $ClO_3^-$ |
| X | $NO_3^-$ | nitrate | $NO_3^-$ |
| X | $IO_4^-$ | metaperiodate | $IO_4^-$ |
| X | $[H_4IO_6]^-$ | tetrahydroxidodioxidoiodate(VII) | $H_4IO_6^-$ |

## Five-fold cross-validation fold assignments

Table S4 lists the five-material validation set for each of the five cross-validation folds. Fold assignments were made by random sampling (random seed 42), yielding validation folds that each contain materials from multiple coarse X-site families where possible and training folds that retain perchlorate, nitrate, and periodate/orthoperiodate examples.

**Table S4.** Five-fold cross-validation fold assignments. Each fold holds out five materials as the validation set; the remaining twenty are used for training.

| Fold | Validation materials |
|---|---|
| 0 | PAN-H2, DAP-M4, DAP-5, PAP-4, PAP-1 |
| 1 | DAP-2, DAP-1, DAP-6, PAN-2, PAP-H4 |
| 2 | DAP-7, PAN-M2, DAI-2, DAP-O4, PAP-H5 |
| 3 | DAP-O2, DAI-4, PAP-M4, DAN-2, PAP-M5 |
| 4 | DAP-3, DAP-4, DAI-1, DAI-X1, PAP-5 |

# Cluster representation

## Representative stoichiometric clusters: training set

The 25 training materials each yield up to three geometrically distinct stoichiometric clusters (variants $n_1$, $n_2$, $n_3$; see also Table S5). Atom colours: H (light grey), C (dark grey), N (blue), O (red), Cl (green); axis triads in each panel indicate the *x*/*y*/*z* directions of the 100 Å non-periodic simulation box.

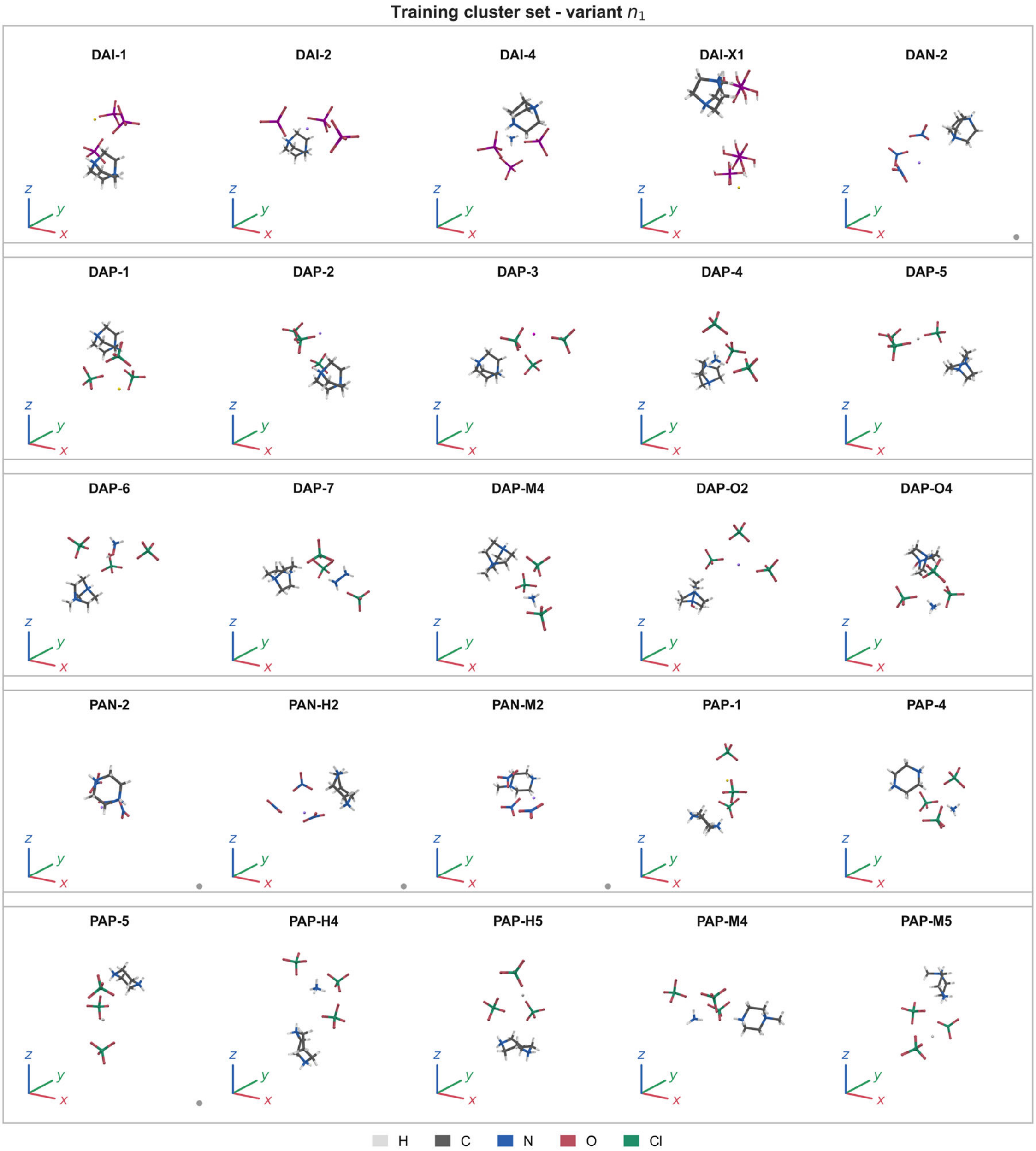

**Figure S1.** Training-set stoichiometric clusters: variant $n_1$ (seed-spread index 0). Each panel is the simplest stoichiometric unit extracted from a 2×2×2 (or larger) supercell by vacancy construction centred on the chosen seed molecule. Material name above each panel. A grey dot in the lower-right corner of a panel indicates that this variant is geometrically identical to either $n_2$ or $n_3$ for that material (cf. Table S5).

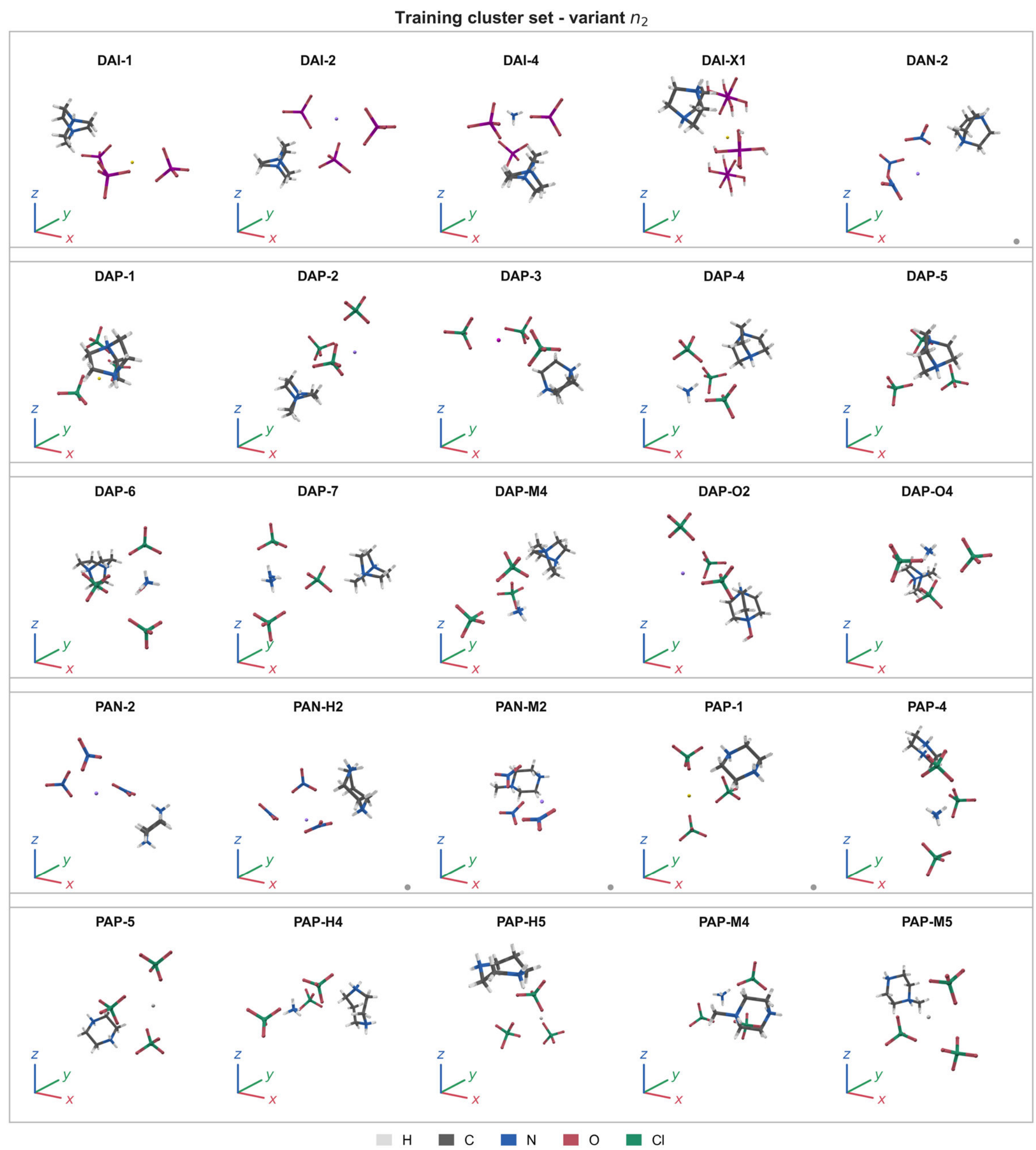

**Figure S2.** Training-set stoichiometric clusters: variant $n_2$ (seed-spread index 1). A grey dot in the lower-right corner of a panel indicates that this variant is geometrically identical to either $n_1$ or $n_3$ for that material (cf. Table S5).

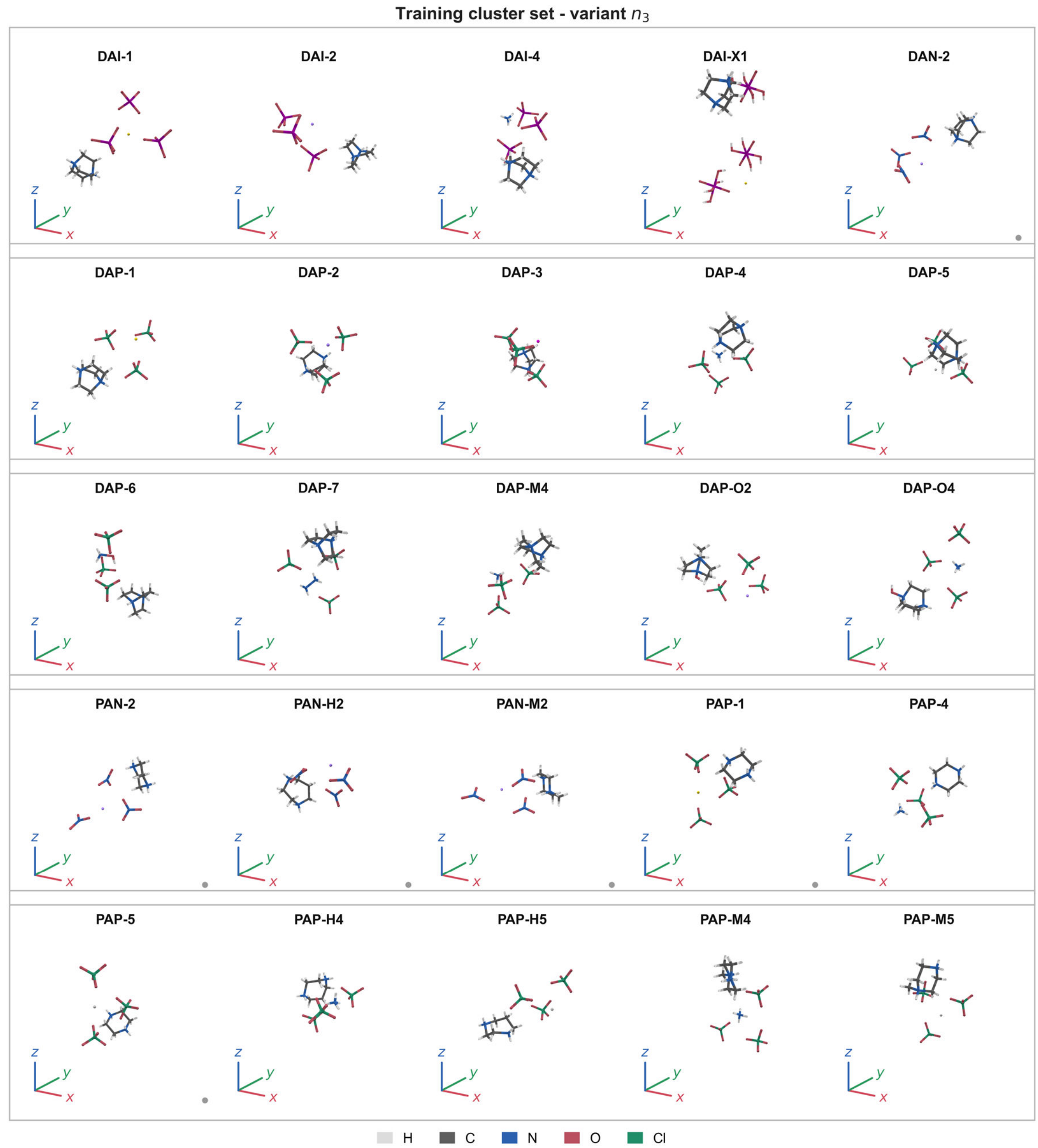

**Figure S3.** Training-set stoichiometric clusters: variant $n_3$ (seed-spread index 2). A grey dot in the lower-right corner of a panel indicates that this variant is geometrically identical to either $n_1$ or $n_2$ for that material (cf. Table S5).

## Representative stoichiometric clusters: OOD materials

The eight OOD materials are split into two reporting categories used throughout the SI ablation tables and OOD-holdout diagnostic figure (Table S18, Extended Data Fig. 3). **OOD-holdout** contains five real, externally synthesized energetic materials: DAC-4, TAP-2, DAI-$1_{0.5}4_{0.5}$ (previously reported as "DPPE-1"), EAP-4, and DEP (previously reported as "SY"); **OOD-new** contains three template-derived $ABX_4$ analogues (PEP, PEP-M, PEP-H) generated for extrapolation testing. All three cluster variants ($n_1$, $n_2$, $n_3$) are shown per material, built from the experimentally resolved (or template-derived) crystal structures using the same vacancy-construction pipeline as the training set.

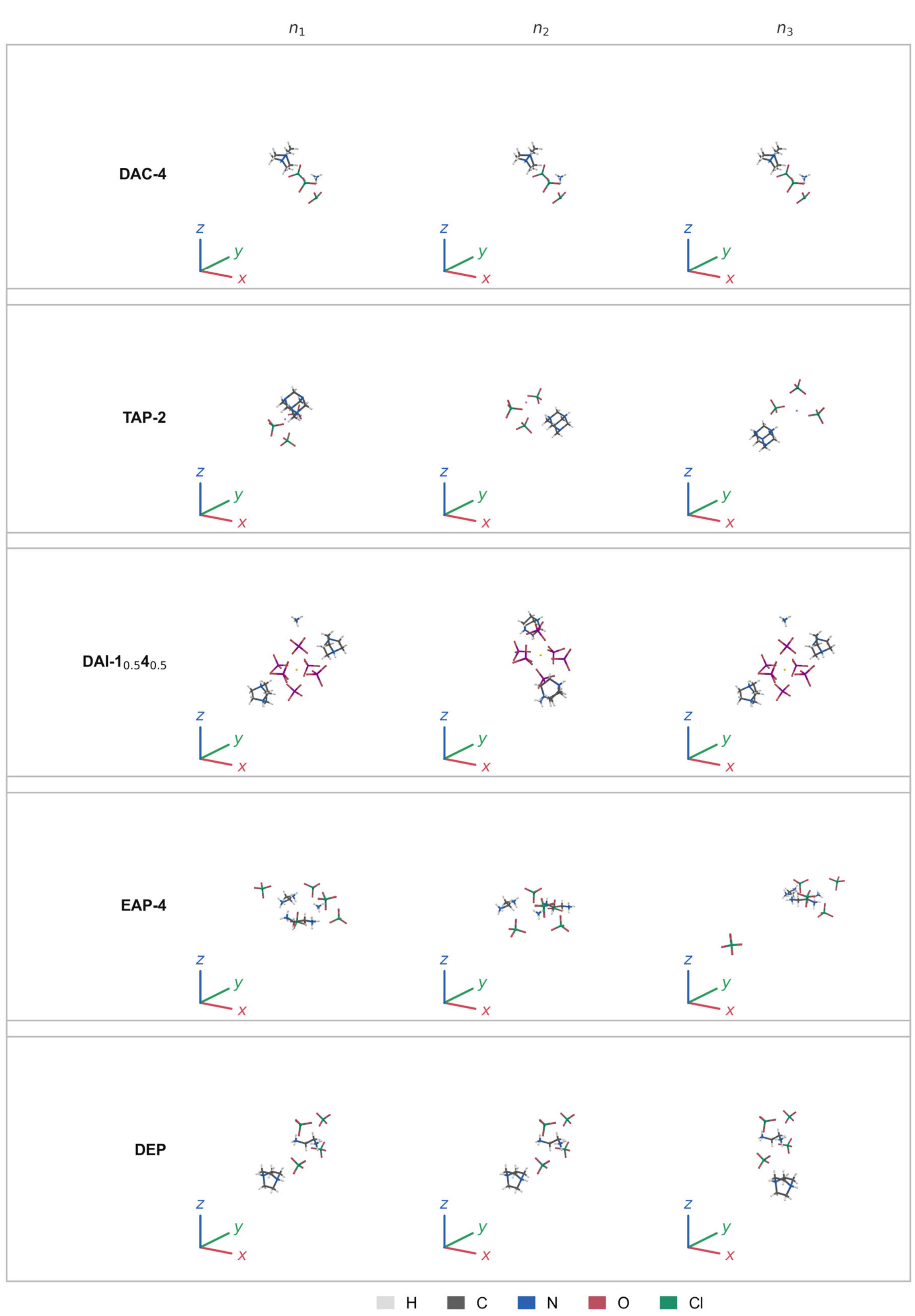

$n_1$
$n_2$
$n_3$
DAC-4
TAP-2
DAI-$1_{0.5}4_{0.5}$
EAP-4
DEP
z
y
x
H
C
N
O
Cl

**Figure S4.** OOD-holdout stoichiometric clusters. Rows are materials (DAC-4, TAP-2, DAI-$1_{0.5}4_{0.5}$, EAP-4, DEP), and columns are variants $n_1$, $n_2$, $n_3$. Atom colours are as in the training-set figures. Exact degeneracies in this panel set are as follows. DAC-4 has $n_1=n_2=n_3$. TAP-2 has $n_1=n_3$. DAI-$1_{0.5}4_{0.5}$ has $n_1=n_2=n_3$. EAP-4 has two geometrically distinct cluster variants among the displayed variants. DEP has $n_1=n_2$.

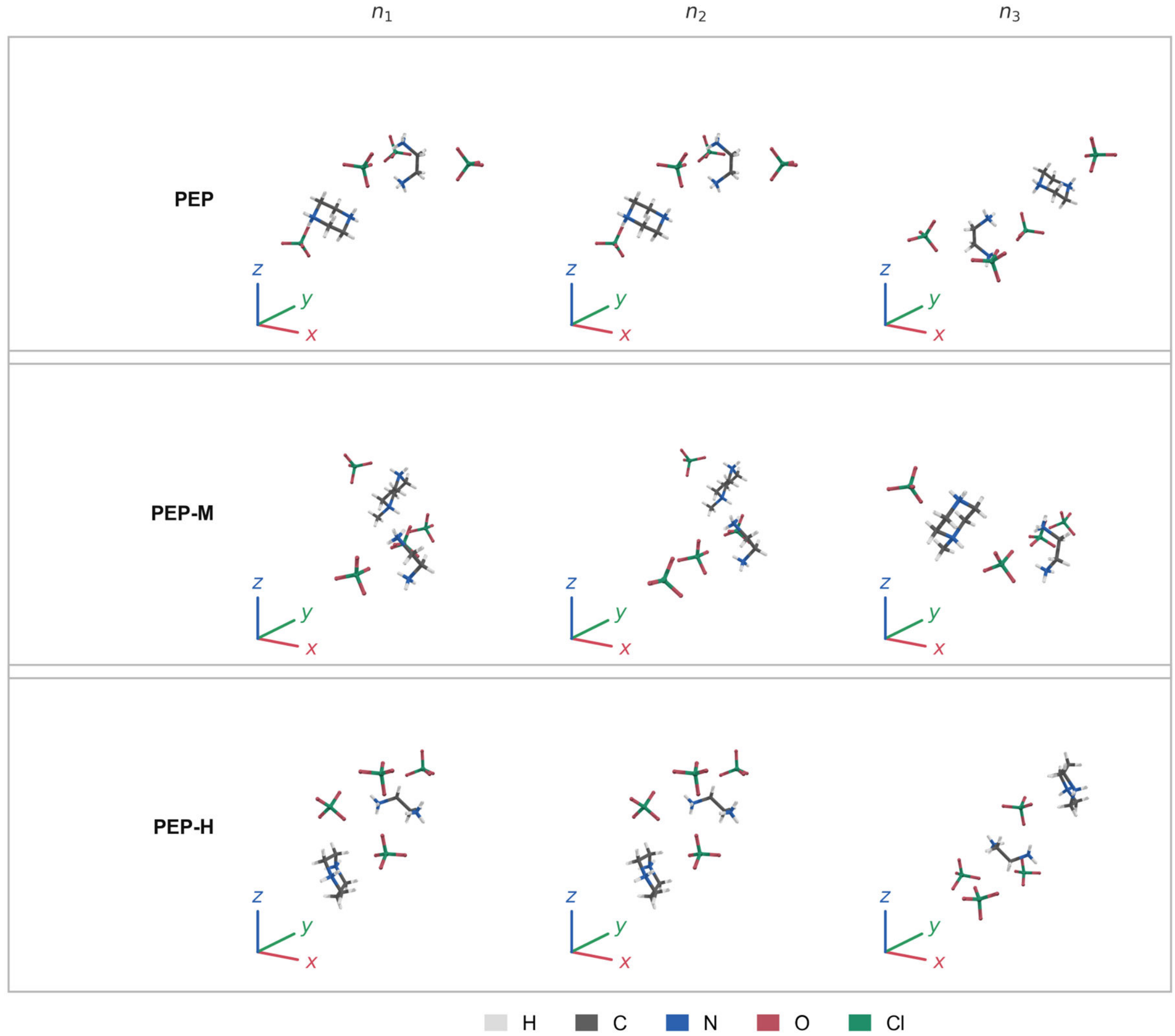

**Figure S5.** OOD-new stoichiometric clusters. Rows are materials (PEP, PEP-M, PEP-H), and columns are variants $n_1$, $n_2$, $n_3$. Atom colours are as in the training-set figures.

## Cluster perturbations used in the packing-compatibility probe

The Fig. 4d perturbation ladder groups operations by how much chemical organization they preserve. Shell-preserving operations keep fragment identities intact and modify either the whole cluster or centre-of-mass separations (rigid rotation, rigid translation, template substitution, uniform scaling, and A–X/B–X/A–B stretches). Fragment-level swaps exchange A-, B-, or X-site fragment centre-of-mass positions while preserving internal molecular geometries. Geometry-destroying operations remove the original fragment organization by permuting atom coordinates or replacing coordinates with random, spherical, or line-like arrangements.

The perturbation grids below render representative structures for the six additional operations used in Fig. 4d. The three stretch figures use scale factor 1.10, while the three fragment-swap figures use seed 42.

B-X stretch, scale 1.10
DAI-1
DAI-2
DAI-4
DAI-X1
DAN-2
DAP-1
DAP-2
DAP-3
DAP-4
DAP-5
DAP-6
DAP-7
DAP-M4
DAP-O2
DAP-O4
PAN-2
PAN-H2
PAN-M2
PAP-1
PAP-4
PAP-5
PAP-H4
PAP-H5
PAP-M4
PAP-M5
z
y
x
H
C
N
O
Cl

**Figure S6.** B–X stretch perturbation structures used in the packing-compatibility probe, shown at scale factor 1.10. Each panel starts from the training-set $n_1$ stoichiometric cluster and rigidly displaces each X-site fragment along its nearest B–X centre-of-mass axis.

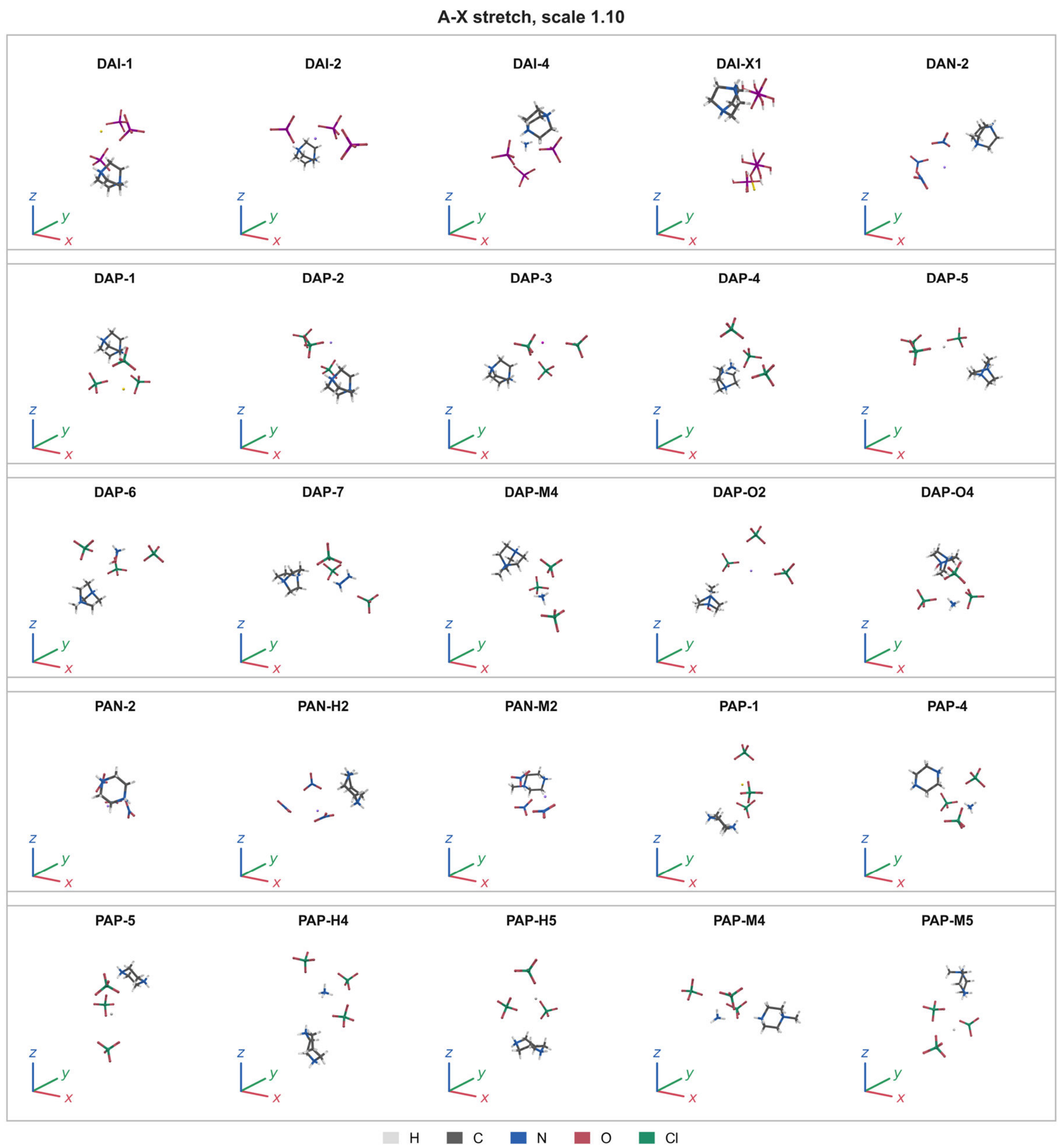

**Figure S7.** A–X stretch perturbation structures used in the packing-compatibility probe, shown at scale factor 1.10. Each X-site fragment is rigidly displaced along its nearest A–X centre-of-mass axis.

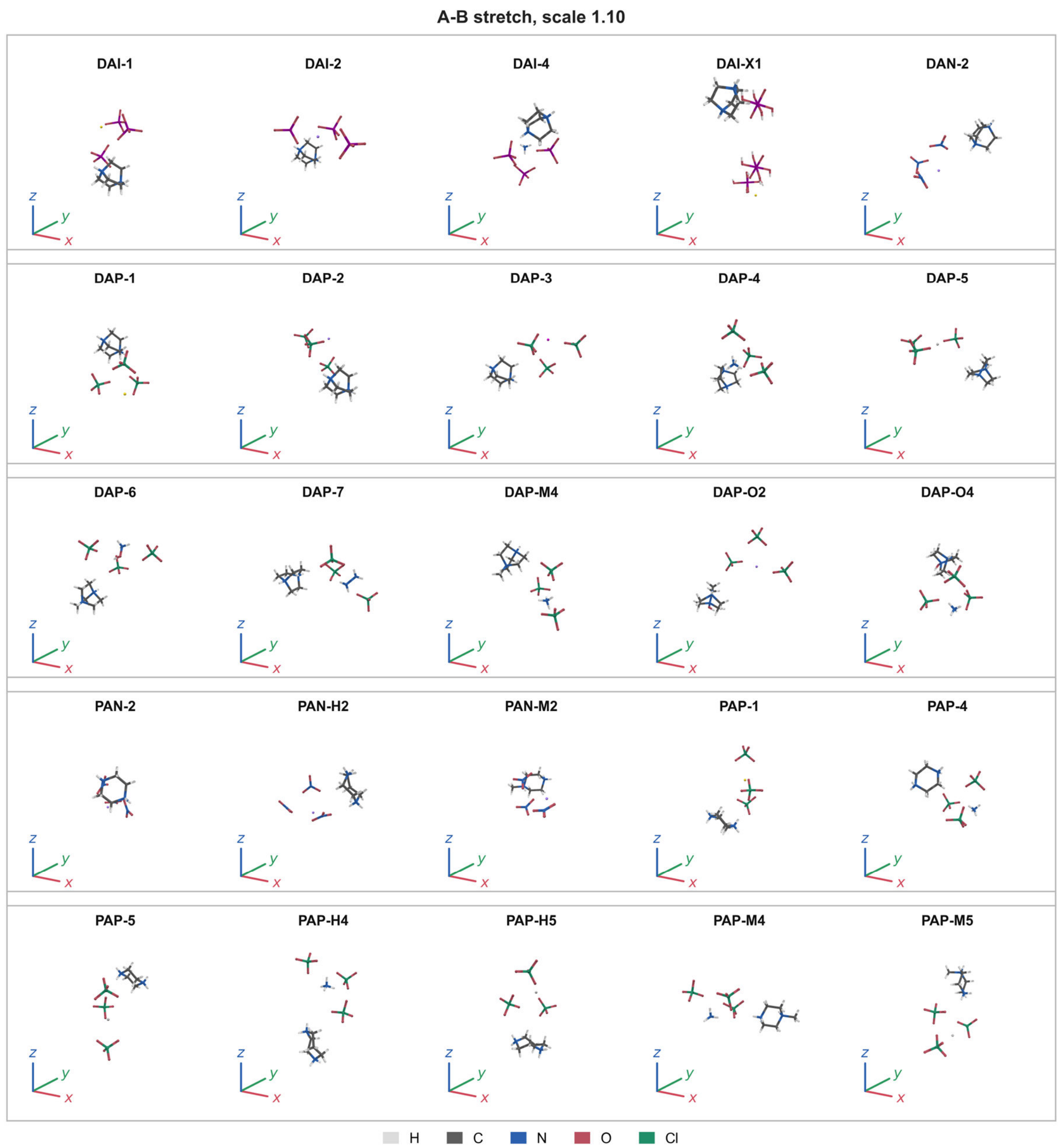

**Figure S8.** A–B stretch perturbation structures used in the packing-compatibility probe, shown at scale factor 1.10. Each A-site fragment is rigidly displaced along its nearest A–B centre-of-mass axis while preserving internal molecular geometry.

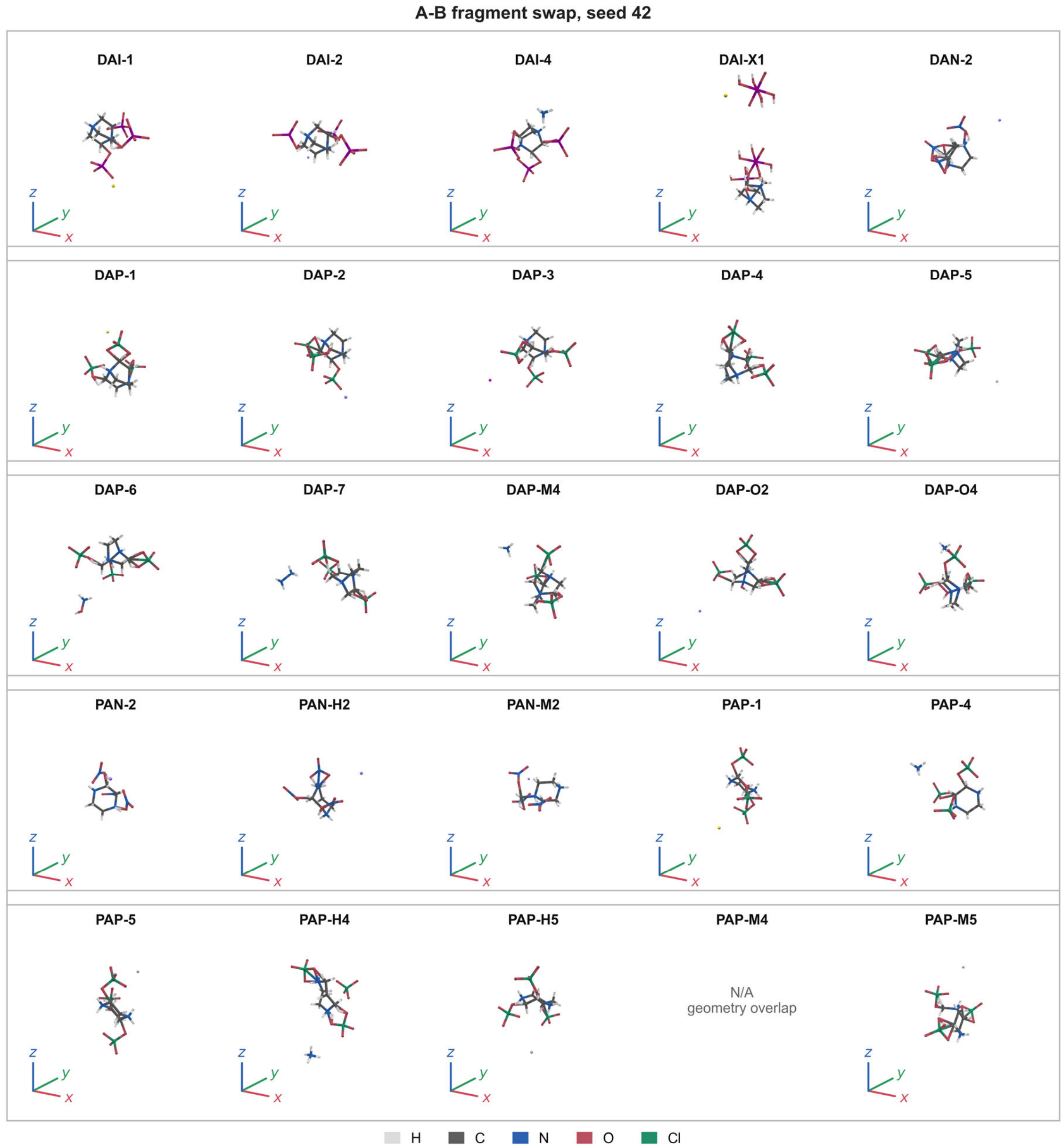

**Figure S9.** A–B fragment-swap perturbation structures used in the packing-compatibility probe, shown for seed 42. One A-site and one B-site fragment exchange centre-of-mass positions while each fragment remains internally rigid; the N/A tile marks the single material for which this representative swap creates an unresolved atom overlap.

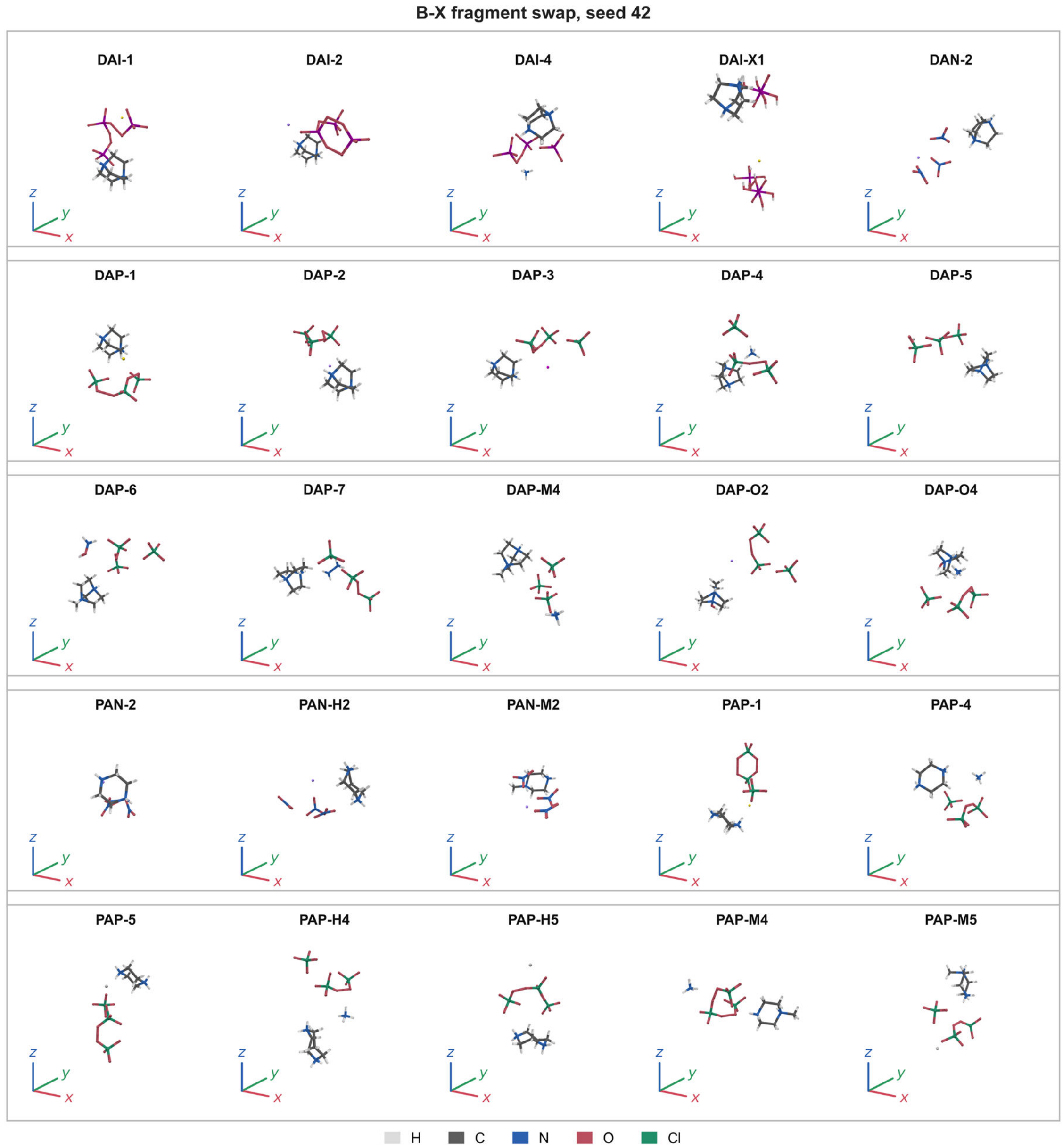

**Figure S10.** B–X fragment-swap perturbation structures used in the packing-compatibility probe, shown for seed 42. One B-site and one X-site fragment exchange centre-of-mass positions while preserving internal fragment geometry.

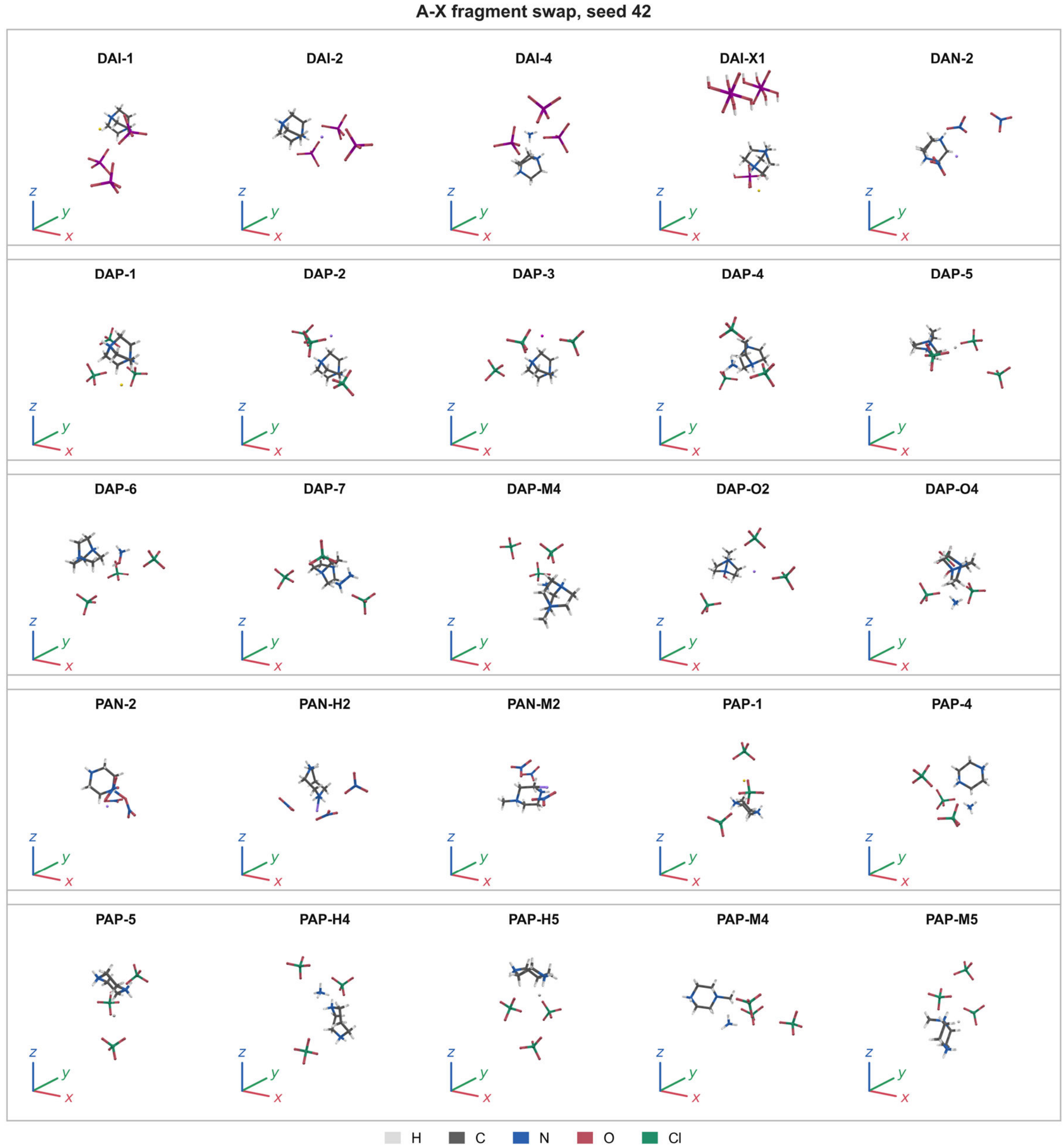

**Figure S11.** A–X fragment-swap perturbation structures used in the packing-compatibility probe, shown for seed 42. One A-site and one X-site fragment exchange centre-of-mass positions while preserving internal fragment geometry.

## Distinct cluster variants per training material

Up to three cluster variants ($n_1$, $n_2$, $n_3$) were generated per material by selecting maximally spread seed molecules in the supercell. For high-symmetry crystals with fewer than three structurally distinct seed positions, two or all three variants are geometrically identical. Across the 25-material training set this yields 50 distinct clusters across 75 nominal training frames (7 materials with 1 distinct variant, 11 with 2, and 7 with 3). Distinctness was assessed by comparing sorted pairwise distance matrices (tolerance $10^{-4}$ Å).

**Table S5.** Number of geometrically distinct cluster variants per training material. "Distinct" counts the number of distinct structures among $n_1$, $n_2$, $n_3$.

| Material | Distinct variants | Identical pairs | Note |
|---|---|---|---|
| DAI-1 | 3 | N/A | all distinct |
| DAI-2 | 3 | N/A | all distinct |
| DAI-4 | 2 | $n_1$=$n_2$ | |
| DAI-X1 | 2 | $n_1$=$n_3$ | |
| DAN-2 | 1 | $n_1$=$n_2$=$n_3$ | single distinct variant |
| DAP-1 | 3 | N/A | all distinct |
| DAP-2 | 2 | $n_1$=$n_2$ | |
| DAP-3 | 2 | $n_1$=$n_2$ | |
| DAP-4 | 2 | $n_1$=$n_3$ | |
| DAP-5 | 3 | N/A | all distinct |
| DAP-6 | 1 | $n_1$=$n_2$=$n_3$ | single distinct variant |
| DAP-7 | 2 | $n_1$=$n_2$ | |
| DAP-M4 | 1 | $n_1$=$n_2$=$n_3$ | single distinct variant |
| DAP-O2 | 3 | N/A | all distinct |
| DAP-O4 | 3 | N/A | all distinct |
| PAN-2 | 1 | $n_1$=$n_2$=$n_3$ | single distinct variant |
| PAN-H2 | 1 | $n_1$=$n_2$=$n_3$ | single distinct variant |
| PAN-M2 | 1 | $n_1$=$n_2$=$n_3$ | single distinct variant |
| PAP-1 | 2 | $n_1$=$n_2$ | |
| PAP-4 | 3 | N/A | all distinct |
| PAP-5 | 1 | $n_1$=$n_2$=$n_3$ | single distinct variant |
| PAP-H4 | 2 | $n_1$=$n_2$ | |
| PAP-H5 | 2 | $n_1$=$n_2$ | |
| PAP-M4 | 2 | $n_1$=$n_2$ | |
| PAP-M5 | 2 | $n_1$=$n_2$ | |
| **Total** | **50** | | 7 with 1 distinct; 11 with 2; 7 with 3 |

## Cluster-construction algorithm

The full procedure is implemented in an open-source materials-cluster construction utility; the pseudocode below summarizes the four logical stages.

**Stage 1: parse and group molecules.** The ordered CIF is parsed into molecular components by connecting all atom pairs whose distance is below a covalent threshold, with explicit overrides (Table S6) preventing ionic alkali-/silver-/iodine–oxygen contacts from being merged into covalent fragments. Molecular species are then grouped by chemical formula and graph isomorphism, so that conformers and translation copies of the same molecule are assigned to a single species id.

**Stage 2: simplest stoichiometric unit.** Let $\boldsymbol{s}$=($s_1$,…,$s_K$) be the per-species molecule counts in the unit cell. The simplest stoichiometric unit defines the target spec for one cluster instance:

$$u = \frac{s}{\gcd(s)}. \quad \text{(S1)}$$

**Stage 3: spread-seed selection.** For each cluster realization $n_k$∈$n_1$,$n_2$,$n_3$ the unit cell is expanded over the fixed schedule ((2,2,2), (3,3,3), (2,3,3), (3,3,4), (4,4,4)) until at least three structurally distinct, maximally spread seed molecules of the rarest target species can be chosen. Starting from the seed-species centroid closest to the supercell origin, additional seeds are added by farthest-point sampling on minimum-image fractional centroids. Once |*seeds*|≥3, the seed for variant $n_k$ is selected at the deterministic offset 0, 1, or 2.

**Stage 4: stoichiometric removal by nearest-neighbour growth.** Starting from the chosen seed and the residual species quota, the cluster is grown by greedily adding the molecule that minimizes its minimum-image distance to the current removal set, restricted to candidates whose species still has unmet quota. The full algorithm is summarized below.

**Algorithm S1** Stoichiometric vacancy-cluster construction.

**Require:** ordered CIF C, supercell schedule S, variant offset o ∈ 0,1,2, bond thresholds T (Table S6)
**Ensure:** non-periodic cluster C centred in a 100 Å box
1: M ← molecules from C using thresholds T
2: **s** ← per-species counts in unit cell, grouped by formula+graph isomorphism
3: **u** ← **s** / gcd(**s**) *// simplest stoichiometric unit*
4: **for** $(n_a, n_b, n_c)$ ∈ S **do**
  5: $M^\star$ ← supercell expansion of M by $(n_a, n_b, n_c)$
  6: seed_species ← rarest species in **u** with at least $u_{rare}$ candidates in $M^\star$
  7: F ← farthest-point-sampled centroids in seed_species under min-image distance
  8: **if** | F | ≥ 3 **then break**
  9: **end if**
10: **end for**
11: seed ← F[o]
12: removal ← seed;   needed ← **u** - $\mathbf{1}_{seed.species}$
13: **while** needed > 0 **do**
  14: cand ← m ∈ $M^\star$ \ removal : needed[m.species] > 0
  15: $m^\star \leftarrow \arg\min_{m \in cand} \min_{r \in removal} d_{min\text{-}img}(m, r)$
  16: removal ← removal ∪ $m^\star$;   needed[$m^\star$.species] -= 1
17: **end while**
18: C ← minimum-image wrap of removal centred in a 100 Å cubic box, periodic boundary disabled
19: **return** C

The supercell schedule, spread-seed quotas, and per-variant deterministic offsets used in production are listed in Table S7; no coordinate perturbation is applied between variants, so $n_1$, $n_2$, and $n_3$ differ only in which spread seed is selected at line 10 of Algorithm S1.

**Table S6.** Explicit pairwise bond-distance limits used for PEM molecule identification when heavy or metallic atoms are present. All values are maximum distances in Å; reverse-order pairs use the same limit.

| Pair class | Elements | Maximum distance (Å) |
|---|---|---|
| I–C/H/N/Cl | I with C, H, N, Cl | 2.10 |
| Na–C/H/N/Cl | Na with C, H, N, Cl | 2.30 |
| K–C/H/N/Cl | K with C, H, N, Cl | 2.50 |
| Rb/Ba–C/H/N/Cl | Rb or Ba with C, H, N, Cl | 2.60 |
| Ag–C/H/N/Cl | Ag with C, H, N, Cl | 2.30 |
| I–O | I with O | 2.05 |
| Na/Ag–O | Na or Ag with O | 2.20 |
| K–O | K with O | 2.30 |
| Rb/Ba–O | Rb or Ba with O | 2.40 |
| Heavy–heavy | I, Na, K, Rb, Ba, Ag pairs | 3.20 |

**Table S7.** Supercell and seed-selection parameters used for stoichiometric vacancy-cluster extraction.

| Parameter | Value |
|---|---|
| Supercell schedule | (2,2,2), (3,3,3), (2,3,3), (3,3,4), (4,4,4) |
| Minimum spread seeds | 3 |
| Candidate spread seeds for diversity guard | 10 |
| Cluster random seeds | $n_1 = 101$, $n_2 = 202$, $n_3 = 303$ |
| Default seed offsets | $n_1 = 0$, $n_2 = 1$, $n_3 = 2$ |
| Geometric identity tolerance | $10^{-4}$ Å on sorted pairwise distances |
| Export box | 100 Å cubic, non-periodic |

# Model evaluation

## Adaptation strategies and representation variants

The main text reports MT-FT as the production model; Extended Data Table 1 summarizes the primary adaptation strategies, representation controls, split controls, and full-data production check used below. MT-FT (two heads: DFT energy and detonation-velocity property) jointly trains the shared descriptor on 67 DFT energy–force systems and on 20 in-fold PEM materials per fold (60 cluster systems, three cluster variants per material); the energy head inherits its fitting-net weights from the pretrained backbone, the property head is randomly initialized, and the sampling probability is 0.50 for each head. An MT-FT high-LR variant (MT-FT-highLR, Extended Data Table 1) keeps the same two-head cluster architecture but raises the initial learning rate to $1\times10^{-4}$. MT-FT-aux (three heads) adds a third head trained on the 10,623-system CHNO auxiliary dataset, with equal normalized sampling across the energy, auxiliary, and target heads. ST-FT finetunes only a single property head from the pretrained backbone, with no energy regularization. ST-TFS trains the same single-head architecture from random initialization, providing a lower bound on the contribution of the pretrained backbone. MT-FT-crystal keeps the two-head baseline recipe but replaces non-periodic cluster inputs with resolved periodic crystal cells under periodic boundary conditions. The default multi-task cluster and crystal strategies use an exponential learning-rate schedule that starts at $2\times10^{-5}$, decays by a factor of two every 5 000 steps, asymptotes to $10^{-8}$, and runs for 400 000 training steps total; the single-task strategies use the same start and stop rates but run for 200 000 steps with 2 500-step decay. The auxiliary-head variant uses the same schedule architecture but runs for 600 000 steps with 7 500-step decay at the default learning rate, so that the property head sees the same number of effective decay periods (80) given the equal per-head sampling. The higher-learning-rate auxiliary-head sweep starts at $1\times10^{-4}$ and decays every 1 500 steps to the same $10^{-8}$ floor.

## Per-seed and per-fold MAE distribution

**Table S8.** Per-fold MAE (m s$^{-1}$), mean MAE with bootstrap 95% confidence interval (10,000 resamples at the material level, percentile method, $n$=25 materials), and paired Wilcoxon signed-rank $p$ value against MT-FT. Resampling unit is the per-material absolute error averaged across the three held-out cluster variants per material to respect non-independence between variants sharing a common experimentally-derived $V_{det}$. For ST-FT, the fold-mean MAE of 323.4 m s$^{-1}$ is 18% higher than MT-FT (273.2 m s$^{-1}$), but the paired Wilcoxon $p$=0.83 is non-significant at $n$=25 because the mean difference is dominated by the fold-4 DAI-X1 outlier (single-task absolute error 2086 m s$^{-1}$); the median paired difference is only +4.6 m s$^{-1}$. Paired Wilcoxon $p$ values are computed against MT-FT; the baseline row is therefore marked as the reference. The MT-FT-aux row corresponds to the three-head variant defined in Extended Data Table 1.

| Variant | F0 | F1 | F2 | F3 | F4 | Mean (95% CI) | Wilcoxon $p$ |
|---|---|---|---|---|---|---|---|
| MT-FT | 264.9 | 333.1 | 318.3 | 201.1 | 248.6 | 273.2 [201, 350] | (reference) |
| MT-FT-aux | 254.7 | 315.7 | 323.3 | 192.4 | 237.7 | 264.8 [197, 337] | 0.13 |
| ST-FT | 172.4 | 276.6 | 233.0 | 161.5 | 773.3 | 323.4 [198, 506] | 0.83 |
| ST-TFS | 298.6 | 413.1 | 324.6 | 215.3 | 234.5 | 297.2 [225, 376] | 0.56 |

## Discussion on single-task collapse on DAI-X1

DAI-X1 is the only material in the present benchmark carrying the $[H_4IO_6]^-$ X-site. The single-task pretrained variant exhibits a fold-4 outlier of 773.3 m s$^{-1}$ (Table S8), more than four times its typical per-fold MAE (161.5–276.6 m s$^{-1}$ on folds 0–3) and roughly three times the multi-task baseline fold-4 MAE (248.6 m s$^{-1}$). A reproduction run on fold 4 with the higher-learning-rate schedule (start rate $10^{-4}$, decay steps 1000) confirmed the collapse is reproducible. The dominant contributor is DAI-X1 ($H_2dabco^{2+}$ / $Na^+$ / $[H_4IO_6]^-$, absolute error 2086 m s$^{-1}$ for the single-task pretrained variant vs. 554 m s$^{-1}$ for the multi-task baseline), whose $[H_4IO_6]^-$ X-site is absent from the corresponding training split. The orthoperiodate ion $[H_4IO_6]^-$ is chemically distinct from the $IO_4^-$ metaperiodate form carried by the other three periodate-family training materials (DAI-1, DAI-2, DAI-4). The multi-task auxiliary force-field head constrains the representation of Na, O, and I through DFT force supervision, providing a residual structural prior that the single-task variant lacks; this explanation is observational, consistent with Zhang *et al.*[36].

## Discussion on auxiliary-head variant of MT-FT

The auxiliary-head variant slightly improves in-distribution accuracy relative to the two-head baseline, reaching a fold-mean MAE of 264.8 $m\,s^{-1}$ versus 273.2 $m\,s^{-1}$ for the baseline in Table S8, but the paired difference is not significant at $n$=25 (paired Wilcoxon $p$=0.13). The auxiliary-head variant is also slightly lower on OOD-new (86.1 versus 92.2 $m\,s^{-1}$), but this difference is small compared with the uncertainty expected from three materials. That variant also degrades the externally curated OOD-holdout aggregate from 34.5 to 67.1 $m\,s^{-1}$, a 1.9-fold increase in MAE after excluding the DAI-$1_{0.5}4_{0.5}$ failure case reported separately. Because the OOD-holdout block is the closer stress test for literature-level chemical transfer, this degradation outweighs the modest in-distribution and OOD-new gains for selecting a production model. The auxiliary-head variant also introduces an additional 10,623-system CHNO training dependency and a third fitting head, whereas the two-head baseline keeps the few-shot PEM adaptation closer to the stated experimental-validation setting. We therefore use the two-head multi-task baseline as the production model and treat MT-FT-aux as a CHNO-domain-mismatch ablation.

# Representation and mechanism diagnostics

## Perturbation robustness

Three perturbations are applied to each held-out cluster. *Rigid rotation* applies a random SO(3) rotation to all atomic coordinates. *Rigid translation* translates each identified molecule by 0.5 Å in a random direction. *Template substitution* replaces the cluster built from the material's own crystal structure with a cluster built from the DAP-4 template crystal by ion substitution, preserving stoichiometry but altering the local packing geometry. Table S9 reports the fold-averaged ΔMAE.

**Table S9.** Perturbation sensitivity for all four adaptation variants. Baseline MAE is the fold-mean MAE of the multi-task baseline from Table S8. ΔMAE is the change in MAE induced by each perturbation, averaged over the three cluster variants.

| Variant | Perturbation | Baseline MAE (m s$^{-1}$) | ΔMAE (m s$^{-1}$) |
|---|---|---|---|
| Multi-task baseline | Rotation | 273.2 | +58.6 |
| Multi-task baseline | Translation | 273.2 | +56.9 |
| Multi-task baseline | Template subst. | 273.2 | -17.2 |
| Multi-task auxiliary-head | Rotation | 264.8 | +66.6 |
| Multi-task auxiliary-head | Translation | 264.8 | +67.8 |
| Multi-task auxiliary-head | Template subst. | 264.8 | -12.6 |
| Single-task pretrained | Rotation | 323.4 | +36.2 |
| Single-task pretrained | Translation | 323.4 | +26.6 |
| Single-task pretrained | Template subst. | 323.4 | +50.1 |
| Single-task scratch | Rotation | 297.2 | -50.6 |
| Single-task scratch | Translation | 297.2 | -43.0 |
| Single-task scratch | Template subst. | 297.2 | -5.3 |

## Geometry-scrambling rank preservation

Five geometry-destroying transformations are applied to test whether predictions reflect chemical information rather than incidental geometric features: coordinate swap (permuting atom positions within the cluster), full random scrambling (replacing all coordinates with random positions), random sphere placement, sorted-line arrangement, and B-site swap (replacing the B-site cation with a different species while preserving all other coordinates). The Spearman rank correlation between predictions on the original and scrambled clusters is computed per transformation across all 25 materials. A high rank correlation under B-site swap but low correlation under full scrambling indicates that the model reads chemical identity rather than geometric detail.

**Table S10.** Rank preservation under geometry and template perturbations. $\rho$ is the Spearman rank correlation between original and perturbed predictions across 25 materials; $|\Delta|$ is the mean absolute prediction shift in m s$^{-1}$.

| Perturbation | Multi-task baseline | | Single-task pretrained | |
|---|---|---|---|---|
| | $\rho$ | $\|\Delta\|$ | $\rho$ | $\|\Delta\|$ |
| Coordinate swap | 0.15 | 1428 | 0.51 | 1442 |
| Random coordinates | 0.13 | 2449 | -0.07 | 6031 |
| Random sphere | 0.06 | 3294 | -0.03 | 7587 |
| Sorted line | 0.37 | 4190 | 0.05 | 8282 |
| B-site swap | 0.83 | 148 | 0.98 | 36 |
| Template substitution | 0.98 | 51 | 0.91 | 149 |

## Uniform scaling sensitivity

All interatomic distances are uniformly scaled by 11 factors from 0.70 to 1.30, and the predicted detonation velocity is recorded at each scale. The sensitivity metric is the fractional change in prediction between scale 0.80 and scale 1.20, normalized by the prediction at scale 1.00. A model that encodes density information directly will show a large sensitivity; a model that has been regularized by the DFT energy head will show a dampened response.

**Table S11.** Uniform-scaling sensitivity of the two pretrained variants. Sensitivity is the fractional prediction change between scale 0.80 and 1.20 normalized by the prediction at scale 1.00.

| Variant | Mean sensitivity | Across-material s.d. | Materials |
|---|---|---|---|
| Multi-task baseline | 0.46 | 0.61 | 25 |
| Single-task pretrained | 2.91 | 0.69 | 25 |

## Probe protocols

**Material-level probe.** Per-atom 256-dimensional descriptors from the shared backbone are averaged across atoms and across the held-out cluster realizations to obtain one material-level feature vector per material. The targets are six material-level quantities: $V_{\mathrm{det}}$, crystal density, oxygen balance, nitrogen mass fraction, oxygen mass fraction, and cluster atom count. The linear probe is Ridge ($L_2$-regularized least-squares) regression. Its regularization strength is selected once by five-fold cross-validation over a logarithmic penalty grid spanning $10^{-3}$ to $10^{4}$ (eight values per decade step); the final cross-validated $R^2$ is then evaluated by leave-one-out cross-validation (LOOCV) at the selected penalty. The nonlinear control standardizes each input feature to zero mean and unit variance using statistics from the current training fold, projects to at most 20 principal components, and fits a kernel-ridge regressor with a Gaussian (RBF) kernel. Outer validation for the nonlinear control uses the same five-fold split structure as the model-training folds rather than LOOCV; the ridge penalty (same grid as the linear probe) and the RBF bandwidth ($\gamma \in 10^{-4}, 10^{-3}, 10^{-2}, 0.1, 1.0$) are jointly selected by an inner cross-validation with $min(3, n_{\mathrm{train}}-1)$ folds.

**Site-pooled probe.** Per-atom descriptors are grouped by ionic role (A-site, B-site, and X-site) and averaged within each role to form one site-level vector per material. These site-pooled feature vectors are used to probe either the target property $V_{\mathrm{det}}$ or the corresponding site identity, depending on the diagnostic panel. The regression probe is the same LOOCV Ridge protocol used for the material-level probe.

## Linear and nonlinear probe analyses

Both probes follow the protocols defined above. The nonlinear–linear $R^2$ gap is a qualitative test of whether target information requires a nonlinear readout.

**Table S12.** Linear and nonlinear leave-one-out probe $R^2$ values from material-level descriptors to chemical and energetic targets.

| Target | Multi-task baseline | | Single-task pretrained | |
|---|---|---|---|---|
| | Linear | Nonlinear | Linear | Nonlinear |
| $V_{\mathrm{det}}$ | 0.93 | 0.94 | 0.95 | 0.86 |
| Density | 0.83 | 0.75 | 0.60 | 0.56 |
| Oxygen balance | 0.89 | 0.93 | -0.10 | 0.83 |
| N mass fraction | 0.92 | 0.92 | 0.92 | 0.70 |
| O mass fraction | 0.94 | 0.78 | -0.06 | 0.67 |
| Cluster atom count | 0.88 | 0.70 | 0.73 | -0.04 |

## Cross-fold descriptor stability and template invariance

Material-level descriptors are extracted as the mean per-atom embedding from each of the five fold models. The mean pairwise cosine similarity across all $C(5,2)=10$ fold pairs is then computed per material. The template-invariance test uses the DAP-4 perchlorate cluster as a universal template: for each of the 25 in-distribution training materials except DAP-4 itself, ionic species in the DAP-4 cluster are substituted to match the target material's stoichiometry using the same site-substitution procedure as the OOD cluster construction. Predictions are taken from the held-out fold model that did not see the target material during training. The mean absolute difference in $V_{det}$ predictions ($|\Delta|$) between the template cluster and the material's original cluster is computed per model across all 24 test materials.

**Table S13.** Descriptor stability across five fold models.

| Variant | Mean cosine | s.d. | Median cosine | Materials |
|---|---|---|---|---|
| Multi-task baseline | 0.961 | 0.005 | 0.962 | 25 |
| Single-task pretrained | 0.982 | 0.001 | 0.983 | 25 |
| Single-task scratch | 0.995 | 0.002 | 0.995 | 25 |

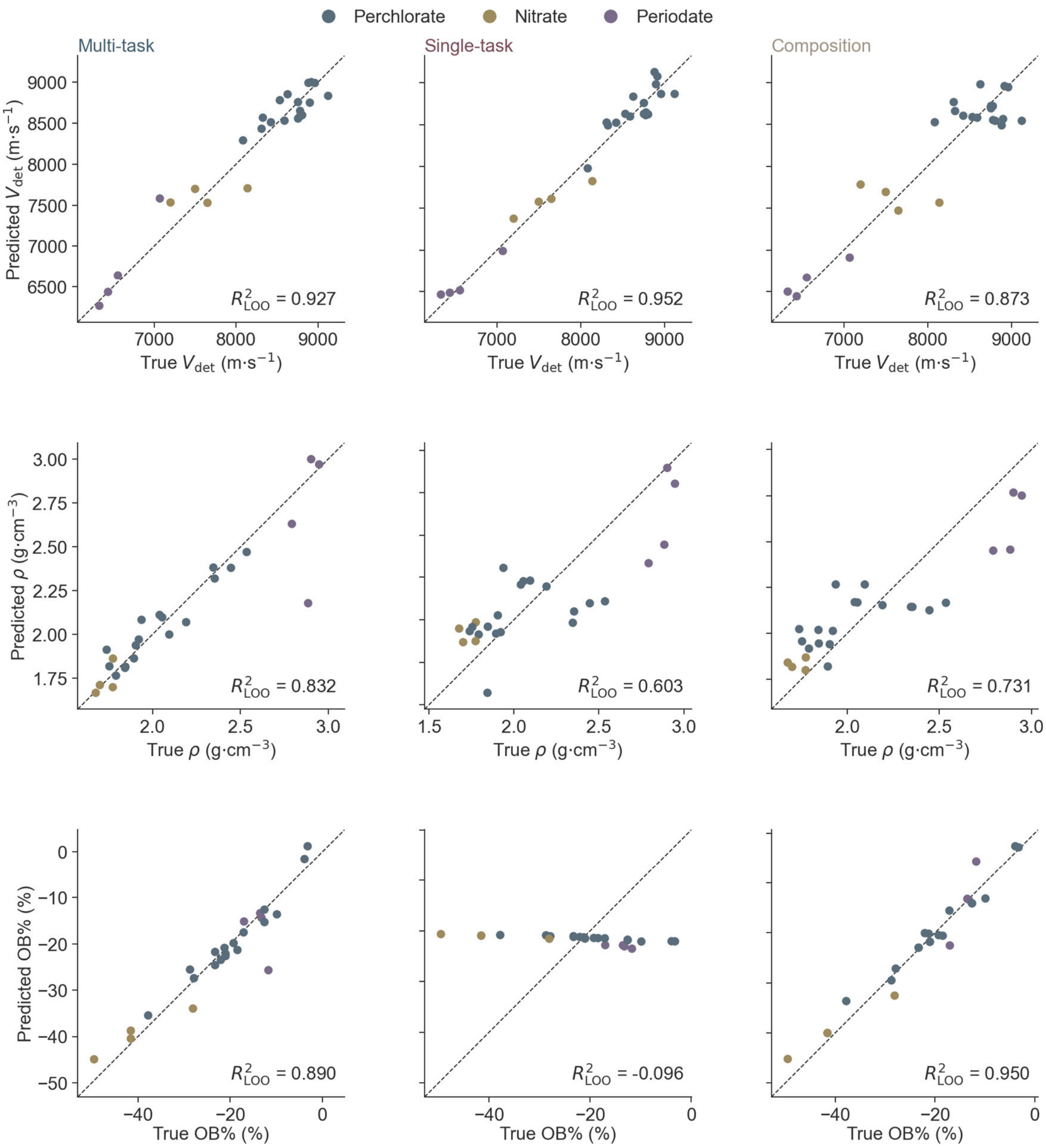

**Figure S12.** Material-level probe parity plots for $V_{det}$, density, and oxygen balance. Rows correspond to target quantities and columns compare the multi-task descriptor, single-task pretrained descriptor, and composition-only baseline.

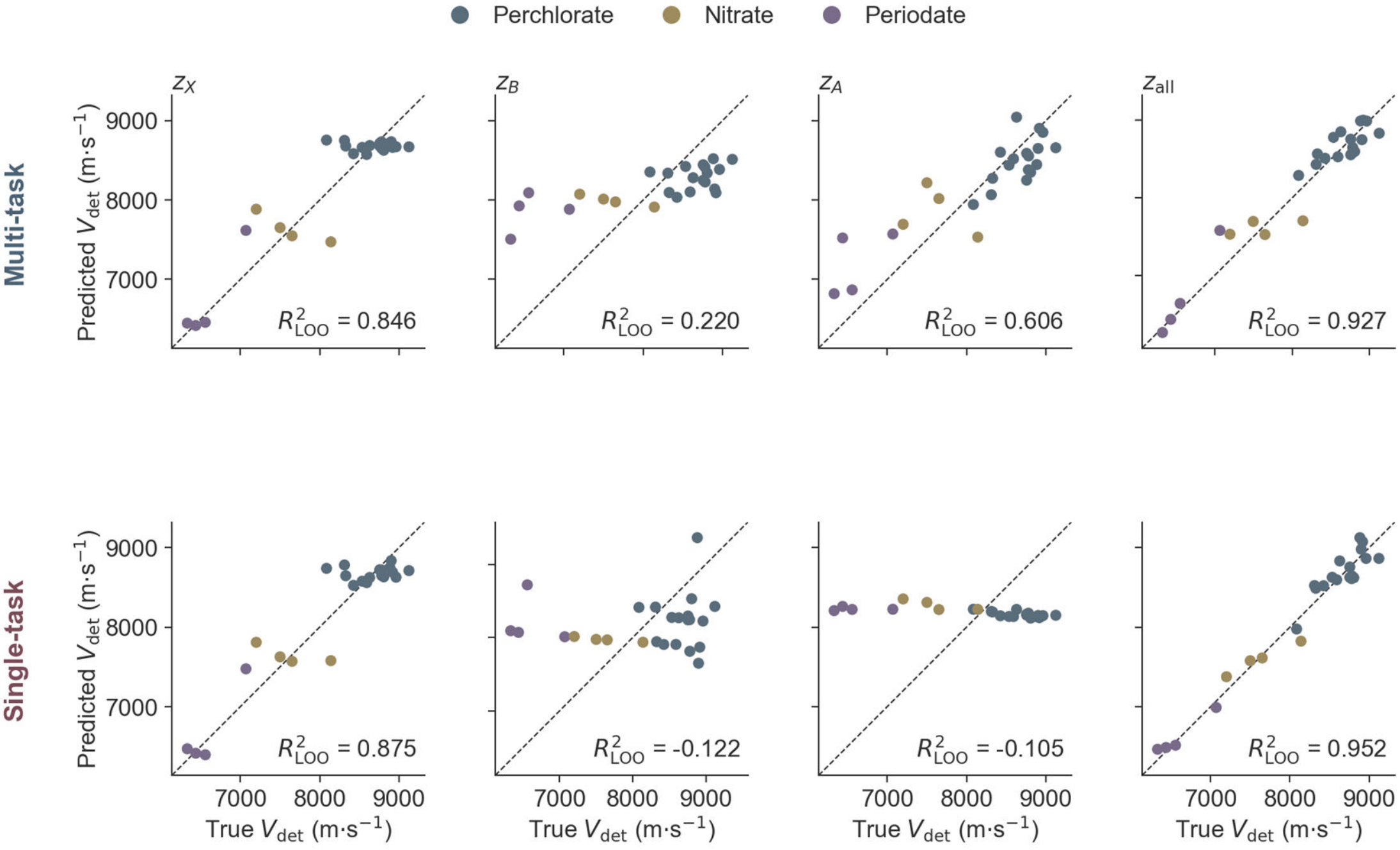

**Figure S13.** Site-pooled probe parity plots for $V_{det}$. Columns compare X-site, B-site, A-site, and all-site pooled descriptors; rows compare the multi-task baseline and the single-task pretrained variant.

## Site-resolved atomic embedding analysis

Per-atom descriptor vectors are extracted and pooled separately for A-site, B-site, and X-site atoms in each material cluster. UMAP projections use two components with $n_{neighbors}$=15, min_dist=0.1, and random seed 42. The silhouette values reported in the main text are computed in the original 256-dimensional descriptor space using heavy atoms only; the 2D UMAP-space silhouette is retained only as a visual check.

One-way ANOVA is performed on the material-level $V_{det}$ values grouped by A-site type, B-site type, and X-site type, yielding one-way effect-size estimates ($\eta^2$) for each compositional factor; the Type II partial $\eta^2$ is computed as a robustness check from an OLS model that includes all three site factors simultaneously, with the contradictory A versus B ordering reconciled in the LOSO confound analysis below. Ridge regression probes are fitted from each site-specific embedding pool to $V_{det}$ by leave-one-out cross-validation, and the per-site $R^2$ is compared across sites and models. Principal component analysis of the full material embeddings is used to identify the dominant axes of variation and their correlation with $V_{det}$ and X-site identity.

### ANOVA decomposition: one-way and Type II partial $\eta^2$

The one-way ANOVA reported in the main text uses the standard $\eta^2$ (between-group sum of squares divided by total sum of squares). The X-site dominance is robust to the choice of metric (one-way $\eta^2$=0.897, partial $\eta^2$=0.950); the A versus B ordering, however, differs between the two metrics: one-way puts B above A (0.227 versus 0.122) while Type II partial puts A above B (0.605 versus 0.479). This disagreement is a direct consequence of the LOSO confounding pattern documented below: at $n$=25 the A and B factors are mutually nested with the dominant X factor (Table S17), so partial $\eta^2$ inflates both A and B by transferring X-attributable variance through the collinear residual. We therefore report *both* estimates side by side and read the X  A, B dominance as the only quantitative claim supported at this sample size; the relative A versus B ordering is reported as qualitative.

**Table S14.** ANOVA effect sizes for the three ionic site factors. One-way $\eta^2$ uses the standard between-group ratio; Type II partial $\eta^2$ controls for collinearity among site factors via an OLS model that includes all three site factors simultaneously. At $n$=25 the partial $\eta^2$ for A and B is inflated by the confounds reported in Table S17; the one-way $\eta^2$ is therefore retained as the primary main-text metric.

| Factor | $F$ (one-way) | $p$ (one-way) | $\eta^2$ (one-way) | Partial $\eta^2$ (Type II) |
|---|---|---|---|---|
| X-site | 91.3 | $4.4\times10^{-11}$ | 0.897 | 0.950 |
| B-site | 1.76 | 0.191 | 0.227 | 0.479 |
| A-site | 0.66 | 0.626 | 0.122 | 0.605 |

## Leave-one-site-out analysis

The leave-one-site-out (LOSO) analysis tests how well the model generalizes when an entire ionic site type is withheld from training. In each split, all materials sharing a given A-site cation, B-site cation, or X-site anion are removed from the training set and used as the validation set.

**Table S15.** Leave-one-site-out split definitions. $n_{train}$ and $n_{val}$ give the number of training and validation materials in each split.

| Level | Withheld species | $n_{train}$ | $n_{val}$ | Note |
|---|---|---|---|---|
| X | $ClO_4^-$ | 8 | 17 | dominant X-site family |
| | $NO_3^-$ | 21 | 4 | |
| | $IO_4^-$ | 21 | 4 | |
| A | $H_2dabco^{2+}$ | 10 | 15 | largest A-site family |
| | $H_2pz^{2+}$ | 21 | 4 | |
| | $H_2hpz^{2+}$ | 22 | 3 | |
| | $MeHpz^{2+}$ | 22 | 3 | |
| B | $K^+$ | 18 | 7 | |
| | $Na^+$ | 21 | 4 | |
| | $NH_4^+$ | 19 | 6 | |
| | $Ag^+$ | 21 | 4 | |

### Leave-one-site-out MAE by split

The level-averaged MAE hierarchy (X > A ≈ B) is retained only as a qualitative diagnostic, because the sparse factorial design creates coupled site-level holdouts (Table S17).

**Table S16.** Leave-one-site-out mean absolute error ($m\ s^{-1}$) for the multi-task baseline. Level averages are computed over all splits at the same site level.

| Level | Withheld species | MAE ($m\ s^{-1}$) | Level mean ($m\ s^{-1}$) |
|---|---|---|---|
| X | $ClO_4^-$ | 1129.7 | 1219.7 |
| | $NO_3^-$ | 699.7 | |
| | $IO_4^-$ | 1829.6 | |
| A | $H_2dabco^{2+}$ | 504.5 | 345.0 |
| | $H_2pz^{2+}$ | 316.9 | |
| | $H_2hpz^{2+}$ | 155.7 | |
| | $MeHpz^{2+}$ | 403.2 | |
| B | $K^+$ | 567.0 | 299.4 |
| | $Na^+$ | 243.0 | |
| | $NH_4^+$ | 231.8 | |
| | $Ag^+$ | 155.8 | |

The LOSO comparison should not be read as an apples-to-apples comparison between model classes, because the multi-task variants also see 67 DFT energy–force systems whereas single-task property-only variants do not. The useful LOSO message is the within-model level hierarchy: X-site holdout is much harder than A- or B-site holdout, consistent with the ANOVA hierarchy above.

### LOSO confound matrix

The 25-material training set is not a fully crossed A×B×X factorial design, so withholding one site identity can simultaneously withhold materials of other site identities. Two splits are fully nested: B = $K^+$ removes *all four* $NO_3^-$-containing training materials (DAN-2, PAN-2, PAN-H2, PAN-M2), and A = $H_2$dabco-family removes *all four* materials in the $IO_4^-$/$[H_4IO_6]^-$ family (DAI-1, DAI-2, DAI-4, DAI-X1).

**Table S17.** X-site composition of the training set after each leave-one-site-out holdout. Original training counts are $ClO_4^-$ = 17, $NO_3^-$ = 4, $IO_4^-$-family (including $[H_4IO_6]^-$) = 4. Bold cells flag full secondary-site removal.

| Level | Withheld species | $ClO_4^-$ | $NO_3^-$ | $IO_4^-$-family | Note |
|---|---|---|---|---|---|
| X | $ClO_4^-$ | 0 | 4 | 4 | primary holdout |
| | $NO_3^-$ | 17 | 0 | 4 | primary holdout |
| | $IO_4^-$-family | 17 | 4 | 0 | primary holdout |
| A | $H_2$dabco family | 7 | 3 | **0** | $IO_4^-$-family fully removed |
| | $H_2$pz | 14 | 3 | 4 | minor |
| | $H_2$hpz | 15 | 3 | 4 | minor |
| | MeHpz | 15 | 3 | 4 | minor |
| | $NH_4^+$ | 12 | 4 | 3 | minor |
| B | $K^+$ | 15 | **0** | 3 | $NO_3^-$ fully removed |
| | $Na^+$ | 15 | 4 | 2 | half of $IO_4^-$-family |
| | $Ag^+$ | 13 | 4 | 4 | minor |

The B = $K^+$ split is reported as a B-site holdout but the test set is dominated by materials whose actual extrapolation lies at the X-site ($NO_3^-$ training coverage drops to zero), and the A = $H_2$dabco-family split is similarly dominated by $IO_4^-$-family extrapolation ($IO_4^-$-family training coverage drops to zero). These two splits therefore upper-bound the per-level average MAE for B and A, respectively, by quantities that are mechanistically attributable to X-site extrapolation rather than to B- or A-site extrapolation. The level-averaged hierarchy (X > A ≈ B) agrees qualitatively with the ANOVA decomposition but cannot be read quantitatively given these confounds.

## Training-variant ablations

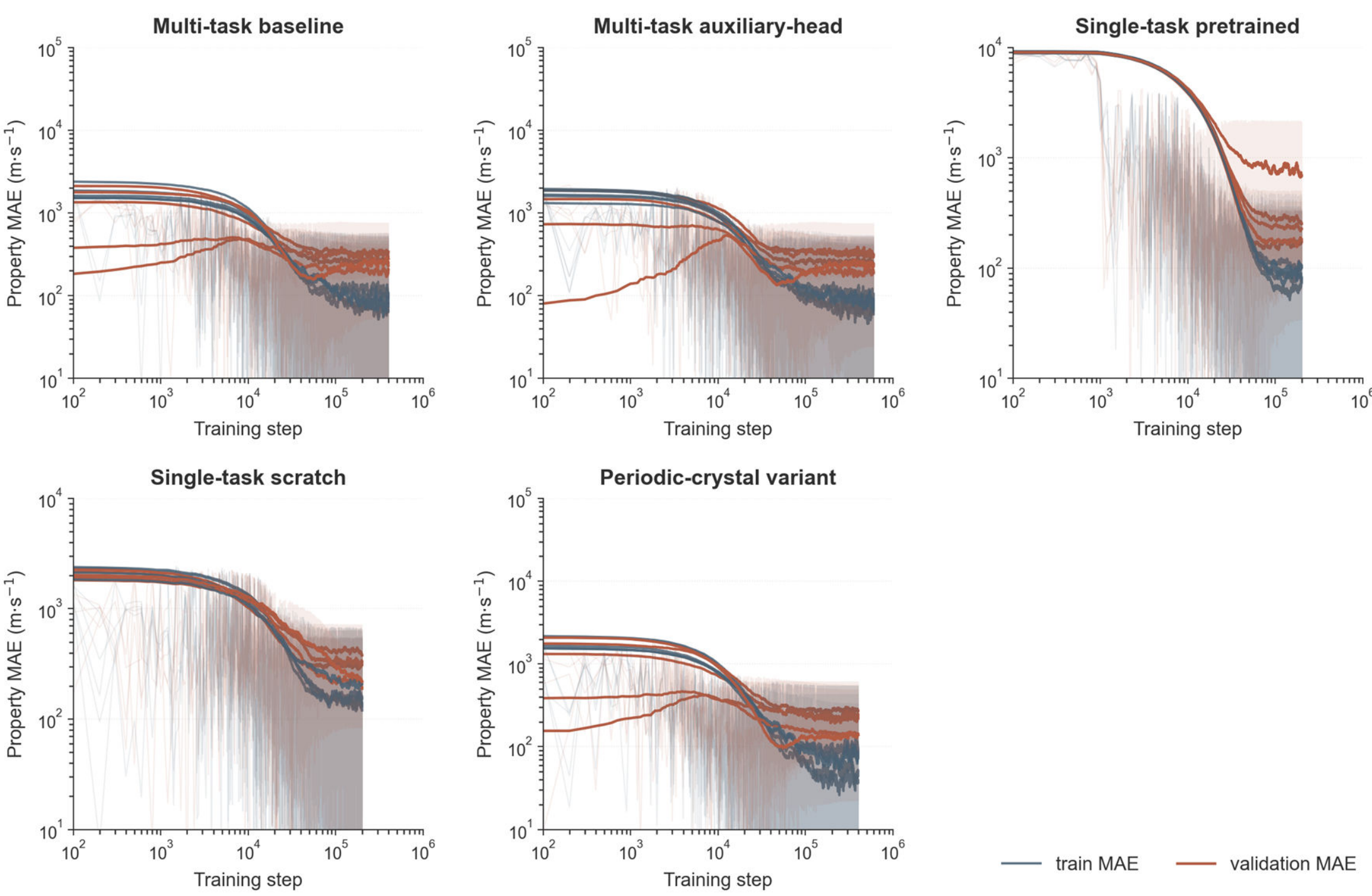

**Figure S14.** Training curves for all five folds of every adaptation variant. Translucent traces show the raw property-head training (navy) and validation (rust) MAE extracted from the DeepMD learning-curve logs; bold lines show exponential moving averages ($\alpha$=0.01) of the five overlaid per-fold raw traces. The periodic-crystal variant is included as a matched-input control for direct crystal-input finetuning.

MT-FT-crystal replicates the MT-FT architecture and hyperparameters exactly but substitutes periodic crystal inputs (with periodic boundary conditions) for the cluster inputs, enabling a direct representation comparison. It reaches 201.4 m s$^{-1}$ MAE when trained and evaluated on matched periodic crystal inputs, below the 273.2 m s$^{-1}$ cluster-to-cluster value for MT-FT. However, cross-representation transfer is asymmetric: in the paired cluster $n_1$ diagnostic the cluster-trained baseline gives 405.4 m s$^{-1}$ MAE on periodic crystal inputs, whereas the crystal-trained variant gives 1314.3 m s$^{-1}$ MAE on cluster inputs. The same crystal-trained models give 1623.9 m s$^{-1}$ MAE on the three **OOD-new** $ABX_4$ cluster inputs (Extended Data Fig. 2). This control therefore separates structural fidelity from synthesis-facing deployability.

## Hyperparameter and split-robustness ablations

Beyond the architecture-level variants compared above, we ran a second sweep that fixes the multi-task baseline (and, where useful, the single-task pretrained variant) and perturbs only the four most influential hyperparameters: the learning-rate schedule, the total number of training steps, the random seed used to assign materials to cross-validation folds, and the exponential learning-rate decay ratio. Every cell of the sweep is a complete five-fold run; each value reported is the cluster-pooled prediction averaged across the held-out cluster realizations of every fold (in-distribution) or across all five fold models (out-of-distribution).

We score each variant on three disjoint material sets:

- **IND**: the 25 perovskite-type energetic materials (PEMs) used to train the property head, scored on the fold in which each material is held out from training.

- **OOD-holdout**: real, externally synthesized energetic materials (DAC-4, TAP-2, EAP-4[S8], and the DEP $ABX_4$ material[38,39]) that were never part of the training pool but for which experimentally-derived K–J detonation velocities are available. DAI-$1_{0.5}4_{0.5}$ is reported within the same data block but is excluded from the four-material aggregate MAE as a confidently-wrong failure case (see Section "Failure analysis: DAI-$1_{0.5}4_{0.5}$ double perovskite" below).
- **OOD-new**: three template-derived $ABX_4$-series compounds (PEP, PEP-M, PEP-H) generated for extrapolation testing.

Aggregate mean absolute errors across all variants are summarized in Table S18; per-material OOD-holdout absolute errors, including the excluded DAI-$1_{0.5}4_{0.5}$ failure case, are visualized in Extended Data Fig. 3. Those OOD-holdout materials projected into the multi-task baseline atomistic UMAP are shown in Fig. S15.

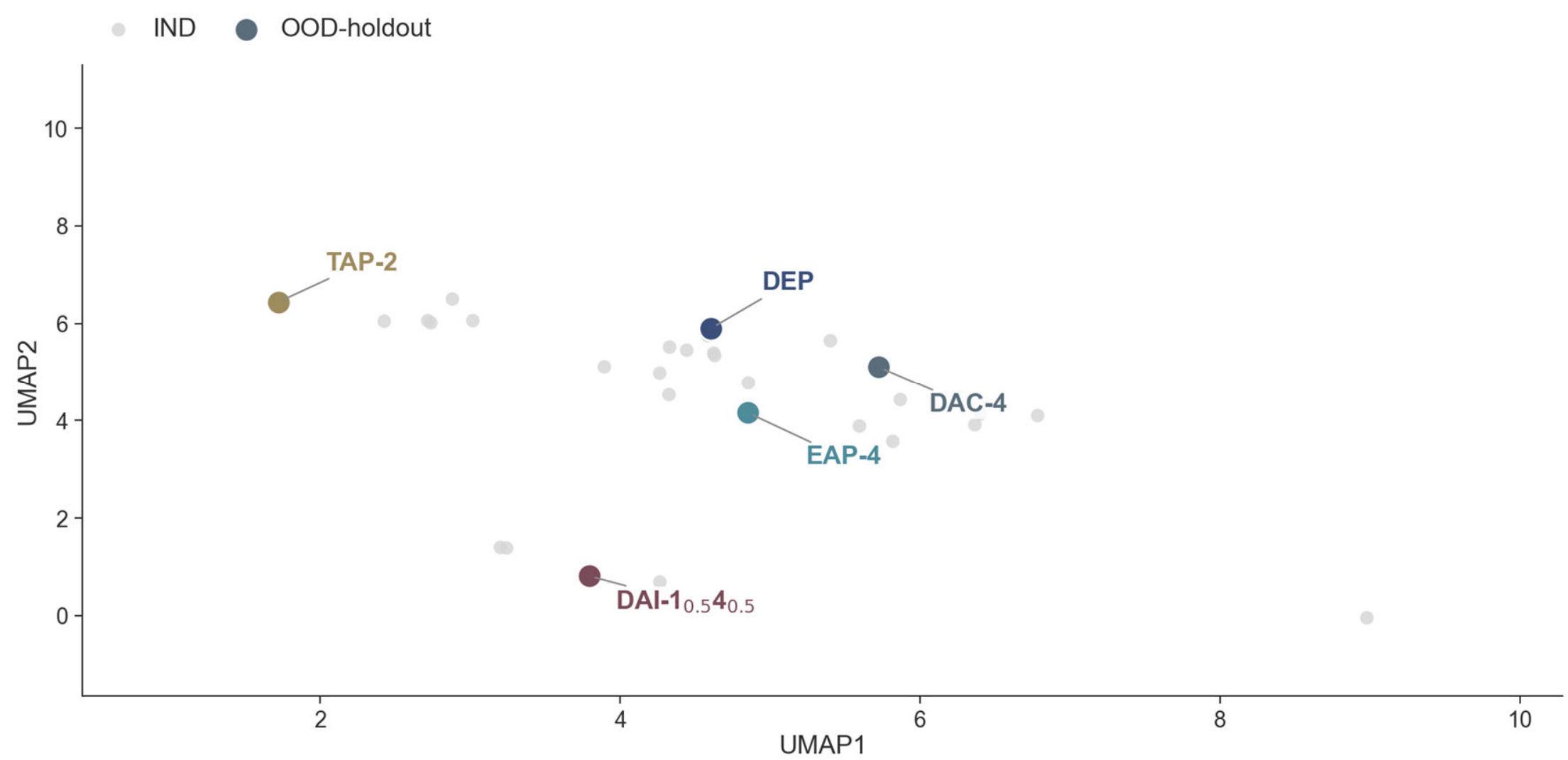

**Figure S15.** OOD-holdout materials projected into a multi-task baseline atomistic UMAP space computed from the same descriptor family as main-text Fig. 5f. Each marker is the centroid of fold-averaged atomistic cluster embeddings for one material, with the 25 **IND** materials shown as grey references.

**Table S18.** Aggregate mean absolute error (m s$^{-1}$) on IND (n=25, cross-validation), OOD-holdout (n=4; DAI-$1_{0.5}4_{0.5}$ excluded as a failure case), and OOD-new (n=3) for each ablation variant. The four default-LR rows reproduce the architecture variants of Table S8; subsequent groups perturb a single hyperparameter at a time on top of those baselines.

| Variant | IND MAE | OOD-holdout MAE | OOD-new MAE |
|---|---|---|---|
| **Default learning-rate baselines** | | | |
| MT-FT | 273.2 | 34.5 | 92.2 |
| MT-FT-aux | 264.8 | 67.1 | 86.1 |
| ST-FT | 323.3 | 606.6 | 584.1 |
| ST-TFS | 297.2 | 203.9 | 184.0 |
| **Learning-rate schedule** | | | |
| MT-FT-highLR | 235.2 | 90.1 | 164.1 |
| MT-FT, $5\times10^{-6}$ | 266.5 | 136.9 | 97.4 |
| MT-FT-aux, $1\times10^{-4}$ | 220.8 | 77.2 | 150.7 |
| MT-FT-aux, $5\times10^{-6}$ | 261.3 | 160.7 | 83.2 |
| ST-FT, $1\times10^{-4}$ | 238.9 | 497.0 | 306.9 |
| ST-TFS, $1\times10^{-4}$ | 229.3 | 144.3 | 139.3 |
| **Total training length** | | | |
| MT-FT, 200 k steps | 267.5 | 103.5 | 90.0 |
| MT-FT, 800 k steps | 281.0 | 69.3 | 146.4 |
| **Cross-validation split seed** | | | |
| MT-FT, split seed 7 | 266.9 | 31.8 | 91.4 |
| MT-FT, split seed 13 | 286.8 | 44.5 | 79.8 |
| ST-FT, split seed 7 | 323.0 | 616.6 | 554.3 |
| ST-FT, split seed 13 | 313.8 | 606.4 | 547.6 |
| **Learning-rate decay ratio (decay_steps / numb_steps)** | | | |
| MT-FT, ratio 1/200 | 273.6 | 37.2 | 90.5 |
| ST-FT, ratio 1/200 | 323.7 | 608.7 | 587.2 |

**Single-task narrowing is specific to this low-data setting.** The single-task pretrained variant is uniformly the worst on the two OOD blocks (OOD-holdout MAE 497–617, OOD-new MAE 547–587), and *none* of the perturbations rescues it: changing the learning rate by an order of magnitude, doubling the decay-step granularity, or repartitioning the cross-validation folds with two independent seeds all leave the OOD MAEs in the same 500–620 m s$^{-1}$ band on OOD-holdout and 540–590 m s$^{-1}$ on OOD-new. We interpret this as representational narrowing under extreme few-shot fine-tuning. The probe diagnostics support this reading: single-task fine-tuning preserves the $V_{det}$ direction but makes auxiliary physical quantities such as oxygen balance and oxygen fraction nearly inaccessible, whereas the multi-task objective keeps those signals available for transfer.

**Default learning rate sits at the IND/OOD trade-off optimum.** Increasing the multi-task learning rate to $1\times10^{-4}$ improves IND MAE by roughly 40–50 m s$^{-1}$ (from 273/265 down to 235/221), but inflates OOD-new MAE by a factor of 1.8 (from 92/86 up to 164/151) and OOD-holdout MAE by a factor of 1.1–2.6 (from 35/67 up to 90/77); the gain on the in-distribution fold is bought by overfitting away the cluster-level generalisation signal. Decreasing the rate to $5\times10^{-6}$ leaves IND essentially unchanged but pushes OOD-holdout MAE to 137–161 m s$^{-1}$ (vs 35–67 at the default), because the property head never fully reaches the in-fold optimum within the allotted training budget.

**Total training length is saturated near 400 k steps.** Halving (200 k) or doubling (800 k) the training length on the multi-task baseline produces only a ±8 m s$^{-1}$ shift in IND MAE. On OOD-holdout the 200 k run is the worst (104 m s$^{-1}$ vs 35 / 69 for default / 800 k), but on OOD-new the 800 k variant is markedly worse (146 vs 90), indicating mild overfitting of the property head to the in-fold training systems past the default schedule.

**Random partitioning of the 5-fold CV is robust.** Across three independent split seeds (the published baseline plus seeds 7 and 13), multi-task IND MAE varies in ±13 m s$^{-1}$ (267–287), OOD-holdout in ±7 m s$^{-1}$ (32–45), and OOD-new in ±12 m s$^{-1}$ (80–92). The seed-13 OOD-new dip (PEP AE 166 vs 219, PEP-M 4.7 vs 30.5) reflects favourable fold placement; the same seeds applied to the single-task variant change OOD-holdout by less than 11 m s$^{-1}$, leaving the collapse intact.

**Learning-rate decay ratio is essentially irrelevant.** Both '1/80' (default) and '1/200' decay-step ratios produce nearly indistinguishable IND/OOD MAEs on both the multi-task baseline ($\Delta$<3 m s$^{-1}$ on all three blocks) and the single-task pretrained variant ($\Delta$<3 m s$^{-1}$).

## Failure analysis: DAI-$1_{0.5}4_{0.5}$ double perovskite

DAI-$1_{0.5}4_{0.5}$ is an $A_2BB'X_6$ double perovskite with alternating $Na^+$ and $NH_4^+$ B-site cations and $IO_4^-$ X-site anions (reference velocity 5156 m s$^{-1}$, Explo5/BKW-EOS). Its absolute error sits in 1074–1460 m s$^{-1}$ across every ablation cell (architecture, learning rate, training length, split seed, decay ratio), with the multi-task baseline predicting ~6488 m s$^{-1}$. We exclude it from the four-material OOD-holdout aggregate and retain its per-material error in Extended Data Fig. 3 for transparency. The failure reflects two compounding training-set gaps.

**Composition-domain mismatch.** The four $IO_4^-$ materials in the training set (DAI-1 through DAI-4, plus DAI-X1 with ortho-periodate $H_4IO_6^-$) span experimentally-derived K–J velocities of 6331–7070 m s$^{-1}$ (mean ≈6600 m s$^{-1}$).

DAI-$1_{0.5}4_{0.5}$'s reference velocity of 5156 m s$^{-1}$ lies ≈1175 m s$^{-1}$ below the lowest $IO_4^-$ training value. The model's prediction of ~6488 m s$^{-1}$ is nearly indistinguishable from the training $IO_4^-$ mean, indicating that no signal in the training data motivates a downward correction of this magnitude.

**Topological motif absent from training.** The training set contains one structural analogue to DAI-$1_{0.5}4_{0.5}$'s ordered $Na^+$/$NH_4^+$ double-perovskite B-site: DPE-1, which has the same $A_2BB'X_6$ arrangement but with $ClO_4^-$ X-site anions and an experimentally-derived velocity of 8858 m s$^{-1}$. Because DPE-1 has a high velocity, the double-perovskite topology alone does not provide a low-velocity signal. The down-correction required for DAI-$1_{0.5}4_{0.5}$ would require the model to recognise the co-occurrence of an ordered $Na^+$/$NH_4^+$ B-site *and* $IO_4^-$ X-site as jointly placing the velocity well below the $IO_4^-$ training range. This binary combination is entirely absent from the 25-material training set: no training material pairs $IO_4^-$ with a double-perovskite B-site arrangement.

**Protocol mismatch caveat.** DAI-$1_{0.5}4_{0.5}$'s reference velocity is from Explo5 (BKW-EOS), whereas most $IO_4^-$ training materials use different thermochemical workflows. Typical cross-protocol offsets are 200–300 m s$^{-1}$, far smaller than the ~1300 m s$^{-1}$ error, so protocol mismatch is a minor contributor at most.

**Why ensemble UQ does not catch this case.** The five-fold cross-fold prediction variance for DAI-$1_{0.5}4_{0.5}$ is small ($\sigma$=32 m s$^{-1}$ for the multi-task baseline), placing DAI-$1_{0.5}4_{0.5}$ in the high-confidence region of the IND $\sigma$-vs-AE plane (cf. Spearman $\rho$=0.84 on **IND**; Fig. 3e in the main text), even though its absolute error is 1332 m s$^{-1}$. Every fold's training set is missing the same combination of features ($IO_4^-$ X-site coexisting with a $Na^+$/$NH_4^+$ ordered double-perovskite B-site), so the bias is shared across folds and is not randomised by resampling. Cross-fold variance captures epistemic uncertainty arising from training-set sampling, but cannot detect systematic bias from training-set composition, a confident-extrapolation failure mode that ensemble dispersion alone cannot flag. Applicability-domain checks may flag such cases prospectively.

A weaker version of this structural-domain gap applies to PEP (AE 166–230 m s$^{-1}$ across multi-task variants), which also introduces a $ClO_4^-$ combination with a stoichiometry unseen in training.

## Uncertainty visualization companions

The following companion plots show the distributions underlying selected aggregate MAE summaries.

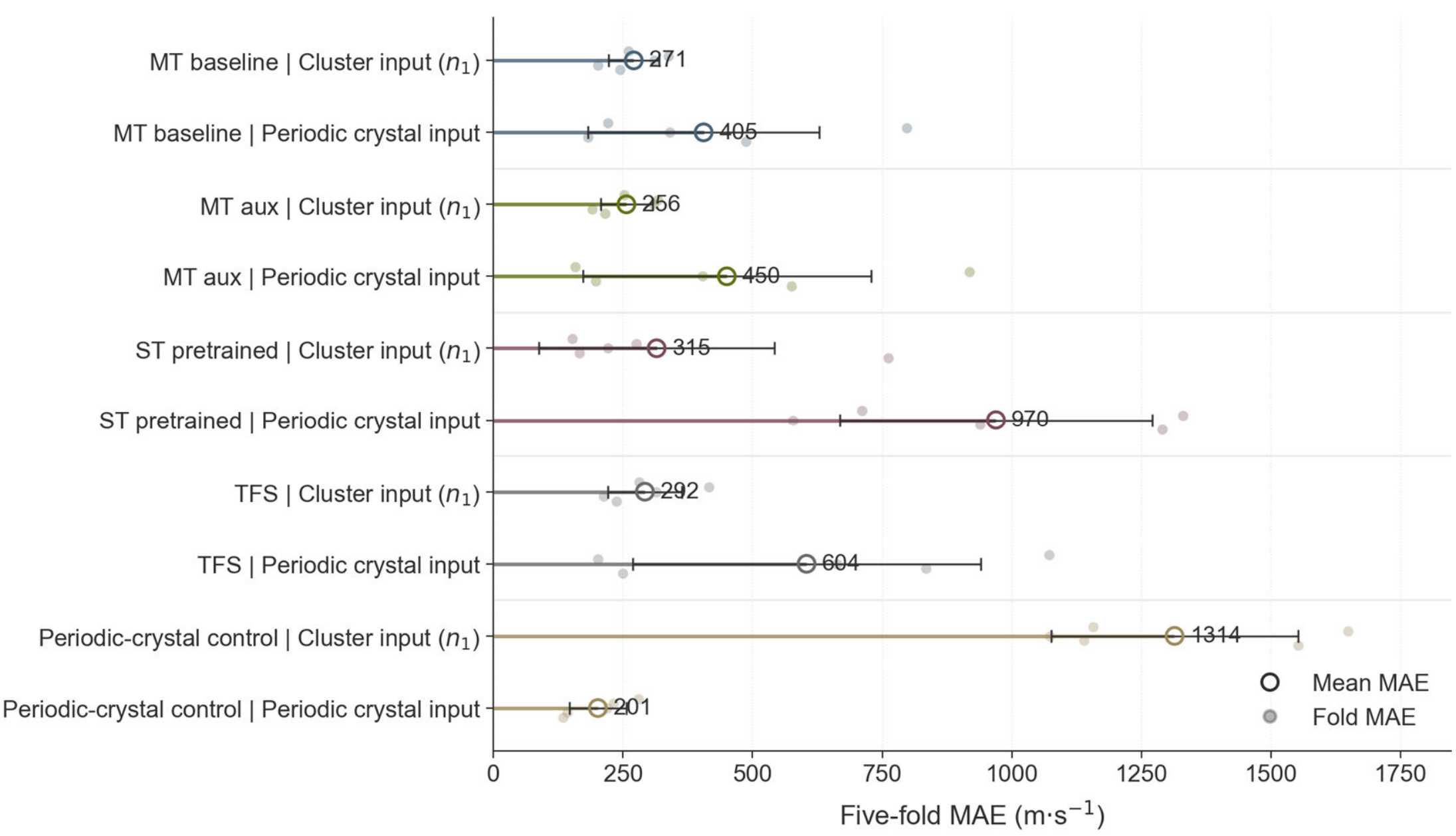

**Figure S16.** Fold-level MAE distributions for the periodic-control comparison.

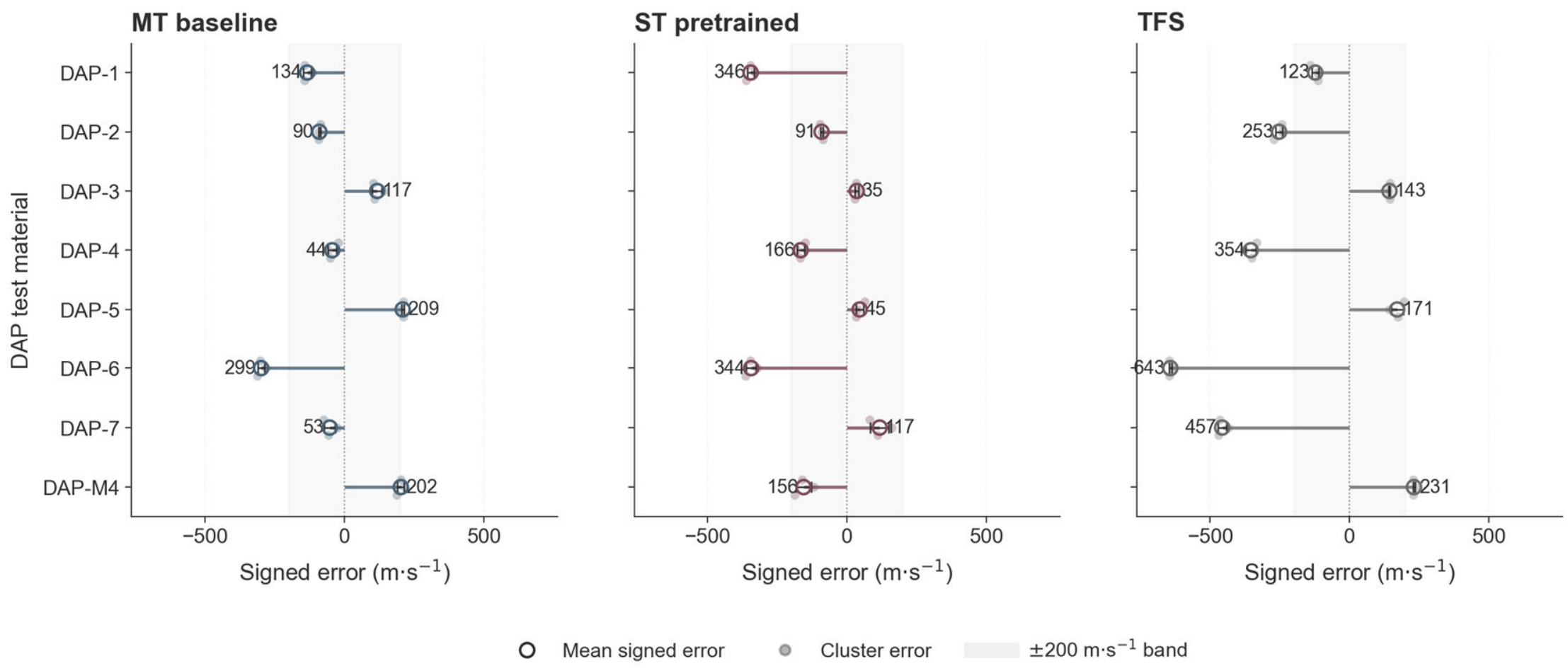

**Figure S17.** DAP-core pretrained-domain signed-error distributions.

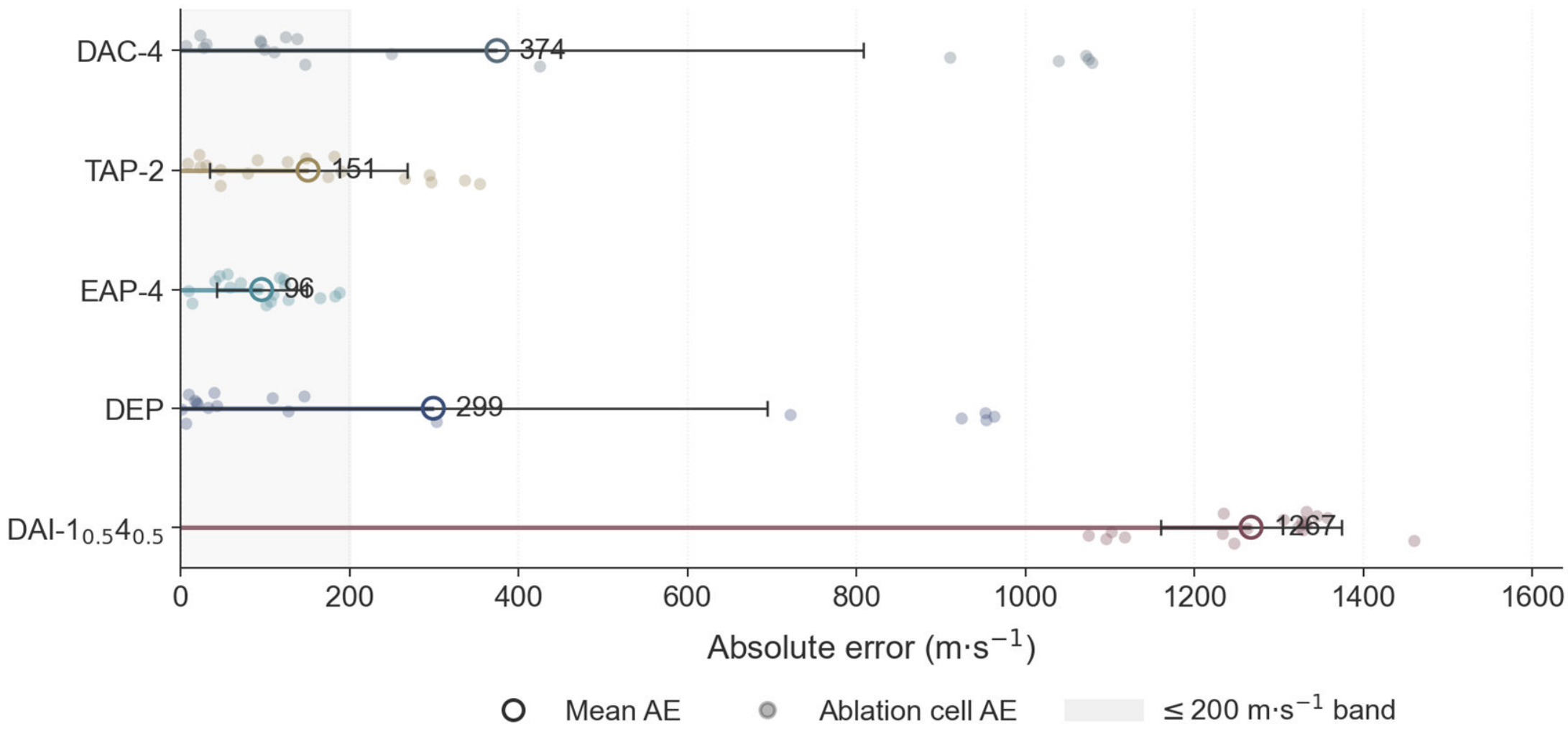

**Figure S18.** OOD-holdout absolute-error distributions across ablation variants.

## Wet-lab characterization of the $H_2en^{2+}$-based $ABX_4$ branch

### Synthesis, phase purity, and crystallography

Safety note. PEP, PEP-M, PEP-H, and DEP are energetic perchlorate salts and should be handled only by trained personnel on small scale under institutional energetic-materials safety procedures, with appropriate shielding and personal protective equipment, and with friction, impact, electrostatic discharge, and unintended heating minimized. PEP, PEP-M, and PEP-H were prepared by acidifying the corresponding A-site diamine in water with perchloric acid and then adding ethylenediamine as the B-site precursor. In a representative synthesis, 8 mmol of 70–72% $HClO_4$ solution was added dropwise with stirring to 5 mL of water containing 2 mmol piperazine, *N*-methylpiperazine, or homopiperazine. A 2 mL aqueous ethylenediamine solution was then added dropwise under stirring, producing a white precipitate immediately. The suspension was stirred for about 2 h, filtered, washed three times with ethanol, and dried to give the target $ABX_4$ materials in about 80% isolated yield. Colorless block crystals suitable for single-crystal X-ray diffraction grew from the mother liquor over several days. The products are friction-sensitive; friction during synthesis and post-synthetic handling was minimized, and mother liquors were not stored for extended periods. Powder X-ray diffraction comparison against simulated patterns from the single-crystal models confirmed phase-pure bulk products for PEP, PEP-M, PEP-H, and DEP. The measured-versus-simulated overlays for PEP, PEP-M, and PEP-H are reported in main-text Fig. 5c; the remeasured DEP reference overlay is reproduced in Fig. S19. For consistency with the new branch data, DEP was remeasured under the same conditions as the three new materials. The DEP detonation reference used in the OOD-holdout analysis is therefore the K–J value recalculated from this remeasured and refined structure ($D_c$=1.869 g cm$^{-3}$, $V_{det}$=8.867 km s$^{-1}$), rather than a value copied directly from the original report.

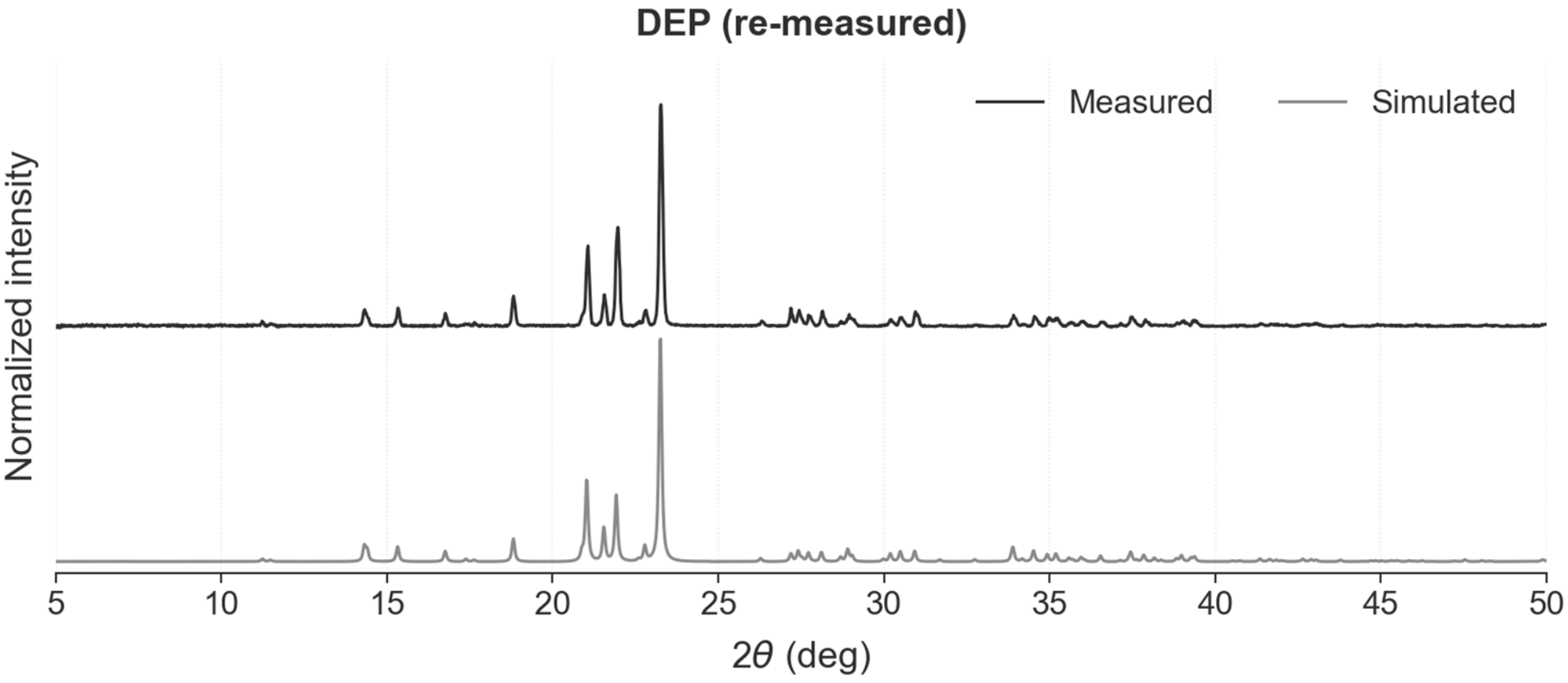

**Figure S19.** Powder X-ray diffraction validation of the remeasured DEP reference sample. Measured diffraction intensities are compared with the simulated pattern from the single-crystal model.

**Table S19.** Crystallographic summary for the $H_2en^{2+}$-based $ABX_4$ branch.

| Parameter | PEP | PEP-M | PEP-H | DEP |
|---|---|---|---|---|
| Formula | $C_6H_{22}Cl_4N_4O_{16}$ | $C_7H_{24}Cl_4N_4O_{16}$ | $C_7H_{24}Cl_4N_4O_{16}$ | $C_8H_{24}Cl_4N_4O_{16}$ |
| Formula weight | 548.07 | 562.10 | 562.10 | 574.11 |
| $T$ (K) | 298(2) | 298(2) | 293(2) | 298(2) |
| Crystal system | orthorhombic | monoclinic | monoclinic | orthorhombic |
| Space group | Pbca | $P2_1/c$ | $P2_1/c$ | $Cmc2_1$ |
| $a$ (Å) | 23.9304(2) | 12.7872(5) | 7.8381(3) | 8.09490(10) |
| $b$ (Å) | 10.32263(11) | 10.3107(3) | 24.9901(8) | 24.7197(3) |
| $c$ (Å) | 31.4590(4) | 15.9171(5) | 10.4647(4) | 10.19550(10) |
| $\beta$ (°) | 90 | 99.987(4) | 93.649(3) | 90 |
| $V$ (Å$^3$) | 7771.12(15) | 2066.79(12) | 2045.62(13) | 2040.16(4) |
| $Z$ | 16 | 4 | 4 | 4 |
| $D_c$ (g cm$^{-3}$) | 1.874 | 1.806 | 1.824 | 1.869 |
| Reflections collected | 48851 | 18524 | 15275 | 10327 |
| Independent reflections | 8137 | 4203 | 4158 | 2258 |
| $R_{int}$ | 0.0545 | N/A | 0.0637 | 0.0367 |
| $R_1$ [$I$>2$\sigma$($I$)] | 0.0545 | 0.0675 | 0.0851 | 0.0545 |
| $wR_2$ [$I$>2$\sigma$($I$)] | 0.1602 | 0.2201 | 0.2653 | 0.1364 |
| $R_1$ (all data) | 0.0626 | 0.0728 | 0.0979 | 0.0555 |
| $wR_2$ (all data) | 0.1675 | 0.2222 | 0.2849 | 0.1376 |
| GOF on $F^2$ | 1.058 | 1.118 | 1.099 | 1.058 |
| Completeness | 1.00 | 0.99 | 0.99 | 1.00 |
| CCDC | 2574803 | 2574804 | 2574805 | 2575016 |

## Thermal analysis and sensitivities

Differential thermal analysis measurements were collected on a DTA 552-Ex explosion-proof differential thermal analyzer at a heating rate of 5 °C min$^{-1}$ using about 6 mg sample per run, placing the onset decomposition temperatures of the $ABX_4$ branch between 292.6 and 359.6 °C. The order is PEP-M (292.6 °C) < PEP (310.6 °C) < PEP-H (324.1 °C) < DEP (359.6 °C). A compact benchmark comparison against TNT, RDX, HMX, CL-20, and representative MIX star materials is provided in Fig. S21; DAP-4 is plotted by onset temperature (358 °C[16S1]) with its peak temperature (385 °C[16]) marked separately, and EAP-4 is plotted by onset temperature (208 °C[S8]). Impact sensitivities are 9, 20, and 17.5 J for PEP, PEP-M, and PEP-H, respectively. Friction sensitivities are 12, 9, and 12 N for PEP, PEP-M, and PEP-H, whereas DEP is reported at ≤5 N. Impact and friction sensitivities were measured according to BAM standards using BFH 10 BAM impact and FSKM 10 BAM friction sensitivity instruments.

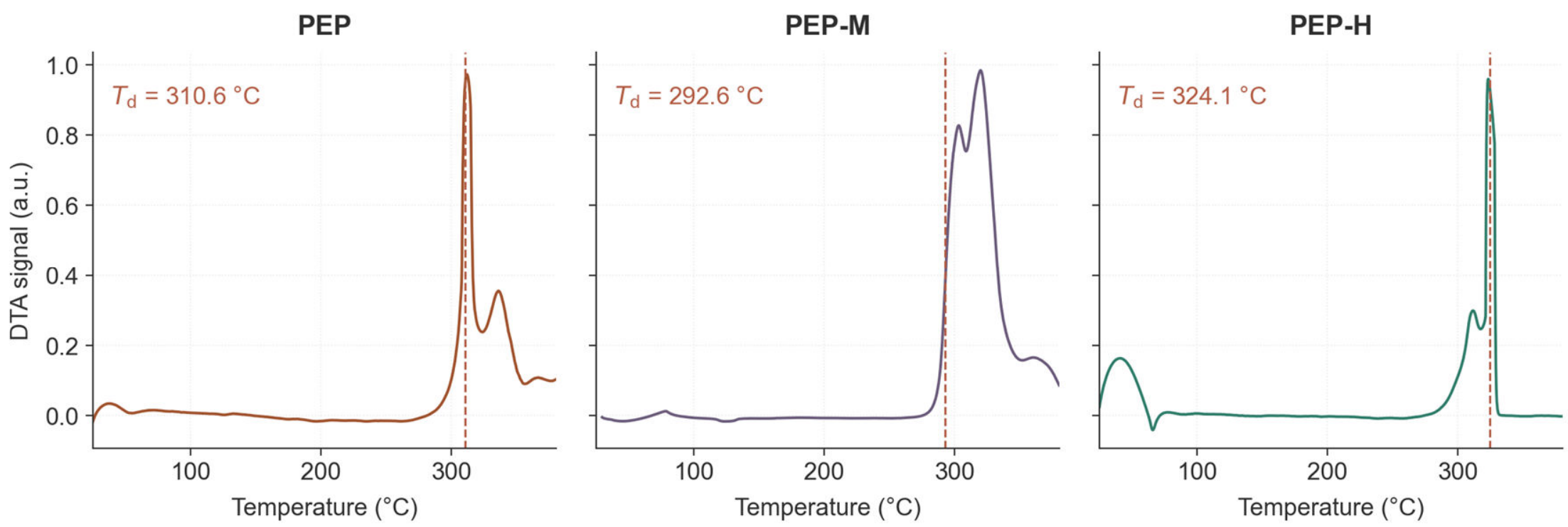

**Figure S20.** Differential thermal analysis traces for PEP, PEP-M, and PEP-H. Dashed lines mark the onset decomposition temperatures tabulated in Table S20.

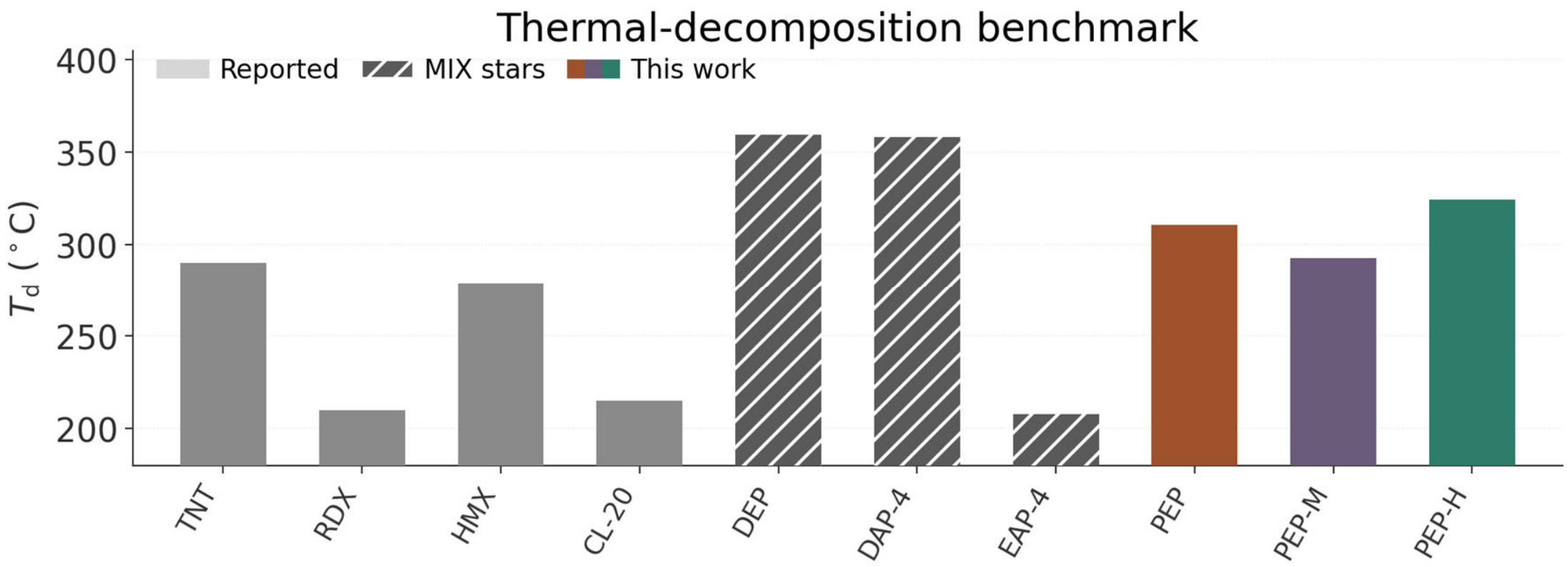

**Figure S21.** Thermal-decomposition benchmark for the three newly synthesized $ABX_4$ materials, representative MIX star materials, and traditional energetic benchmarks. Bars report onset temperatures except for the DAP-4 triangle, which marks the reported peak temperature.

**Table S20.** Thermal and sensitivity descriptors for the $H_2en^{2+}$-based $ABX_4$ branch.

| Material | $T_d$ (°C) | Impact sensitivity (J) | Friction sensitivity (N) |
|---|---|---|---|
| PEP | 310.6 | 9.0 | 12 |
| PEP-M | 292.6 | 20.0 | 9 |
| PEP-H | 324.1 | 17.5 | 12 |
| DEP | 359.6 | N/A | ≤5 |

## Calculated detonation parameters

Branch-specific detonation parameters were estimated by combining structure-refined densities with density-functional-theory reaction energies and K–J equations. The resulting detonation velocities span 8.729–9.090 km s$^{-1}$, and the detonation pressures span 34.0–37.6 GPa.

**Table S21.** Calculated detonation descriptors for the $H_2en^{2+}$-based $ABX_4$ branch.

| Material | $Q$ (kJ g$^{-1}$) | $V_{det}$ (km s$^{-1}$) | $P$ (GPa) | $I_{sp}$ (s) | OB (%) |
|---|---|---|---|---|---|
| PEP | 6.10 | 9.090 | 37.6 | 264.2 | –14.6 |
| PEP-M | 5.86 | 8.729 | 34.0 | 256.4 | –22.8 |
| PEP-H | 5.87 | 8.764 | 34.4 | 256.8 | –22.8 |
| DEP | 6.03 | 8.867 | 35.7 | 257.6 | –27.8 |

## Kamlet–Jacobs calculation protocol

This protocol applies to the four $ABX_4$ materials (PEP, PEP-M, PEP-H, DEP) only. Training-set K–J velocities were taken from the literature sources cited in Table S2. Densities are rounded in Table S22, and the DEP entry uses the remeasured single-crystal density of 1.869 g cm$^{-3}$, tabulated as 1.87 g cm$^{-3}$.

The detonation velocity $D$ (km s$^{-1}$) and detonation pressure $P$ (GPa) were calculated using the Kamlet–Jacobs empirical equations.[S9]

$$D = 1.01\ \Phi^{1/2}(1 + 1.30\rho)$$

$$P = 1.558\ \Phi\rho^2$$

$$\Phi = 31.68\ N(MQ)^{1/2} \quad \text{(S2)}$$

where $\rho$ is the crystal density in g cm$^{-3}$, $N$ is the molar amount of gaseous detonation products per gram of explosive, $M$ is the average molecular weight of the gaseous products, and $Q$ is the heat of detonation in kcal g$^{-1}$.

**Reaction balance: $CO_2$-priority rule.** For a generic energetic salt with elemental composition $C_aH_bN_cCl_dO_e$, the detonation products were assigned in the following fixed priority order, which we refer to throughout this work as the $CO_2$-priority rule:

1. all $c$ nitrogen atoms form $c/2$ $N_2$(g);
2. all $d$ chlorine atoms form $d$ HCl(g), consuming $d$ hydrogen atoms;
3. the remaining $b'=b-d$ hydrogen atoms form $b'/2$ $H_2O$(g), consuming $b'/2$ oxygen atoms;
4. the remaining $e'=e-b'/2$ oxygen atoms oxidize carbon to $CO_2$(g): $min(a,e'/2)$ carbon atoms form $CO_2$(g);
5. any unreacted carbon remains as solid C(s).

Water is treated as gas-phase $H_2O$(g) in the K–J product inventory, consistent with the original Kamlet–Jacobs formulation. For example, balancing PEP = $(C_4H_{12}N_2)[(NH_3CH_2CH_2NH_3)(ClO_4)_4]$ (composite formula $C_6H_{22}Cl_4N_4O_{16}$) under this rule gives

$$C_6\,H_{22}\,Cl_4\,N_4\,O_{16} \rightarrow 4\,HCl(g) + 9\,H_2\,O(g) + 3.5\,CO_2\,(g) + 2.5\,C(s) + 2\,N_2(g). \quad (S3)$$

The same rule reproduces the balanced reactions used for PEP-M, PEP-H, and DEP (Table S22). The $CO_2$-priority rule is fixed across all four entries because a CO-priority variant shifts $D$ by several hundred m s$^{-1}$ at the same $\rho$.

DFT total energies for reactants and the gaseous products in the balanced reaction above were calculated with the DMol$^3$ module in BIOVIA Materials Studio 17.1 using GGA-PBE, DNP basis set, DSPP pseudopotentials, a fine integration grid, restricted spin, 2×1×1 $k$-point sampling, SCF convergence of $1.0\times10^{-6}$ Ha, energy convergence of $1.0\times10^{-5}$ Ha, gradient convergence of $2.0\times10^{-3}$ Ha Å$^{-1}$, displacement convergence of $5.0\times10^{-3}$ Å, and up to 500 geometry-optimization steps. The computed energy difference $\Delta E_{det}$ was converted to heat of detonation using $\Delta H_{det}=1.127\,\Delta E_{det}+0.046$. The oxygen balance based on $CO_2$ was computed as OB[%]=$1600[e-2a-(b-d)/2]/M_W$ for $C_aH_bN_cCl_dO_e$.

**Table S22.** Inputs and outputs used for Kamlet–Jacobs calculations for the $ABX_4$ branch.

| Material | $\rho$ (g cm$^{-3}$) | $Q$ (kJ g$^{-1}$) | $D$ (km s$^{-1}$) | $P$ (GPa) | $\Delta H_f$ (kJ mol$^{-1}$) | $I_{sp}$ (s) | OB (%) |
|---|---|---|---|---|---|---|---|
| PEP | 1.88 | 6.10 | 9.090 | 37.6 | –625.74 | 264.2 | –14.6 |
| PEP-M | 1.82 | 5.86 | 8.729 | 34.0 | –721.37 | 256.4 | –22.8 |
| PEP-H | 1.83 | 5.87 | 8.764 | 34.4 | –715.75 | 256.8 | –22.8 |
| DEP | 1.87 | 6.03 | 8.867 | 35.7 | –541.05 | 257.6 | –27.8 |

## Structural topology and OOD interpretation

At minimum, the $H_2en^{2+}$-based $ABX_4$ branch is stoichiometrically OOD with respect to the $ABX_3$-dominated training set because it changes the composition class from $ABX_3$ to $ABX_4$. The branch also differs topologically from the training structures. In the $ABX_3$ training materials, the A-site dication is enclosed in a cage formed by corner-sharing $BX_3$ octahedra, following the molecular perovskite topology. In the $ABX_4$ branch, the additional X-site anion per formula unit disrupts this cage geometry: the $H_2en^{2+}$ B-site cation is smaller and more flexible than the $NH_4^+$ or alkali-metal B-site cations in the training set, and the resulting packing is stabilized by an extended hydrogen-bonding network rather than by the rigid perovskite framework. Accurate transfer across this topological change thus tests the representation beyond stoichiometric extrapolation alone.

## Template-based pre-synthesis screening

To test whether the same surrogate can screen this branch before crystal growth, we also constructed template-based $ABX_4$ candidates without using the experimentally refined $ABX_4$ crystal structures. Each OOD-new target was built from a DAP-4 $ABX_3$ scaffold by substituting in the target A-site molecule and an $H_2en^{2+}$ B-site molecule, then applying the same stoichiometric vacancy-cluster construction and fold-ensemble inference used for the main OOD-new test.

The template-based and own-CIF OOD-new evaluation is shown in Extended Data Fig. 4.

Under this template-based protocol, the DAP-4-templated inputs give a mean absolute deviation of approximately 100 $m \cdot s^{-1}$ relative to the K–J references, close to the own-CIF value reported in the main text. This agreement indicates that the surrogate retains useful signal when only the target building blocks and a chemically related template are available.

The full-data multi-task model (MT-FT-full, Extended Data Table 1) and the five-fold MT-FT ensemble give consistent rankings on this OOD assessment, whereas ST-FT remains less stable. We therefore use the five-fold MT-FT for the main uncertainty-bearing results and treat MT-FT-full solely as a production-style consistency check.